\pdfoutput=1

\documentclass[11pt]{article}
\usepackage{jheppub}

\usepackage{amsmath}
\usepackage{mathtools}
\usepackage{amssymb}
\usepackage{graphicx,color,slashed,braket}
\usepackage{bbm}
\usepackage{ifpdf}
\usepackage{comment}

\usepackage{booktabs}
\usepackage{multirow}
\usepackage{makecell}
\usepackage{float}
\usepackage{verbatim}
\usepackage{caption}
\usepackage{cancel}
\usepackage{enumerate}
\usepackage[font=small,labelfont=bf]{caption}
\usepackage{tensor}
\usepackage{xcolor}
\usepackage{tikz}
\usepackage[framemethod=tikz]{mdframed}
\usepackage{soul}
\usepackage{comment}

\usepackage{titlesec} 
\titleformat{\paragraph}
{\normalfont \normalsize \itshape}{\theparagraph}{2ex}{}
\titlespacing*{\paragraph}
{0pt}{3.25ex plus 1ex minus .2ex}{1.5ex plus .2ex}

\counterwithin{table}{section}

\newcommand{\be}{\begin{equation}}
\newcommand{\ee}{\end{equation}}
\newcommand{\ba}{\begin{eqnarray}}
\newcommand{\ea}{\end{eqnarray}}

\newcommand{\lp}{\left(}
\newcommand{\rp}{\right)}

\renewcommand{\d}{\textrm{d}}

\renewcommand{\a}{\alpha}
\renewcommand{\b}{\beta}

\def\rmi{{\rm i}}

\newcommand{\ab}{|}

\newcommand{\de}{\mathrm{d}}
\newcommand{\NS}{\ket{\mathrm{NS}}}
\newcommand{\aR}{\ket{a}_{\mathrm{R}}}
\newcommand{\daR}{\ket{\dot{a}}_{\mathrm{R}}}

\newcommand{\sdil}{\phi}

\allowdisplaybreaks

\title{Misaligned Supersymmetry and Open Strings}

\author[a]{Niccol\`o Cribiori,}
\author[b]{Susha Parameswaran,}
\author[b]{Flavio Tonioni}
\author[c,a]{and Timm Wrase}

\affiliation[a]{Institute for Theoretical Physics, TU Wien,\\
Wiedner Hauptstrasse 8-10/136, A-1040 Vienna, Austria}
\affiliation[b]{Department of Mathematical Sciences, University of Liverpool,\\
Mathematical Sciences Building Liverpool, L69 7ZL, UK}
\affiliation[c]{Department of Physics, Lehigh University,\\
16 Memorial Drive East, Bethlehem, PA 18018, USA}

\emailAdd{niccolo.cribiori@tuwien.ac.at}
\emailAdd{susha@liv.ac.uk}
\emailAdd{flavio.tonioni@liv.ac.uk}
\emailAdd{timm.wrase@lehigh.edu}

\notoc

\abstract{

\noindent
The study of non-supersymmetric string theories is shedding light on an important corner of the string landscape and might ultimately explain why, so far, we did not observe supersymmetry in our universe. We review how misaligned supersymmetry in closed-string theories leads to a cancellation between bosons and fermions even in non-supersymmetric string theories. We then show that the same cancellation takes place for open strings by studying an anti-D$p$-brane placed on top of an O$p$-plane in type II string theory. Misaligned supersymmetry consists in cancellations between bosons and fermions at \emph{different} energy levels, in such a way that the averaged number of states grows at a rate dominated by a factor $\mathrm{e}^{C_{\mathrm{eff}}\sqrt{n}}$, with $C_{\mathrm{eff}}<C_{\mathrm{tot}}$, where $C_{\mathrm{tot}}$ is the inverse Hagedorn temperature. We prove the previously conjectured complete cancellation, i.e. we prove that $C_{\mathrm{eff}}=0$, for a vast class of models.

}

\begin{document}

\maketitle

\newpage

\section{Introduction}

Understanding the breaking of supersymmetry within string theory is of the utmost importance, if one aims to describe the real world from a microscopic point of view. In this respect, it is essential to distinguish models in which the spectrum is non-supersymmetric only below a certain energy scale, from those in which supersymmetry cannot be restored, since either it is not present at all or it is broken at the string scale. These models, which can experience only the non-supersymmetric phase, are the subject of the present work. Two notable examples, which we discuss in what follows, are the non-supersymmetric heterotic $\mathrm{SO}(16) \!\times\! \mathrm{SO}(16)$-theory \cite{AlvarezGaume:1986jb,Dixon:1986iz} and 
the open-string system made by an anti-D$p$-brane on top of an O$p$-plane in type II string theory. This second class of models provides also a realization of the general phenomenon called ``brane supersymmetry breaking'' \cite{Sugimoto:1999tx,Antoniadis:1999xk,Angelantonj:1999jh,Aldazabal:1999jr,Angelantonj:1999ms,Dudas:2000ff,Dudas:2000nv,Pradisi:2001yv}, studied also in \cite{Blumenhagen:1998uf,Blumenhagen:1999ns} (see \cite{Mourad:2017rrl} for a recent review).

A central question is how string theory is capable of maintaining finiteness and stability, even in the absence of supersymmetry in the target space. A proposal to answer this question has been given in \cite{Dienes:1994np} and further developed in \cite{Dienes:1995pm,Dienes:2001se,Angelantonj:2010ic,Abel:2015oxa,Abel:2017rch,Abel:2018zyt}. It relies on the idea of misaligned supersymmetry, which explains the finiteness of string theory as a consequence of exponentially growing oscillations between the net number of bosons and fermions at each mass level, leading to cancellations when considering the entire, infinite, tower of states as a whole. Instead, standard supersymmetric spectra lead to finite quantities as a consequence of cancellations taking place at each individual mass level, due to an exact matching in the number of fermionic and bosonic degrees of freedom.

The identification of an almost exact cancellation between spacetime bosons and fermions in the asymptotic density of states in the world sheet theory dates back to the work of \cite{Kutasov:1990sv}. It was subsequently realized \cite{Dienes:1994np} that such a cancellation does not need to be necessarily level by level, as for standard supersymmetry, hence the name \emph{misaligned supersymmetry}. A first extension to open-string models was made in \cite{Niarchos:2000kw}, where it is argued that misaligned supersymmetry (called there ``asymptotic supersymmetry'') is needed to decouple the open-string sector from a closed-string tachyon. Hence, it is again related to stability.

Sufficient conditions for misaligned supersymmetry in closed-string theories were found to be modular invariance and the absence of physical tachyons \cite{Dienes:1994np}. Indeed, it is well known that modular invariance is crucial for closed strings and in particular it dictates how left- and right- moving sectors are coupled. In closed-string theories, the cancellations implied by misaligned supersymmetry occur precisely in accordance with the way modular invariance fixes the couplings among sectors: changing these couplings would in general spoil modular invariance and prevent such cancellations from occurring. In this work, we review the original literature, discussing in detail the non-supersymmetric heterotic $\mathrm{SO}(16) \!\times\! \mathrm{SO}(16)$-theory and showing how the predicted cancellations occur. Then, we turn our attention to open-string models and, in particular, we consider the system in which an anti-D$p$-brane is placed on top of an O$p$-plane in type II string theory. Here, the orientifold involution generically breaks modular invariance down to a subgroup. Nevertheless, we find that misaligned supersymmetry is again at work. In particular, we develop a systematic procedure to show the occurrence of the required cancellations at all orders. This very same procedure can be applied also to the heterotic example, allowing us to prove a previous conjecture. Our results also indicate that brane supersymmetry breaking and misaligned supersymmetry present similar features. This hints at a deeper connection that would substantially improve our understanding of non-supersymmetric type II string theory compactifications.

The interplay between anti-D$p$-branes and O$p$-planes plays an important role in phenomenologically important type II string theory constructions with broken supersymmetry. In particular, an anti-D3-brane in a Calabi-Yau background with an O3-plane is at the core of the KKLT and LVS proposals for de Sitter vacua in string theory \cite{Kachru:2003aw, Conlon:2005ki}. The anti-D3-brane can be placed on top of an O3-plane that is located at the bottom of a warped throat \cite{Kallosh:2015nia,Garcia-Etxebarria:2015lif}. We are studying here an analogous flat-space model and discover a surprising connection to misaligned supersymmetry. More generically, non-supersymmetric branes lead to four-dimensional low energy effective theories with broken supersymmetry, which constitute a much wider class of models than supersymmetric ones \cite{Cribiori:2020bgt} and could lead to phenomenologically interesting non-supersymmetric realisations of the Standard Model \cite{Cascales:2003wn,Parameswaran:2020ukp}. These are the motivations why we decided to revisit and investigate further anti-D$p$-branes and O$p$-planes in string theory, uncovering their connection to misaligned supersymmetry. 

To identify misaligned supersymmetry, we need to study the physical states' net boson-fermion degeneracies encapsulated in the partition function of the theory. These net physical degeneracies can be obtained from a Hardy-Ramanujan-Rademacher sum. To the best of our knowledge, so far misaligned supersymmetry has been demonstrated only by looking at the leading exponentials in the Hardy-Ramanujan-Rademacher sum but here we will study all terms. In order to do that, we specialise our discussion to the case in which the partition function can be written entirely in terms of products of Dedekind $\eta$-functions and, for such a subclass of theories, we show analytically how misaligned supersymmetry is at work at any order of the Hardy-Ramanujan-Rademacher expansion and this allows us to prove that the state degeneracies vanish in an averaged sense.  Once more, the heterotic $\mathrm{SO}(16) \!\times\! \mathrm{SO}(16)$-theory and the type II models with an anti-D$p$-brane on top of an O$p$-plane will serve as two explicit examples. We believe this can be an important step in understanding misaligned supersymmetry as a general property of string theory and of string phenomenology in particular. Heterotic string models exhibiting misaligned supersymmetry have been previously analysed in \cite{Abel:2015oxa, Abel:2017rch, Abel:2017vos, Nibbelink:2015vha, Faraggi:2020wej, Faraggi:2020fwg, Faraggi:2020hpy}.

This work is organized as follows. In section \ref{misaligned supersymmetry in closed strings}, we recall the general approach to misaligned supersymmetry in closed-string theories. Then, we review in detail the non-supersymmetric heterotic ${\rm SO}(16) \!\times\! {\rm SO}(16)$-theory, showing that misaligned supersymmetry is present at leading order. In section \ref{misaligned supersymmetry in open strings}, we then turn our attention to the investigation of misaligned supersymmetry in open-string models. First, we show that the spectrum of an anti-D$p$-brane on top of an O$p$-plane presents several hints that supersymmetry is not just broken but also misaligned. Then, we perform an analysis of the one-loop partition function of an anti-D$p$-brane on top of an O$p$-plane, showing that the previous intuition is indeed correct. In section \ref{sec:sussman}, we then review the Hardy-Ramanujan-Rademacher expansion for partitions functions in the form of Dedekind $\eta$-quotients. Based on this, a discussion of misaligned supersymmetry beyond leading order is presented in section \ref{misaligned SUSY beyond leading order}, in which we also discuss in detail the heterotic $\mathrm{SO}(16) \!\times\! \mathrm{SO}(16)$-theory and an anti-D$p$-brane on top of an O$p$-plane. In section \ref{open-string cosmological constant}, we calculate the one-loop cosmological constant of an anti-D$p$-brane on top of an O$p$-plane and find that it is finite for $0\leq p\leq 6$. In section \ref{conclusions}, we present an outlook on future research directions.

\subsection{Summary of the results}
We present an explicit example of open-string misaligned supersymmetry by showing that the spectrum of an anti-D$p$-brane on top of an O$p$-plane exhibits an oscillating behaviour between bosonic and fermionic states (see Fig. \ref{anti-D/O plot}). The leading exponential growth of both the bosonic and the fermionic states at mass level $n$ is $\mathrm{e}^{C_{\mathrm{tot}}\sqrt{n}}$ with $C_{\mathrm{tot}}=2\pi\sqrt{2}$. Since this is the same, it cancels in an averaged sense and the averaged net number of states grows more slowly, with a coefficient $C_{\mathrm{eff}}<C_{\mathrm{tot}}$. 

The Hardy-Ramanujan-Rademacher sum gives the exact number of states for any $n$ and allows us to study subleading corrections to the leading approximate number of states $\mathrm{e}^{C_{\mathrm{tot}}\sqrt{n}}$. We discover a peculiar property of the phase factor in the Hardy-Ramanujan-Rademacher sum (see our lemma \eqref{closure}). This property translates into oscillations around the leading growth for the number of states in each sector (see Fig. \ref{corrections to anti-D/O plot}). These oscillations are such that they exactly cancel in each sector separately when averaged appropriately. This allows us to prove that $C_{\mathrm{eff}}=0$ in a wide class of models that includes the anti-D$p$-brane on top of an O$p$-plane as well as the heterotic $\mathrm{SO}(16) \!\times\! \mathrm{SO}(16)$-theory.

\section{Misaligned supersymmetry in closed strings} \label{misaligned supersymmetry in closed strings}

In this section, we review the concept of misaligned supersymmetry in closed-string models, following mainly the discussion in \cite{Dienes:1994np,Dienes:2001se}. Then, we analyze the heterotic $\mathrm{SO}(16) \!\times\! \mathrm{SO}(16)$-theory \cite{AlvarezGaume:1986jb,Dixon:1986iz} as an instructive example in which misaligned supersymmetry is at work.

\subsection{The many faces of misaligned supersymmetry}
\label{sec:gendisc}
Misaligned supersymmetry can be understood from different perspectives. It can be formulated as the occurrence of exponentially growing oscillations in the net number of bosonic minus fermionic physical states at each energy level. Equivalently, it can be related to the presence of unphysical tachyons in the partition function, namely virtual excitations with negative squared mass, which are not dangerous as long as they remain off-shell. Alternatively, it can be deduced by looking at the asymptotic behaviour of an appropriately defined sector-averaged number of states that is argued to grow more slowly than the various state degeneracies as the energy increases. Below, we present how all of these concepts are intertwined.

Our starting point is the one-loop torus partition function. For a generic closed-string theory with $D$ non-compact dimensions this can be written in the form 
\begin{equation}
\label{Ztorus}
Z (\tau, \bar \tau) = ({\rm Im} \,\tau)^{1-\frac{D}{2}} \sum_{i,\bar\jmath = 1}^N N_{i\bar\jmath}\,\chi_i (q) \bar\chi_{\bar\jmath}(\bar q), \qquad q = \mathrm{e}^{2\rmi \pi\tau}.
\end{equation}
In consistent physical models, the matrix $N_{i\bar \jmath}$ is restricted by modular invariance. The quantities $\chi_i(q)$ and $\bar \chi_{\bar \jmath}(\bar q)$ are the characters of the left-moving and right-moving highest-weight sectors respectively and form an $N$-dimensional representation of the modular group with modular weight $\kappa\in \mathbb{Z}/2$:
\begin{equation}
    \chi_i(M \tau) = (c\tau+d)^\kappa \sum_{j=1}^N M_{ij}\chi_j(\tau), \qquad M\tau =\frac{a\tau +b}{c\tau +d},
\end{equation}
where $M_{ij}$ is a $N \times N$ matrix representing the modular group in the basis of the $\chi_i$. They can be expressed as power-series in the variable $q$
\begin{equation}
\label{chi}
\chi_i(q) = q^{H_i}\sum_n a_n^{(i)}q^n, \qquad \qquad H_i = h_i - \frac{c}{24},
\end{equation}
where $h_i\geq 0$ is the highest weight and $H_i$ the vacuum energy of the $i^{\mathrm{th}}$-sector, while $c$ is the central charge of the world sheet conformal field theory. The coefficients $a_n^{(i)}$, which are assumed to be non-negative, count the degeneracy of states of the $i^{\rm th}$-sector at the excited level $n$. They are the objects of our primary interest in the present work. 

As long as one can define a set of characters that is closed under the modular $\mathrm{SL}(2,\mathbb{Z})$-group, one can write the general expression of these coefficients $\smash{a_n^{(i)}}$ by means of a Hardy-Ramanujan-Rademacher series as \cite{Hardy:1919, Kani:1989im,Dienes:1994np}
\begin{equation} \label{hardy-ramanujan formula}
    a_n^{(i)} = a_0^{(i)} \sum_\alpha \dfrac{2 \pi}{\alpha} \sum_{j=1}^N Q(\alpha; n)_{ij} f_j (\alpha, i; n),
\end{equation}
where
\begin{subequations}
\begin{align}
    Q(\alpha; n)_{ij} & = \mathrm{e}^{\frac{\rmi \pi \kappa}{2}} \sum_{\substack{0 \leq \beta < \alpha, \\ \mathrm{gcd} (\alpha, \beta) = 1}} (M_{\alpha \beta}^{-1})_{ij} \, \mathrm{e}^{2 \pi \rmi \left( H_j \frac{\beta'}{\alpha} - H_i \frac{\beta}{\alpha}\right)} \, \mathrm{e}^{- 2 \pi \rmi n\frac{\beta}{\alpha}}, \label{Q} \\
    f_j (\alpha, i; n) & = \Bigl[ \dfrac{H_j}{n + H_i} \Bigr]^{\frac{1-\kappa}{2}} \, J_{\kappa-1} \, \Bigl[ \dfrac{4 \pi}{\alpha} \bigl[ H_j (n+H_i) \bigr]^{\frac{1}{2}} \Bigr],
\end{align}
\end{subequations}
with $J_\nu = J_\nu (x)$ being the Bessel functions of the first kind. $a_0^{(i)}$ is the net number of states at $n=0$. $(M_{\alpha \beta})_{ij}$ is the representation of the modular group acting on the characters $\chi_i(q)$ and associated to the $\mathrm{SL}(2,\mathbb{Z})$ element
\begin{equation}
    M_{\alpha \beta} = \left( \begin{array}{cc}
        -\beta' & (1 + \beta \beta')/\alpha \\
        -\alpha & \beta
    \end{array} \right),
\end{equation}
where $\beta'$ is an arbitrary integer parameterising the freedom of acting on $M_{\alpha \beta}$ by acting on the left with the $T$-matrix. In the summation over $\alpha$, each term $\alpha > 1$ gives subleading corrections to the leading-order $\alpha=1$ (to see this, notice e.g. that for $H_j < 0$ the exponential growth is of the form $\mathrm{e}^x$, with $x \approx 4 \pi \ab H_j \ab^{1/2}/\alpha$). For $\alpha=1$, the functions $f_j$ have an asymptotic behaviour
\begin{equation}
\label{fasymp}
    f_j(1,i;n) \overset{n \to \infty}{\approx} f_{j}(1;n)= \left\lbrace\begin{array}{lcl}
        \dfrac{1}{\sqrt{2}} \, \dfrac{1}{2 \pi} \, \ab H_j \ab^{\frac{1-2\kappa}{4}} \, n^{\frac{2\kappa-3}{4}} \, \mathrm{e}^{4 \pi [\ab H_j \ab n]^{\frac{1}{2}}}, & \qquad & H_j < 0; \\[2.5ex]
        (2 \pi n)^{\kappa-1}, & & H_j = 0; \\[1.0ex]
        \dfrac{\sqrt{2}}{2 \pi} \, H_j^{\frac{1-2\kappa}{4}} \, n^{\frac{2\kappa-3}{4}}, & & H_j>0.
    \end{array}\right.
\end{equation}
For the choice $\beta'=0$, one finds $M_{10} = S^{-1}$, so that $Q(1;n)_{ij} = \mathrm{e}^{\rmi \pi \kappa/2} S_{ij}$, $S_{ij}$ being the matrix representing the modular transformation $S$ in the representation of the modular group acting on the characters $\chi_i$. The leading-order ($\alpha=1$) behaviour of the degeneracies $a_n^{(i)}$ as $n \to \infty$ can thus be expressed in the form
\begin{equation}
    a_n^{(i)} \overset{n \to \infty}{\approx} \dfrac{a_0^{(i)}}{\sqrt 2} \, \mathrm{e}^{\frac{\rmi \pi \kappa}{2}} \, n^{\frac{2\kappa-3}{4}} \sum_{j: \; H_j<0} \ab H_j \ab^{\frac{1-2\kappa}{4}} \, S_{ij} \, \mathrm{e}^{4 \pi [\ab H_j \ab n]^{\frac{1}{2}}}.
\end{equation}
For cases where there is only one sector with negative vacuum energy this reduces to
\begin{equation} \label{asymptotic a_n^i}
a_n^{(i)} \overset{n \to \infty}{\approx} A_i \, n^{-B} \mathrm{e}^{C \sqrt{n}},
\end{equation}
where $C = 1/T_H$ is the inverse Hagedorn temperature and
\begin{equation}
A_i = \dfrac{a_0^{(i)}}{\sqrt 2} \, \mathrm{e}^{\frac{\rmi\pi \kappa}{2}} S_{i1} \, \Bigl[ \dfrac{c}{24} \Bigr]^{\frac{1-2\kappa}{4}}, \quad \quad B = \dfrac{3}{4} - \dfrac{\kappa}{2}, \quad \quad C = 4 \pi \Bigl[ \dfrac{c}{24} \Bigr]^{\frac{1}{2}}.
\end{equation}
Without loss of generality, the sectors have been ordered such that $i=1$ corresponds to the identity sector, having $h_1 = 0$ and $H_1 = -c/24 < 0$, where $c$ is the central charge. An important and well-known result is that the entries $S_{i1}$ are always non-vanishing, namely $S_{i1}\neq 0$.\footnote{This can be seen, for example, from the Verlinde formula \cite{Verlinde:1988sn}
\begin{equation}
N_{ijk} = \sum_l \frac{S_{il}S_{jl}S_{lk}}{S_{l1}},
\end{equation}
which expresses the structure constants $N_{ijk}$ appearing in the fusion rules in terms of $S_{ij}$ and it requires $S_{i1}\neq 0$ for consistency.} As a consequence, the (real) coefficients $A_i$ will be non-vanishing as well and, within each sector, the degeneracy of states grows exponentially with the energy (as long as $H_1 < 0$). All sectors experience the same exponentially growing behaviour, since they are all related to the identity sector by the coefficients $S_{i1}\neq 0$.

To understand better how misaligned supersymmetry works, we need to look more closely at the partition function. To this purpose, we can insert \eqref{chi} into \eqref{Ztorus} and obtain
\begin{equation}
    Z (\tau, \bar \tau) = (\mathrm{Im} \,\tau)^{1-\frac{D}{2}} \sum_{i,{\bar \jmath}} \, N_{i \bar \jmath} \sum_{m=H_i}^\infty \sum_{n=H_{\bar \jmath}}^\infty \, a_{m-H_i}^{(i)} \bar{a}_{n-H_{\bar \jmath}}^{(\bar \jmath)} \, q^m \bar{q}^n.
\end{equation}
Notice that now the indices $m-H_i$ and $n-H_{\bar \jmath}$ of the left and right states degeneracies depend on the sectors $i$, $\bar \jmath$ being considered. Then, provided they are defined (i.e. provided both $\smash{a_{n-H_i}^{(i)}}$ and $\smash{\bar{a}_{n-H_j}^{(\bar \jmath)}}$ exist in the original series for the given $n$), the net physical degeneracies of the theory at the mass level $M_n^2 = 4n/\alpha'$ are
\begin{equation} \label{level-2n degeneracy}
    a_{nn} = \sum_{i,{\bar{\jmath}}} \, N_{i \bar{\jmath}} \, a_{n-H_i}^{(i)} \bar{a}_{n-H_j}^{(\bar \jmath)} = \sum_{i,{\bar{\jmath}}} \, N_{i \bar{\jmath}} \, a_{nn}^{(i \bar{\jmath})}.
\end{equation}
These coefficients are important in order to define the partition function computed only over the physical states, i.e.
\begin{equation} \label{physical partition function}
    \int_{-\frac12}^\frac12 \de {\rm Re} \,\tau \; Z(\tau,\bar \tau) = ({\rm Im} \,\tau)^{1-\frac{D}{2}} \sum_{n} \, a_{nn} \, q^n \bar{q}^n.
\end{equation}
Indeed, we recall that physical states are given by $m=n$, while $m\neq n$ correspond to unphysical states, which drop out from physical quantities when imposing the level matching condition.

From the Hardy-Ramanujan-Rademacher formula \eqref{hardy-ramanujan formula}, we see that tachyonic states with $H_j<0$ have an exponentially growing behaviour, while states with $H_j \geq 0$ are power-law suppressed, since $\kappa=1-D/2<0$, for $D>2$. Therefore, the fastest growing contribution in the sum \eqref{level-2n degeneracy} is given by the terms $l=\bar l=1$ for all the sectors, which have $H_1 = H_{\bar 1} = - c/24$. In fact, thanks to the expansion \eqref{asymptotic a_n^i} (notice that the shift $n\to n+H_i$ performed in \eqref{level-2n degeneracy} does not affect the asymptotic behaviour in the large-$n$ limit, in which $n + H_i \approx n$), we find
\begin{equation}
\label{annii2}
a_{nn}^{(i\bar \jmath )} \overset{n\to \infty}{\approx} \dfrac{1}{2} a_0^{(i)} \bar{a}_0^{(\bar{\jmath})} S_{i1} \bar{S}_{\bar{1} \bar{\jmath}} \; n^{\frac{3}{2}-\kappa} \; \mathrm{e}^{C_{\mathrm{tot}} \sqrt n}, 
\end{equation}
and thus we can write
\begin{equation}
\label{ctotdef}
C_{\mathrm{tot}} = 4 \pi \lp |H_1|^{\frac{1}{2}} + |H_{\bar 1}|^{\frac{1}{2}} \rp = \lim_{n\to\infty} \frac{\mathrm{log} \, |a_{nn}^{(i\bar\jmath )}|}{\sqrt n}.
\end{equation}
The definition of $C_{\mathrm{tot}}$ does not make a distinction in the indices $i, \bar\jmath$, since all sectors grow with the same exponential behaviour, as explained before. Misaligned supersymmetry predicts then that these asymptotic exponential behaviours cancel when summing over all sectors in a specific way, leading to an effective exponential growth governed by $C_{\mathrm{eff}}< C_{\mathrm{tot}}$, as we are going to explain now.

The presence of exponentially growing numbers of states can lead to divergences in physical quantities. Misaligned supersymmetry avoids this by predicting the occurrence of certain cancellations, taking place in a specific generalization of the quantities $a^{(i\bar\jmath)}_{nn}$, namely a sector-averaged number of state which grows more slowly than \eqref{annii2}. To construct this quantity, we formally replace the asymptotic state degeneracies $a_{nn}^{(i\bar \jmath )}$ with the functional forms $\Phi^{(i\bar \jmath)}(n)$:
\begin{equation}
\label{Phirepl}
a_{nn}^{(i\bar \jmath)} \;\; \rightarrow \;\; \Phi^{(i\bar \jmath)}(n).
\end{equation} 
The crucial point here is that, while the index $n$ in $a_{nn}^{(i\bar \jmath )}$ tacitly depends also on the sectors $i, \bar \jmath$, as it is clear from \eqref{level-2n degeneracy}, the functional forms are $\Phi^{i\bar\jmath}(n)$ are defined for all positive, real $n$. In other words, the argument $n$ in $\Phi^{i\bar\jmath}(n)$ is independent from $i,\bar \jmath$. For practical purposes, it is convenient to take the variable $n$ to be continuous after the replacement \eqref{Phirepl}. However, while this works at leading order in the Hardy-Ramanujan-Rademacher expansions, one can check that the functions $\Phi^{(i\bar \jmath)}(n)$ are in general not real when subleading orders are considered. For this reason, in section \ref{sec:sussman} we will propose a different procedure in order to formulate the problem in a consistent manner beyond the leading exponentials. Since assuming $\Phi^{(i\bar \jmath)}(n)$ to be continuous functions of $n$ helps in visualising the cancellations at leading order, we will proceed with this assumption for the time being. Once the quantities $\Phi^{(i\bar \jmath)}(n)$ are introduced, we can define the sector-averaged number of states, $\langle a_{nn} \rangle $, as the sum of these functions over all sectors in the theory, namely
\begin{equation}
\label{sad}
\langle a_{nn} \rangle = \sum_{i,\bar\jmath} N_{i \bar\jmath} \Phi^{( i\bar\jmath )}(n),
\end{equation}
then analogously to \eqref{ctotdef} we have
\begin{equation} \label{Ceff}
C_{\mathrm{eff}} = \lim_{n\to\infty} \frac{\mathrm{log} \, |\langle a_{nn} \rangle|}{\sqrt n}.
\end{equation}
It is now clear that, in the case in which 
\begin{equation}
\label{ceffctot}
C_{\mathrm{eff}} < C_{\mathrm{tot}},
\end{equation}
the asymptotic growth of the sector-averaged number of states is slower than that of the state degeneracies within each sector. If this happens, it means that cancellations occur indeed when considering the sum over all sectors and misaligned supersymmetry is at work. In other words, the difference between $C_{\mathrm{tot}}$ and $C_{\mathrm{eff}}$ can be explained as follows: although in each sector the coefficients $a_{nn}^{(i \bar \jmath)}$ grow at a rate fixed by $C_{\mathrm{tot}}$, there are cancellations which make the effective coefficients $\langle a_{nn} \rangle$ grow at the rate fixed by $C_{\mathrm{eff}}$. Notice that the cancellations in $\langle a_{nn} \rangle$ occur among sectors which are misaligned, since each of the $H_i$ can be integer or half integer as well (the only constraint imposed by modular invariance is that $H_i = H_{\bar \jmath} \, {\rm mod}\,1$, if $N_{i\bar\jmath}\neq 0$).

In \cite{Dienes:1994np}, the result \eqref{ceffctot} is proven to hold for generic modular invariant partition functions like \eqref{Ztorus}, having $\kappa<0$. In addition, it is \emph{conjectured} that 
\begin{equation}
\label{Ceff=0}
C_{\mathrm{eff}} = 0,
\end{equation}
which implies that the sector-averaged number of states does not grow exponentially, but at most polynomially. We will refer to \eqref{ceffctot} as the weak form of misaligned supersymmetry, while the conjecture \eqref{Ceff=0} will be denoted as its strong form. Notice that \eqref{Ceff=0} implies the occurrence of cancellations also at all subleading orders. This is a highly non-trivial fact and in the present work we argue that it is indeed the case in a large class of string theory models. In particular, in section \ref{sec:sussman} we will contextualise the specific framework which we will refer to and in section \ref{misaligned SUSY beyond leading order} we will then see the details of how misaligned supersymmetry can work at any subleading order and test it in specific examples, i.e. in the heterotic ${\rm SO}(16) \!\times\! {\rm SO(16)}$-theory and in open-string models with an anti-D$p$-brane on top of an O$p$-plane.

We can understand the cancellations implied by misaligned supersymmetry also from a more physical point of view. Knowing the expressions of $a_n^{(i)}$, with the prescription given above one can calculate the explicit form of the function $\langle a_{nn} \rangle$ at leading order, which is 
\begin{equation}
\label{annav}
\langle a_{nn} \rangle = 4 \pi^2 \sum_{i,\bar\jmath} N_{i\bar \jmath } f_i(1;n) \bar f_{\bar \jmath}(1; n)+\dots.
\end{equation} 
Contrary to $S_{i1}$ in \eqref{asymptotic a_n^i}, nothing prevents now some of the coefficients $N_{i\bar{\jmath}}$ from vanishing. When  $N_{i\bar \jmath }\neq 0$, three different situations can occur. If $H_i\geq 0$ and $H_{\bar \jmath}\geq 0$, then by looking at the asymptotic behaviour given in \eqref{fasymp}, we see that there is no exponential growth at all.  Indeed, an example of such a case is when the partition function is vanishing as a consequence of spacetime supersymmetry, which excludes both physical and unphysical tachyons. On the contrary, the situation in which $H_i< 0$ and $H_{\bar \jmath}< 0$ is clearly unacceptable, since it implies the existence of physical tachyons. Notice that this case corresponds to the situation in which both the left and right moving sector experience asymptotic exponential growth. Finally, in the situation in which either $H_i \geq 0$ or $H_{\bar \jmath} \geq 0$, which is a milder assumption than the first one, we see that the asymptotic growth is slower than \eqref{annii2} and the condition \eqref{ceffctot} is verified. In such a case, physical tachyons are not present, but unphysical ones are. This is the situation leading to misaligned supersymmetry in spacetime.

Finally, let us comment on how \eqref{ceffctot} is related to the presence of oscillations in the net number of bosons minus fermions at each energy level. Of course, for cancellations to take place when summing over all the sectors in \eqref{annav}, some of the coefficients $N_{i\bar \jmath}$ have to be positive, implying the presence of more bosons than fermions, and some have to be negative, implying more fermions than bosons. Along with this logic, we can see directly that \eqref{ceffctot} predicts then an exponentially growing oscillation in the net number of bosons and fermions at each energy level.

\subsection{An heterotic non-supersymmetric string theory model} \label{heterotic SO(16)xSO(16)}
The non-supersymmetric heterotic ${\rm SO}(16) \!\times\! {\rm SO}(16)$-theory, originally constructed in \cite{AlvarezGaume:1986jb,Dixon:1986iz}, is perhaps the prototype example of a non-supersymmetric and yet tachyonic-free closed-string theory. In this section, we show how such a model exhibits the main features of misaligned supersymmetry that we reviewed above.

It is instructive to start by recalling how the heterotic ${\rm SO}(16) \!\times\! {\rm SO}(16)$-theory can be obtained from an orbifold of the heterotic ${\rm E}_8 \times {\rm E}_8$ superstring theory \cite{Gross:1984dd}. The  one-loop torus partition function of the heterotic ${\rm E}_8 \times {\rm E}_8$ model is given by 
\begin{equation}
\label{ZE8E8}
Z_{\mathrm{E}_8\times \mathrm{E}_8} = 
\frac{({\rm Im} \, \tau)^{-4}}{ \eta^8\bar \eta^8} (V_8-S_8)(\bar O_{16}+\bar S_{16})^2.
\end{equation}
Here and in the following we use the definition of ${\rm so}(2n)$ characters given in \cite{Angelantonj:2002ct} and reviewed in appendix \ref{app:thetafunc}. The partition function \eqref{ZE8E8} is vanishing due to the well known Jacobi's \emph{aequatio identica satis abstrusa}, i.e. $V_8 = S_8$. Physically, this is a consequence of spacetime supersymmetry, namely the fact that the number of bosons and fermions is the same at each energy level. The partition function of the heterotic ${\rm SO}(16) \!\times\! {\rm SO}(16)$-theory is obtained by inserting in \eqref{ZE8E8} the projector
\begin{equation}
 P_g = \frac12\left(1+g\right).
\end{equation}
The orbifold generator is $g=(-1)^{F+F_1 + F_2}$, where $F$ is the spacetime fermion number, while $F_{1}$ $(F_2)$ is the fermion number of the first (second) $\mathrm{E}_8$ factor.  The insertion gives
\begin{equation}
    P_g Z_{\mathrm{E}_8 \times \mathrm{E}_8} = \frac12 \big[Z_{\mathrm{E}_8\times \mathrm{E}_8} +  g( Z_{\mathrm{E}_8\times \mathrm{E}_8})\big],
\end{equation}
where
\begin{equation}
    g (Z_{\mathrm{E}_8\times \mathrm{E}_8}) = \frac{({\rm Im} \, \tau)^{-4}}{\eta^8\bar \eta^8} (V_8+S_8)(\bar O_{16}-\bar S_{16})^2
\end{equation}
is obtained by flipping the signs of the $S_8$ and $S_{16}$ sectors in \eqref{ZE8E8}. However, the projected partition function as it stands is not modular invariant. To obtain a modular invariant expression, one acts repeatedly with modular transformations ${T}$ and ${S}$, adding at each step the new terms thus generated, until the final result is modular invariant. Eventually, one finds
\begin{equation}
\begin{split}
\label{ZSO16}
Z_{{\rm SO}(16)\times {\rm SO}(16)} = 
\frac{({\rm Im} \, \tau)^{-4}}{ \eta^8\bar \eta^8} 
\bigg[V_8(\bar O_{16} \bar O_{16} + \bar S_{16}\bar S_{16}) - S_8(\bar O_{16}\bar S_{16}+\bar S_{16}\bar O_{16}) & \\
+ O_8(\bar V_{16}\bar C_{16}+\bar C_{16}\bar V_{16})-C_8(\bar V_{16}\bar V_{16}+\bar C_{16}\bar C_{16}) & \bigg].
\end{split}
\end{equation}
This is the partition function of the heterotic ${\rm SO}(16) \!\times\! {\rm SO}(16)$-theory. It is related to the supersymmetric partition function $Z_{\mathrm{E}_8\times \mathrm{E}_8}$ by 
\begin{equation}
    \frac12\left(Z_{\mathrm{E}_8\times \mathrm{E}_8}+ Z_g\right) = Z_{{\rm SO}(16)\times {\rm SO}(16)},
\end{equation}
where $ Z_g$ is the modular invariant expression
\begin{equation} \label{tildeZg}
\begin{split} 
    Z_g = 
    \frac{({\rm Im} \, \tau)^{-4}}{ \eta^8\bar \eta^8} [(V_8+S_8)(\bar O_{16}-\bar S_{16})^2+(O_8-C_8)(\bar V_{16}+\bar C_{16})^2 \\
    - (O_8+C_8)(\bar V_{16}-\bar C_{16})^2 & ].
\end{split}
\end{equation}
As we will show in the rest of the section, the partition function \eqref{ZSO16} exhibits the features of misaligned supersymmetry. It has been obtained as a projection of a supersymmetric partition function and indeed we will see that we can use a similar logic to construct models with open strings and misaligned supersymmetry. In that case, we will start from an anti-D$p$-brane and enforce on it the projection induced by an O$p$-plane.

\begin{figure}[h]
    \centering
    \begin{tikzpicture}[xscale=0.50,yscale=0.035,bos/.style={draw,circle,minimum size=2.0mm,inner sep=0pt,outer sep=0pt,fill=cyan!85!blue,solid},fer/.style={draw,circle,minimum size=2.0mm,inner sep=0pt,outer sep=0pt,fill=red!40!purple,solid}]
    
    \draw (-1,0) -- (0,0) node[below left]{$0$};
    \draw[-|] (0,0) -- (10,0) node[below]{$10$};
    \draw[-|] (10,0) -- (20,0) node[below]{$20$};
    \draw[->] (20,0) -- (22,0) node[below]{$n$};
    \draw[-|] (0,-100) -- (0,-80) node[left]{$-80$};
    \draw[-|] (0,-80) -- (0,-40) node[left]{$-40$};
    \draw[-|] (0,-40) -- (0,40) node[left]{$40$};
    \draw[-|] (0,40) -- (0,80) node[left]{$80$};
    \draw[->] (0,80) -- (0,100) node[left]{$\pm \mathrm{log} \, (\pm g_n)$};
    
    \draw [dotted] (0,-7.6553906) node[fer]{} -- (1/2,11.900647423601159) node[bos]{} -- (1,-15.366115243076267) node[fer]{} -- (3/2,18.408583000458208) node[bos]{} -- (2,-21.16315375576227) node[fer]{} -- (5/2,23.701531329417083) node[bos]{} -- (3,-26.071600201723697) node[fer]{} -- (7/2,28.306272654497494) node[bos]{} -- (4,-30.42800544006649) node[fer]{} -- (9/2,32.45315896542731) node[bos]{} -- (5,-34.39463173175853) node[fer]{} -- (11/2,36.26276936232364) node[bos]{} -- (6,-38.06581469304425) node[fer]{} -- (13/2,39.81047549708508) node[bos]{} -- (7,-41.50239175867394) node[fer]{} -- (15/2,43.14636015252391) node[bos]{} -- (8,-44.74646208528521) node[fer]{} -- (17/2,46.306206168844504) node[bos]{} -- (9,-47.828654419373954) node[fer]{} -- (19/2,49.31649921058091) node[bos]{} -- (10,-50.77211468669064) node[fer]{} -- (21/2,52.1976071452932) node[bos]{} -- (11,-53.5948603173675) node[fer]{} -- (23/2,54.96556760649477) node[bos]{} -- (12,-56.311255964000686) node[fer]{} -- (25/2,57.6333079356119) node[bos]{} -- (13,-58.932981463653924) node[fer]{} -- (27/2,60.21142533800566) node[bos]{} -- (14,-61.46969148321978) node[fer]{} -- (29/2,62.708746045460195) node[bos]{} -- (15,-63.92947932141723) node[fer]{} -- (31/2,65.13271396300998) node[bos]{} -- (16,-66.31921182246194) node[fer]{} -- (33/2,67.48968008671878) node[bos]{} -- (17,-68.64477677753732) node[fer]{} -- (35/2,69.78511546662668) node[bos]{} -- (18,-70.91126933662045) node[fer]{} -- (37/2,72.02377482645566) node[bos]{} -- (19,-73.12313491322418) node[fer]{} -- (39/2,74.20982199427318) node[bos]{} -- (20,-75.2842804251451) node[fer]{};
    
    \node[bos] at (22,100){};
    \node[right] at (22,100){\, bosons};
    \node[fer] at (22,90){};
    \node[right] at (22,90){\, fermions};

\end{tikzpicture}

\caption{The net number of physical degrees of freedom for the lightest energy levels of the heterotic $\mathrm{SO}(16) \! \times \! \mathrm{SO}(16)$-theory, defined as $(-1)^{F_n} g_n = N_b(n) - N_f(n)$. Each point corresponds to string states with mass $M^2_n = 4n/\alpha'$, for $n=0,1/2,1, \dots, 20$. A positive value indicates a surplus of bosonic states compared to the fermionic ones, and vice versa for negative values. The presence of two misaligned sectors is clearly visible. In particular, red dots are associated to integers values of $n$, while blue dots to half-integers values. As predicted by misaligned supersymmetry, we observe an exponentially growing oscillation between the net number of bosons and fermions.}

\label{fighet1}

\end{figure}
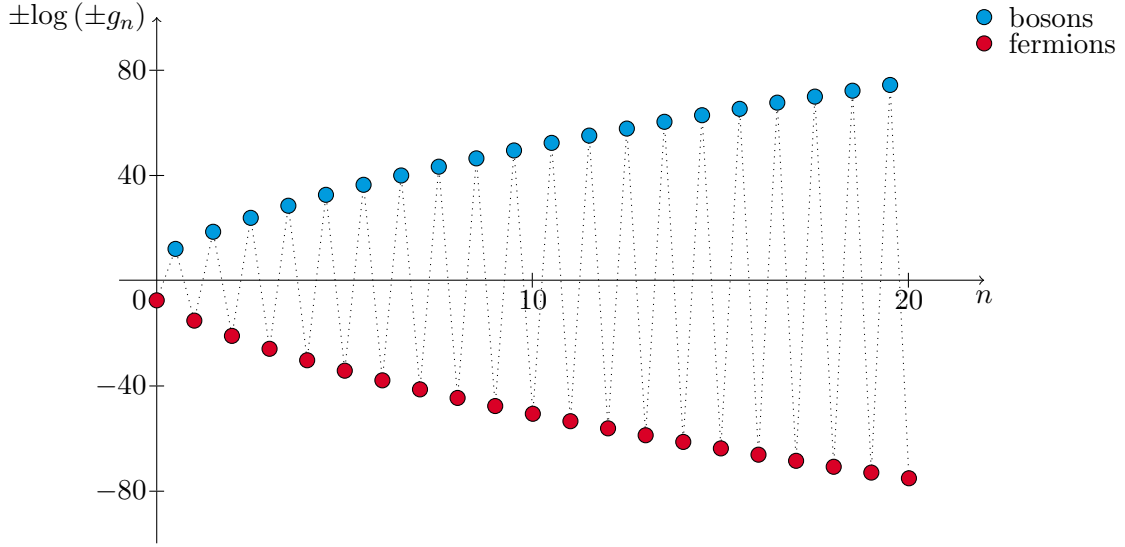

Expanding the partition function \eqref{ZSO16} in powers of $q$ and $\bar{q}$, we can infer the partition function restricted to physical states, i.e. those with equal powers of $q$ and $\bar q$. This can be defined as $\hat{Z}(\tau_2) = \int_{-1/2}^{1/2} \de \tau_1 \; Z(\tau_1,\tau_2)$ and has an expansion that reads
\begin{equation}
\label{Zhetexp}
\begin{split}
    \hat{Z}_{{\rm SO}(16)\times {\rm SO}(16)} = ({\rm Im} \, \tau)^{-4} \Bigl[ - 2112 + 147456 \,(q \bar q)^\frac12 - 4713984\, q\bar q + \mathcal{O}(q \bar{q})^{3/2} \Bigr].
\end{split}
\end{equation}
Therefore, we can easily recognize one of the features of misaligned supersymmetry, i.e. an exponentially growing oscillation in the net number of bosons minus fermions at each physical energy level, as shown in Fig. \ref{fighet1}. Moreover, if one looks more closely at terms with different powers of $q$ and $\bar q$, which are not displayed in the expansion above, one would see that some of them have in fact negative powers of $q$ and/or $\bar q$. These correspond to unphysical tachyons and are yet another signal of misaligned supersymmetry, as discussed previously.

A more quantitative way to study the presence of misaligned supersymmetry is by employing explicitly the formalism presented in the previous section. Looking at the very form of \eqref{ZSO16} and comparing it with the general formula \eqref{Ztorus}, a convenient basis for the left and right moving characters $\chi_i$ and $\bar \chi_{\bar\jmath}$ is
\begin{subequations}
\begin{align}
    \chi_i & = \frac{1}{\eta^8}\left( \begin{array}{c}
    O_8\\
    V_8\\
    S_8\\
    C_8
    \end{array}\right) = 
    \left(
    \begin{array}{c}
    q^{-\frac12} + 36 q^{\frac{1}{2}}+\dots\\
    8+128q+\dots\\
    8+128q+\dots\\
    8+128q+\dots
    \end{array}\right), \label{chiL} \\[1.5ex]
    \bar\chi_{\bar \jmath} & = \frac{1}{\bar \eta^8}\left(\begin{array}{c}
    \bar O_{16} \bar O_{16} +\bar S_{16}\bar S_{16}\\
    \bar V_{16}\bar C_{16}+\bar C_{16}\bar V_{16}\\
    \bar V_{16}\bar V_{16} +\bar C_{16} \bar C_{16}\\
    \bar O_{16}\bar S_{16}+\bar S_{16} \bar O_{16}
    \end{array}\right)
    =\left(
    \begin{array}{c}
    \bar q^{\, -1}+248+\dots\\
    4096 \bar q^{\, \frac{1}{2}}+245760 \bar q^{\, \frac32}+\dots\\
    256+36864\bar q+\dots\\
    256+36864\bar q+\dots
    \end{array}
    \right).
\end{align}
\end{subequations}
This basis is chosen in such a way that the characters are eigenfunctions of $T$, they are covariant under $S$ and have a series expansion with positive coefficients. In this basis, the matrix $N_{i\bar\jmath}$ appearing in \eqref{Ztorus} is given by
\begin{equation}
\label{Nij}
N = \left(\begin{array}{cccc}
0&1&0&0\\
1&0&0&0\\
0&0&0&-1\\
0&0&-1&0
\end{array}
\right),
\end{equation}
while the modular transformations $T$ and $S$ act respectively as
\begin{equation}
\label{modL}
T = \mathrm{diag}(-1,1,1,1), \qquad \qquad
S = \frac12(-\rmi\tau)^{-4}\left(
\begin{array}{cccc}
1&1&1&1\\
1&1&-1&-1\\
1&-1&1&-1\\
1&-1&-1&1\\
\end{array}
\right)\!,
\end{equation}
on the left moving sector $\chi_i$ and
\begin{equation}
\label{modR}
\bar T =  \mathrm{diag}(1,-1,1,1), \qquad \qquad
\bar S=\frac12(-\rmi\bar{\tau})^{-4}\left(
\begin{array}{cccc}
1&1&1&1\\
1&1&-1&-1\\
1&-1&1&-1\\
1&-1&-1&1
\end{array}
\right)\!,
\end{equation}
on the right moving sector $\bar \chi_{\bar\jmath}$. To be consistent with the previous discussion, we have ordered the elements in the basis in such a way that the identity sector in both left and right moving sectors resides in the first component of the characters $\chi_i$ and $\bar \chi_{\bar\jmath}$.

Since $S_{i1}\neq 0$ and $\bar S_{\bar \imath\bar 1}\neq 0$, from the general discussion in subsection \ref{sec:gendisc} we expect that each sector $a_{nn}^{(i\bar\jmath)}$ grows with an exponential behaviour dictated by formula \eqref{asymptotic a_n^i}, i.e.\footnote{Here and in the following, we read the matrices $S_{ij}$ from the $S$-transformations on the characters, omitting powers of $\tau$.}
\begin{equation}
\label{anniihet}
a_{nn}^{(i\bar\jmath)} \overset{n\to \infty}{\approx} A_i A_j \, (2n)^{- \frac{11}{2}} \, \mathrm{e}^{2\pi (\sqrt{2}+2)\sqrt n},
\end{equation}
or in other words the inverse Hagedorn temperature is
\begin{equation}
    C_{\mathrm{tot}} = 2\pi (\sqrt 2+2),
\end{equation}
where we used \eqref{annii2}, with $H_1 = -\frac12$ and $H_{\bar 1} = -1$ since the heterotic string theory has $c_L= 12$ and $c_R =24$.

When considering the sum over all sectors entering the partition function, if misaligned supersymmetry is present, as the oscillations in Fig.~\ref{fighet1} hint, then cancellations are expected in the sector-averaged number of states. To verify that this is indeed the case for the system under investigation, we introduce first the functional forms $\Phi^{(i\bar \jmath)}(n)$ associated to $a_{nn}^{(i\bar\jmath )}$. These are given by 
\begin{equation} \label{Phi^ij - SO16 SO16}
    \Phi^{(i\bar\jmath)} (n) = 4 S_{i1} \bar S_{\bar \jmath \bar 1} \,(2n)^{-\frac{11}{2}} \, \mathrm{e}^{2\pi (\sqrt 2 + 2) \sqrt n}+\phi^{(i\bar\jmath)} (n),
\end{equation}
where the functions $\phi^{(i\bar\jmath)} (n)$ stand for the subleading terms. Then, using the explicit form of matrices $S$ and $\bar S$ in \eqref{modL} and \eqref{modR}, one can check that all the leading exponentials cancel when summing over all sectors, indeed
\begin{equation} \label{cancellation}
\sum_{i,\bar \jmath} N_{i\bar\jmath} S_{i1}\bar S_{\bar\jmath \bar1} = S_{1 1}\bar S_{\bar 2\bar 1} + S_{2 1}\bar S_{\bar 1\bar 1} - S_{3 1}\bar S_{\bar 4\bar 1}-S_{4 1}\bar S_{\bar 3\bar 1} = 0.
\end{equation}
This means that the sector-averaged number of states \eqref{sad} is determined by the subleading terms $\phi^{i \bar \jmath}(n)$, 
\begin{equation}
\begin{split}
    \begin{split}
        \langle a_{nn} \rangle & = \sum_{i,\bar\jmath} N_{i \bar\jmath} \Phi^{( i\bar\jmath )}(n) = \sum_{i,\bar\jmath} N_{i\bar\jmath} \phi^{( i\bar\jmath )}(n),
      \end{split}
\end{split}
\end{equation}
which can have different asymptotic behaviours in different sectors, but whose exponential growth, is fixed by a coefficient which is by definition smaller than $C_{\mathrm{tot}}$. This result shows that, in the heterotic $\mathrm{SO}(16) \!\times\! \mathrm{SO}(16)$-theory, misaligned supersymmetry is present in its weak form, leading to a sector-averaged number of physical states growing at a rate $C_{\mathrm{eff}}<C_{\mathrm{tot}}$.

The next step would be to check if in fact misaligned supersymmetry is present in its strong from, namely if $C_{\mathrm{eff}}=0$. Proving this conjecture requires a careful analysis of the subleading contributions to each sector. As explained previously, this cannot be performed immediately with the functional forms $\Phi^{(i\bar\jmath)}(n)$, as the latter are not a priori real for $\alpha>1$. In section \ref{sec:sussman} we will develop the necessary tools to perform such an analysis under certain conditions and we will come back to this point in section \ref{misaligned SUSY beyond leading order}.

\section{Misaligned supersymmetry in open string theories} \label{misaligned supersymmetry in open strings}
The previous discussion involved explicitly closed strings and a major role was played by the underlying modular invariance of the torus partition function. In the following, we will focus on a particular class of open-string models, in which we identify the features of misaligned supersymmetry. These models are obtained by placing an anti-D$p$-brane on top of an O$p$-plane in type II string theory and are examples of the general framework of brane supersymmetry breaking constructions.

\subsection[The perturbative spectrum of an anti-Dp-brane on top of an Op-plane]{The perturbative spectrum of an anti-D\texorpdfstring{$\boldsymbol{p}$}{$p$}-brane on top of an O\texorpdfstring{$\boldsymbol{p}$}{$p$}-plane} \label{spectrum perturbative analysis}
To motivate our choice of the system with an anti-D$p$-brane sitting on top of an O$p$-plane, we now show that the mass levels of the perturbative spectrum indeed present an increasing oscillation in the (net) number of bosons and fermions. This strongly supports the idea that misaligned supersymmetry is underlying it, as we will show in subsection \ref{sec:DpbonOp}.

\subsubsection{Alternance between fermions and bosons}
We consider the perturbative spectrum of an anti-D$p$-brane sitting on top off an O$p$-plane. In lightcone quantization, we have the NS- and R-vacua $\NS$ and $\aR$, $\daR$, as well as the bosonic and fermionic creation operators $\alpha_{-n}^I$ and $b_{-r}^I$, for $n \in \mathbb{N}^+$ and $r \in \mathbb{N}_0^+ + \phi$, with $\phi=\{\frac{1}{2}, 0\}$ in the NS- and R-sectors, respectively. The index $I$ denotes all directions but the gauge-fixed ones. The generic mass formulae in the NS- and R-sectors read
\begin{subequations}
    \begin{align}
        \alpha^\prime M_{\mathrm{NS}}^2 & = N^{(\alpha)} + N^{(b)}_{\mathrm{NS}} - \dfrac{1}{2}, \\
    \alpha^\prime M_{\mathrm{R}}^2 & = N^{(\alpha)} + N^{(b)}_{\mathrm{R}},
    \end{align}
\end{subequations}
where $\smash{N^{(\alpha)}}$ and $\smash{N^{(b)}}$ are the lightcone level number operators
\begin{equation}
    N^{(\alpha)} = \sum_{m \,\in\, \mathbb{N}^+} \delta_{IJ} \, \alpha^I_{-m} \alpha^J_m , \qquad \qquad \qquad N^{(b)} = \sum_{s \,\in\, \mathbb{N}_0^+ + \phi} s \, \delta_{IJ} \, b^I_{-s} b^J_s.
\end{equation}
The physical states are those invariant under the GSO-projection and the action of the orientifold operator $O$. We perform our analysis by studying the action of the orientifold on the GSO-invariant states, based on the discussion in \cite{BLT}. It can be shown that the orientifold acts on the vacua as\footnote{The orientifold operator does not square to the identity on the NS-vacuum since $O^2 \NS = - \NS$. However, this state is removed by the GSO projection.}
\begin{equation}
    O \NS = \mathrm{e}^{-\frac{\rmi \pi}{2}} \NS, \qquad \qquad \qquad O \ket{\mathrm{R}} = \ket{\mathrm{R}},
\end{equation} 
where $\smash{\ket{\mathrm{R}}}$ is either of the Ramond vacua, and on the creation operators as
\begin{equation}
    O \alpha_{m}^I O^{-1} = (-1)^m \alpha_{m}^I, \qquad \qquad \qquad O b_{r}^I O^{-1} = \mathrm{e}^{\rmi \pi r} b_{r}^I.
\end{equation}
This is sufficient to determine that the pattern followed by the spectrum is a fermionic/bosonic alternance at each massive level. We can prove this as follows.
\begin{itemize}
    \item The generic NS-state of mass $\smash{\alpha^\prime M^2 = n}$ requires a total number of excitations $\smash{n+1/2}$ and can be written as
    \begin{equation}
        \ket{\mathrm{NS}_n} = \alpha^{I_1}_{-n_1} \dots \alpha^{I_k}_{-n_k} b^{J_1}_{-r_1} \dots b^{J_l}_{-r_l} \, \NS, \quad \; \textrm{with} \;\; \sum_{i=1}^k n_i + \sum_{j=1}^l r_j = n + 1/2.
    \end{equation}
    In this way, we can write the orientifold action as
    \begin{equation}
        \begin{split}
            O \ket{\mathrm{NS}_n} =(-1)^{n + 1} \ket{\mathrm{NS}_n}.
        \end{split} 
    \end{equation}
    \item The generic R-state of mass $\smash{\alpha^\prime M^2 = n}$ requires a total number of excitations $\smash{n}$ and can be written as
    \begin{equation}
        \ket{\mathrm{R}_n} = \alpha^{I_1}_{-n_1} \dots \alpha^{I_k}_{-n_k} b^{J_1}_{-r_1} \dots b^{J_l}_{-r_l} \, \ket{\mathrm{R}}, \quad \; \textrm{with} \;\; \sum_{i=1}^k n_i + \sum_{j=1}^l r_j = n,
    \end{equation}
    where $\smash{\ket{\mathrm{R}}}$ is either of the Ramond vacua, depending on the GSO-projection. Then, for an anti-D$p$-brane one can observe the action
    \begin{equation}
        \begin{split}
            O \ket{\mathrm{R}_n} = (-1)^n \ket{\mathrm{R}_n}.
        \end{split} 
    \end{equation}
\end{itemize}
The pattern followed by the states of an anti-D$p$-brane sitting on an orientifold O$p$-plane that we find has thus a pure fermionic/bosonic alternance: in contrast to the locally supersymmetric spectrum of an anti-D$p$-brane at a smooth internal point, levels with mass $\alpha^\prime M^2 = 2n$ contain all the fermions of the spectrum but no bosons, while levels with mass $\alpha^\prime M^2 = 2n+1$ contain all the bosons of the spectrum but no fermions.

The analysis of the spectrum of a D$p$-brane on top of an orientifold plane is identical except for the fact that the orientifold operator acts with an opposite sign on the Ramond vacuum, which implies that the projection on the fermions is opposite. This means that, in contrast to the spectrum of a D$p$-brane at a smooth internal point, levels with mass $\alpha^\prime M_n^2 = 2n+1$ contain all the bosons and the fermions of the spectrum, while levels with mass $\alpha^\prime M_n^2 = 2n$ contain neither the bosons nor the fermions.

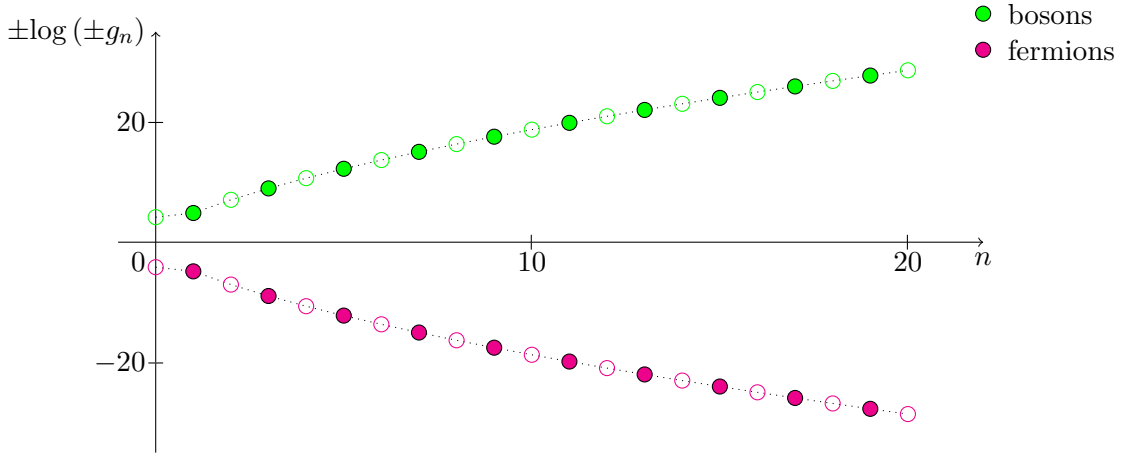
\begin{figure} [h]
    \centering
    \begin{tikzpicture}[xscale=0.5,yscale=0.08,bos/.style={draw,circle,minimum size=2mm,inner sep=0pt,outer sep=0pt,fill=green,solid},fer/.style={draw,circle,minimum size=2mm,inner sep=0pt,outer sep=0pt,fill=magenta,solid},gbos/.style={draw,circle,minimum size=2mm,inner sep=0pt,outer sep=0pt,green,solid},gfer/.style={draw,circle,minimum size=2mm,inner sep=0pt,outer sep=0pt,magenta,solid},zer/.style={draw,circle,minimum size=2mm,inner sep=0pt,outer sep=0pt,black,fill=orange,solid}]
    
    \draw[white] (0,0) -- (34,0);
    
    \draw (-1,0) -- (0,0) node[below left]{$0$};
    \draw[-|] (0,0) -- (10,0) node[below]{$10$};
    \draw[-|] (10,0) -- (20,0) node[below]{$20$};
    \draw[->] (20,0) -- (22,0) node[below]{$n$};
    \draw[-|] (0,-35) -- (0,-20) node[left]{$-20$};
    \draw[-|] (0,-20) -- (0,20) node[left]{$20$};
    \draw[->] (0,20) -- (0,35) node[left]{$\pm \mathrm{log} \, (\pm g_n)$};

    \draw [dotted] (0,4.15888308) node[gbos]{} -- (1,4.852030263919617) node[bos]{} -- (2,7.049254841255837) node[gbos]{} -- (3,8.946374826141717) node[bos]{} -- (4,10.648088014684989) node[gbos]{} -- (5,12.208310140470365) node[bos]{} -- (6,13.659502153634902) node[gbos]{} -- (7,15.023099217200523) node[bos]{} -- (8,16.31417809100231) node[gbos]{} -- (9,17.543847620246574) node[bos]{} -- (10,18.720586679634717) node[gbos]{} -- (11,19.851049252007453) node[bos]{} -- (12,20.940576396593297) node[gbos]{} -- (13,21.99353602270788) node[bos]{} -- (14,23.013556682034682) node[gbos]{} -- (15,24.003693365249816) node[bos]{} -- (16,24.966548083601058) node[gbos]{} -- (17,25.90435950977719) node[bos]{} -- (18,26.81907090081016) node[gbos]{} -- (19,27.71238242905967) node[bos]{} -- (20,28.585792100987767) node[gbos]{};
    
    \draw [dotted] (0,-4.15888308) node[gfer]{} -- (1,-4.852030263919617) node[fer]{} -- (2,-7.049254841255837) node[gfer]{} -- (3,-8.946374826141717) node[fer]{} -- (4,-10.648088014684989) node[gfer]{} -- (5,-12.208310140470365) node[fer]{} -- (6,-13.659502153634902) node[gfer]{} -- (7,-15.023099217200523) node[fer]{} -- (8,-16.31417809100231) node[gfer]{} -- (9,-17.543847620246574) node[fer]{} -- (10,-18.720586679634717) node[gfer]{} -- (11,-19.851049252007453) node[fer]{} -- (12,-20.940576396593297) node[gfer]{} -- (13,-21.99353602270788) node[fer]{} -- (14,-23.013556682034682) node[gfer]{} -- (15,-24.003693365249816) node[fer]{} -- (16,-24.966548083601058) node[gfer]{} -- (17,-25.90435950977719) node[fer]{} -- (18,-26.81907090081016) node[gfer]{} -- (19,-27.71238242905967) node[fer]{} -- (20,-28.585792100987767) node[gfer]{};
    
    \node[bos] at (22,38){};
    \node[right] at (22,38){\, bosons};
    \node[fer] at (22,32){};
    \node[right] at (22,32){\, fermions};

    \end{tikzpicture}
    
    \caption{The number of bosonic and fermionic physical degrees of freedom for the lightest energy levels for a D$p$-brane on top of an O$p$-plane, defined as $(-1)^{F_n} g_n = N_b(n) - N_f(n)$. Each point corresponds to states with mass $M_n^2 = n / \alpha'$, with $n=0,1,\dots,20$. Filled points correspond to states that are invariant under the orientifold projection, whereas empty dots represent states that would be there if the D$p$-brane was at a smooth point but that are projected out by the orientifold. The number of bosonic and fermionic states is the same at each mass level and the partition function vanishes as required by supersymmetry. }
    
    \label{D/O plot}
    
\end{figure}

\begin{figure} [h]
    \centering
    
    \begin{tikzpicture}[xscale=0.5,yscale=0.08,bos/.style={draw,circle,minimum size=2mm,inner sep=0pt,outer sep=0pt,black,fill=green,solid},fer/.style={draw,circle,minimum size=2mm,inner sep=0pt,outer sep=0pt,black,fill=magenta,solid},gbos/.style={draw,circle,minimum size=2mm,inner sep=0pt,outer sep=0pt,green,solid},gfer/.style={draw,circle,minimum size=2mm,inner sep=0pt,outer sep=0pt,magenta,solid},zer/.style={draw,circle,minimum size=2mm,inner sep=0pt,outer sep=0pt,black,fill=orange,solid}]

    \draw[white] (0,0) -- (34,0);
    
    \draw (-1,0) -- (0,0) node[below left]{$0$};
    \draw[-|] (0,0) -- (10,0) node[below]{$10$};
    \draw[-|] (10,0) -- (20,0) node[below]{$20$};
    \draw[->] (20,0) -- (22,0) node[below]{$n$};
    \draw[-|] (0,-35) -- (0,-20) node[left]{$-20$};
    \draw[-|] (0,-20) -- (0,20) node[left]{$20$};
    \draw[->] (0,20) -- (0,35) node[left]{$\pm \mathrm{log} \, (\pm g_n)$};
    
    \draw [dotted,white] (0,4.15888308) node[gbos]{} -- (2,7.049254841255837) node[gbos]{} -- (4,10.648088014684989) node[gbos]{} -- (6,13.659502153634902) node[gbos]{} -- (8,16.31417809100231) node[gbos]{} -- (10,18.720586679634717) node[gbos]{} -- (12,20.940576396593297) node[gbos]{} -- (14,23.013556682034682) node[gbos]{} -- (16,24.966548083601058) node[gbos]{} -- (18,26.81907090081016) node[gbos]{} -- (20,28.585792100987767) node[gbos]{};
    
    \draw [dotted,white] (1,-4.852030263919617) node[gfer]{} -- (3,-8.946374826141717) node[gfer]{} -- (5,-12.208310140470365) node[gfer]{} -- (7,-15.023099217200523) node[gfer]{} -- (9,-17.543847620246574) node[gfer]{} -- (11,-19.851049252007453) node[gfer]{} -- (13,-21.99353602270788) node[gfer]{} -- (15,-24.003693365249816) node[gfer]{} -- (17,-25.90435950977719) node[gfer]{} -- (19,-27.71238242905967) node[gfer]{};

    \draw [dotted] (0,-4.15888308) node[fer]{} -- (1,4.852030263919617) node[bos]{} -- (2,-7.049254841255837) node[fer]{} -- (3,8.946374826141717) node[bos]{} -- (4,-10.648088014684989) node[fer]{} -- (5,12.208310140470365) node[bos]{} -- (6,-13.659502153634902) node[fer]{} -- (7,15.023099217200523) node[bos]{} -- (8,-16.31417809100231) node[fer]{} -- (9,17.543847620246574) node[bos]{} -- (10,-18.720586679634717) node[fer]{} -- (11,19.851049252007453) node[bos]{} -- (12,-20.940576396593297) node[fer]{} -- (13,21.99353602270788) node[bos]{} -- (14,-23.013556682034682) node[fer]{} -- (15,24.003693365249816) node[bos]{} -- (16,-24.966548083601058) node[fer]{} -- (17,25.90435950977719) node[bos]{} -- (18,-26.81907090081016) node[fer]{} -- (19,27.71238242905967) node[bos]{} -- (20,-28.585792100987767) node[fer]{};
    
    \node[bos] at (22,38){};
    \node[right] at (22,38){\, bosons};
    \node[fer] at (22,32){};
    \node[right] at (22,32){\, fermions};

    \end{tikzpicture}
    
    \caption{The number of bosonic and fermionic physical degrees of freedom for the lightest energy levels for an anti-D$p$-brane on top of an O$p$-plane, defined as $(-1)^{F_n} g_n = N_b(n) - N_f(n)$. Each point corresponds to states with mass $M_n^2 = n / \alpha'$, with $n=0,1,\dots,20$. Filled points correspond to states that are invariant under the orientifold projection, whereas empty dots represent states that would be there if the anti-D$p$-brane was at a smooth point but that are projected out by the orientifold. One clearly sees the presence of misaligned supersymmetry.}
    
    \label{anti-D/O plot}
    
\end{figure}
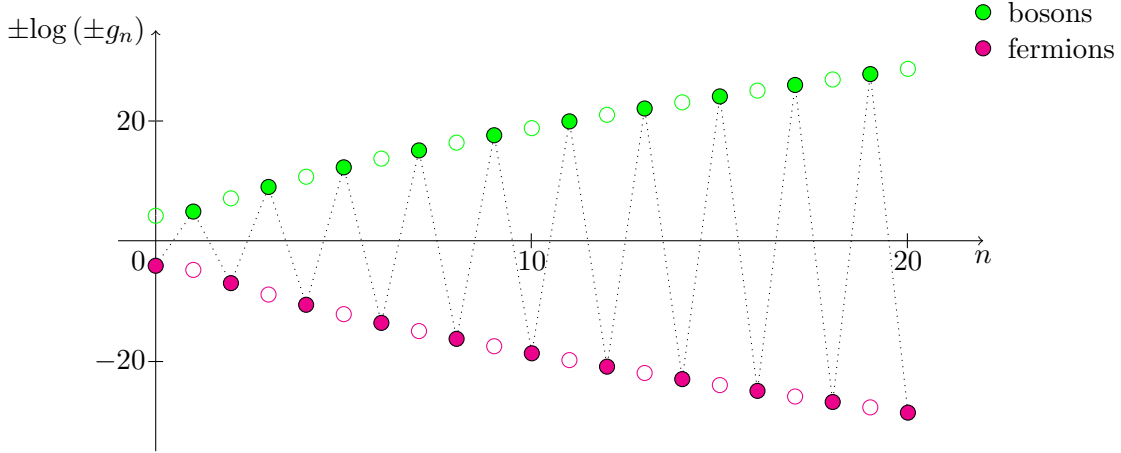

Below, Figs. \ref{D/O plot} and \ref{anti-D/O plot} show the number of physical states for a D$p$- and an anti-D$p$-brane sitting on top on an O$p$-plane, respectively. Notice that, whilst the pattern is clear from the discussion above, it is in the following subsections that we will explain how to compute the degeneracies at each level.

 \vspace{-8pt}

\subsubsection{Number of degrees of freedom by level}
Given the general form of the states in the NS- and R-sectors, a combinatoric analysis allows us to determine the number of the fermionic and bosonic degrees of freedom, $\smash{n_f^{(m)}}$ and $\smash{n_b^{(m)}}$ respectively, at each level.

For each level $\alpha^\prime M^2 = m$, one needs to account for all the possible ways in which it is possible to excite the vacuum giving that mass and to keep in consideration all the symmetrisation and antisymmetrisation factors that are implied by the creation operators. A careful analysis indicates that the number of fermionic degrees of freedom in the Ramond-sector at the level $m$ is expressible as
\begin{equation}
    \dfrac{1}{8} \, N(m) = \sum_{r=0}^m \left\lbrace \sum_{\lambda \,\in\, P(m-r)} \sum_{\mu \,\in\, P(r)} \Biggl[  \prod_{j=1}^{m-r} \prod_{l=1}^{r} \binom{8 + n_j^{(m-r)} - 1}{n_j^{(m-r)}} {\binom{8}{n_l^{(r)}}} \Biggr] \right\rbrace,
\end{equation}
where $P(k)$ denotes the set of all the partitions of the integer number $k$, with $\smash{n_j^{(k)}}$ representing the coefficients in the partition $\lambda$ written as $\smash{k(\lambda) = \sum_{j=1}^k j n_j^{(k)}}$.

At the end of the day, this is just a formal way to summarize the results of a counting which, if needed, can be performed explicitly at the desired level. Thanks to the bosonic-fermionic alternation the net degeneracy at level $m$ turns out to be
\begin{equation}
    (-1)^{F_m} g_m = (-1)^{m+1} N(m).
\end{equation}
From an explicit analysis of the formulae above up to the fourth mass level, we observe that the spectrum gives the degeneracies $g_0=8$, $g_1=128$, $g_2=1152$ and $g_3=7680$. Therefore, we find oscillations in the net numbers of bosons and fermions: a necessary condition for misaligned supersymmetry.

\subsection[Anti-Dp-brane on top of an Op-plane in type II string theory]{Anti-D\texorpdfstring{$\boldsymbol{p}$}{$p$}-brane on top of an O\texorpdfstring{$\boldsymbol{p}$}{$p$}-plane in type II string theory}
\label{sec:DpbonOp}

The one-loop partition function of oriented closed strings is given by the torus, the sole closed orientable Riemann surface with vanishing Euler character. When considering unoriented open (and closed) strings, on the other hand, the situation becomes more interesting. There are indeed three additional Riemann surfaces with vanishing Euler character, with holes, boundaries and crosscaps: the annulus, the M\"obius strip and the Klein bottle. For example, the partition function of a D$p$-brane in flat space is given by the annulus, while that of a (anti-)D$p$-brane on top of an O$p$-plane is encoded in the M\"obius strip. In this subsection, we briefly review the first of these setups and then we discuss in detail the second.

The partition function of a D$p$-brane in flat space is given by the annulus \cite{Polchinski:1996na, Uranga:1999ib, Dudas:2001wd}  
\begin{equation}
\begin{aligned}
    A_p(t) & = \frac{\mathcal{V}_{p+1}}{(2 t)^{\frac{p+1}{2}}} \frac{1}{\eta^8}\left(V_{p-1}O_{9-p} + V_{9-p}O_{p-1} - S_{p-1}S_{9-1}-C_{p-1}C_{9-p}\right)\left[{\rm i} t\right]
    \end{aligned}
\end{equation}
and it can be simplified to
\begin{equation} \label{A_Dp}
  A_p(t) = \frac{\mathcal{V}_{p+1}}{(2 t)^{\frac{p+1}{2}}} \frac{\left(V_8-S_8\right)}{\eta^8}\left[{\rm i} t\right].
\end{equation}
We defined the dimensionless formal D$p$-brane world volume as $\mathcal{V}_{p+1} = V_{p+1}(4 \pi^2 \alpha^\prime)^{-\frac{p+1}{2}}$. Here and in the following, between squared brackets we indicate the argument of the Dedekind function and of the ${\rm so}(2n)$ characters. For the annulus, this is related to the closed-string modulus by $\tau\equiv {\rm i} t$. 
The one-loop amplitude is obtained by integrating over the whole spectrum
\begin{equation}
    \mathcal{A}_p = 2\mathcal{V}_{p+1} \int_0^\infty \frac{dt}{2t} (2 t)^{-\frac{p+1}{2}}\frac{\left(V_8-S_8\right)}{\eta^8}\left[{\rm i} t\right],
\end{equation}
where the overall factor of $2$ is due to the fact that we are summing over the two different orientations of the open string, giving the same contribution. This partition function, and the associated amplitude, are vanishing since $V_8=S_8$. This is a manifestation of the well known fact that a D$p$-brane preserves supersymmetry and thus the net number of bosons minus fermions is vanishing at each energy level. In flat space, there is no real distinction between a D$p$-brane and an anti-D$p$-brane, indeed the partition function is the same. Things will be different after the inclusion of O$p$-planes.

When a O$p^{\pm}$-plane is introduced\footnote{For an odd number of branes, only O$p^-$-planes are allowed. However, we are keeping here the discussion as generic as possible.}, the Riemann surface of interest is not the annulus anymore, but the M\"obius strip. The partition function of a D$p$-brane on a O$p^\pm$-plane is \cite{Uranga:1999ib, Dudas:2001wd}\footnote{We normalize the M\"obius strip parameter ${\rm i} t$ as in \cite{BLT}. This is twice the parameter used in \cite{Dudas:2001wd, Angelantonj:2002ct}.} 
\begin{equation} \label{M_Dp}
    M_{\mathrm{D}p}(t) = \pm \dfrac{1}{2} (2 t)^{-\frac{p+1}{2}}\frac{\big(\hat V_8-\hat S_8\big)}{\hat \eta^8}\left[{\rm i} t\right]\equiv0,
\end{equation}
which vanishes due to supersymmetry, since $\hat V_8=\hat S_8$. As is customary, we wrote the M\"obius strip using hatted characters. These are defined to be manifestly real as
\begin{equation}
\hat \chi_i \left( {\rm i}t \right) = \mathrm{e}^{-{\rm i}\pi H_i} \chi_i\left( {\rm i}t+\frac12\right) =  q^{H_i} \sum_{n=0}^\infty (-1)^n a_n^{(i)} q^n, \qquad q = \mathrm{e}^{-2\pi t}.
\end{equation}
Indeed, the M\"obius strip has $\tau \equiv{\rm i}t+ 1/2$ with a non-vanishing constant real part. On the one hand, this real part is crucial for misaligned supersymmetry, since it introduces relative signs for the number of states at each mass level. On the other hand, as a consequence of the fixed real part of the argument, $\chi({\rm i} t+1/2)$ acquires a phase, $\mathrm{e}^{{\rm i}\pi H}$, which can be conveniently eliminated by defining the manifestly real quantity $\hat \chi({\rm i} t)$. Anyway, one can easily calculate that for the combinations $V_8/\eta^8$ and $S_8/\eta^8$, the phase is trivial, $\mathrm{e}^{{{\rm i} \pi H}}\equiv 1$, therefore in the following we will avoid using the hatted notation if not needed.

When considering an anti-D$p$-brane, the orientifold projection operator in the partition function gives the opposite sign in the Ramond sector, with respect to the D$p$-brane. This is very much similar to what happened in the example of the heterotic string discussed previously, where the orbifold projection was introducing additional minus signs into a supersymmetric (and hence vanishing) partition function. Therefore, the partition function of an anti-D$p$-brane on an O$p^\pm$-plane is 
\begin{equation} \label{M_antiDp}
    M_{\overline{\mathrm{D}p}}(t) = \pm \dfrac{1}{2} (2 t)^{-\frac{p+1}{2}}\frac{\left(V_8+S_8\right)}{\eta^8}\left[{\rm i} t+\frac12\right]
\end{equation}
and, due to the sign flip, it is not vanishing anymore. The associated amplitude is
\begin{equation}
\begin{aligned}
\mathcal{M}_{\overline{\mathrm{D}p}} &=\pm \mathcal{V}_{p+1}\int_0^\infty \frac{dt}{2t}(2 t)^{-\frac{p+1}{2}}\frac{\left(V_8+S_8\right)}{\eta^8}\left[{\rm i} t+\frac12\right].
\end{aligned}
\end{equation}
A first indication of the presence of misaligned supersymmetry can be obtained by expanding the integrand in powers of $q$. In fact, one just needs the $p$-independent factor
\begin{equation} \label{M(t)}
    M (t) = - \frac12 \frac{V_8 + S_8}{ \eta^8}\left[{\rm i}t + \frac12 \right].
\end{equation}
Indeed, for the O$p^-$-plane case we have (the O$p^+$-plane case has just an overall sign difference)
\begin{equation}
\label{ZaDpexp}
   -\frac12 \frac{\left(V_8+S_8\right)}{\eta^8}\left[{\rm i} t+\frac12\right] =-8+128  q-1152 q^2+7680 q^3-42112 q^4 + \mathcal{O}(q^{5})\,,
\end{equation}
with $q=\mathrm{e}^{-2\pi t}$. We notice an increasing oscillation in the (net) number of bosons and fermions at each energy levels, as shown in Fig. \ref{anti-D/O plot}. As anticipated, the alternating sign giving rise to the oscillation is precisely due to the fixed real part in the argument of the characters, as one can check that setting it to zero would result in the same expansion in powers of $q$, but without any sign flip
\begin{equation}
\label{Zexpplus}
   \frac12 \frac{\left(V_8+S_8\right)}{\eta^8}\left[{\rm i} t\right] =8+128 q+1152 q^2+7680 q^3+42112 q^4 + \mathcal{O}(q^5).
\end{equation}
As done for the heterotic string, we can use the formalism of subsection \ref{sec:gendisc} to perform a more quantitative study of the asymptotic growths of the state degeneracies.

\subsection{Asymptotic number of states}

In the general discussion for closed strings in subsection \ref{sec:gendisc} and in the example in subsection \ref{heterotic SO(16)xSO(16)}, a major role is played by the modular properties of the characters $\chi_i$, $\bar \chi_{\bar \jmath}$. In addition, the very proof that misaligned supersymmetry is present in its weak form in the heterotic ${\rm SO}(16)\!\times\!{\rm SO}(16)$-theory, namely formula \eqref{cancellation}, relied on the particular form of the matrix $N_{i\bar\jmath}$ in \eqref{Nij}, which is constrained by modular invariance. 

One can easily check that the combination $V_8+S_8$ is not closed under $S$-transformations, while $V_8 - S_8=0$ trivially is. This observation has implications on the basis of the characters we have to choose in order to study the presence of misaligned supersymmetry in the anti-D$p$-brane, since we need a basis satisfying the properties stated in subsection \ref{sec:gendisc} (see also \cite{Dienes:1994np} for more details): the $\chi_i$ have to be closed under modular transformations, diagonal under $T$-transformations and with non-negative expansion coefficients $a_n^{(i)}$.

As a starting point, since the combination $V_8+S_8$ does not transform covariantly under modular $S$-transformations, we enlarge the set and consider all of the characters $O_8$, $V_8$, $S_8$ and $C_8$, as done for the left-moving sector of the heterotic string in \eqref{chiL}. Then, we would like to separate from the expansion \eqref{ZaDpexp} the contributions with a positive coefficient (bosons) from those with a negative coefficient (fermions). To this purpose and to avoid dealing explicitly with the arguments, it is convenient to restore the hat notation for the time being. We will put a hat on characters whose argument is $({\rm i}t +1/2)$, while characters without hat have just $({\rm i }t)$, i.e.
\begin{equation}
   \hat \chi_V \equiv \frac{V_8}{\eta^8}\left[{\rm i}  t + \frac12 \right], \qquad   \chi_V \equiv \frac{V_8}{\eta^8}\left[{\rm i}  t  \right], \qquad \hat \chi_S \equiv \frac{S_8}{\eta^8}\left[{\rm i}  t + \frac12 \right], \qquad   \chi_S \equiv \frac{S_8}{\eta^8}\left[{\rm i}  t  \right],
\end{equation}
and similarly for the remaining characters $O_8$ and $C_8$. In  addition, from now on we will specify the analysis to the O$p^-$-plane case, but the case with O$p^+$-planes can be easily obtained by changing the appropriate signs. We define then the combinations 
\begin{equation}
\label{chiq}
    \chi_i^{(b)} =\frac12 ( \chi_i -\hat \chi_i), \qquad \chi_i^{(f)} =\frac12( \chi_i \mp \hat\chi_i), \qquad i=\{O,V,S,C\},
\end{equation}
where the upper sign is for a D$p$-brane, while the lower sign is for an anti-D$p$-brane. Introducing the basis vector
\begin{equation}
    \chi_I=q^{H_I} \sum_n a_n^{(I)}q^n=\left(\begin{array}{c}\chi_i^{(b)}\\ \chi_i^{(f)}\end{array}\right), \qquad I=1,\dots,8,
\end{equation}
the partition function can now be rewritten as
\begin{equation}
    M (t) = \sum_{I=1}^8 c_I \chi_I, 
\end{equation}
with only non-vanishing coefficients $c_2=-c_7=1$. For the D$p$-brane, the two vectors \eqref{chiq} are parallel and the partition function vanishes for the chosen coefficients $c_I$ (using again $V_8=S_8$). For the anti-D$p$-brane we recover instead
\begin{equation}
    M (t) = \chi_V^{(b)}-\chi_S^{(f)} = - \frac12 \frac{V_8 + S_8}{ \eta^8}\left[{\rm i}t + \frac12 \right],
\end{equation}
as desired. This notation is slightly redundant, but it is convenient in proving the presence of misaligned supersymmetry.

We have chosen the basis such that the contribution from $\chi_V^{(b)}$ has a $q$ expansion with odd powers, while that from $\chi_S^{(f)}$ has only even powers, but both $\chi_V^{(b)}$ and  $\chi_S^{(f)}$ have non-negative $a_n^{(I)}$, i.e.
\begin{subequations}
    \begin{align}
    \chi_V^{(b)} & = \frac12 \left(\frac{V_8}{\eta^8}\left[{\rm i}t\right] - \frac{V_8}{\eta^8}\left[{\rm i}t+\frac12\right] \right)  =128q+ 7680q^3+\mathcal{O}(q^5), \\
    \chi_S^{(f)} & = \frac12 \left(\frac{S_8}{\eta^8}\left[{\rm i}t\right] +\frac{S_8}{\eta^8}\left[{\rm i}t+\frac12\right]  \right)  =8+1152 q^2 + \mathcal{O}(q^4).
    \end{align}
\end{subequations}
Therefore, this basis has all the necessary properties in order to apply the formalism of subsection \ref{sec:gendisc}. In particular, $\chi_i^{(b)}$ and $\chi_i^{(f)}$ transform under $S$-transformations according to the same matrix $S_{ij}$ as in \eqref{modL}. The vector $\chi_I$ transforms under $S$ with a block diagonal matrix with non-vanishing entries given by such $S_{ij}$\footnote{The presence of vanishing elements in $\mathbb{S}_{IJ}$ means that this is a basis of pseudo-characters, as defined in \cite{Dienes:1994np}. This subtlety will not affect our discussion here.}
\begin{equation}
    \mathbb{S}_{IJ} = S_{ij} \otimes \mathbb{I}_2, \qquad i,j=\{1,2,3,4\}=\{O,V,S,C\}.
\end{equation}
Therefore, the sectors are coupled to the identity sector $i=1=O$ via $S_{i1}\neq 0$ and their leading exponential contribution to the asymptotic number of states is 
\begin{equation}
   a_n^{(I)} \overset{n\to \infty}{\approx} \frac12 (2n)^{-\frac{11}{4}}  \mathrm{e}^{2\sqrt 2 \pi \sqrt n},
\end{equation}
where we used \eqref{asymptotic a_n^i} with $S_{21}=S_{31}=1/2$, $H_1=-1/2$ and $\kappa = -4$.
From this, we deduce directly the inverse Hagedorn temperature
\begin{equation}
C_{\mathrm{tot}} = 2\sqrt 2 \pi.
\end{equation}

When considering the sum over the sectors $I$, if misaligned supersymmetry is present, as the oscillations in Fig. \ref{anti-D/O plot} suggest, we expect cancellations in the sector-averaged number of states, as it happened in the heterotic string previously analyzed. To verify this, we introduce the functional forms $\Phi^{(I)}$ associated to $a_n^{(I)}$
\begin{equation}
    \Phi^{(I)}(n) = s_{I} (2n)^{-\frac{11}{4}}\mathrm{e}^{2\sqrt 2 \pi \sqrt n} + \phi^{(I)}(n),\qquad s_I=\left\{\begin{array}{ccc} \mathbb{S}_{I1} & \textrm{for}& 1\leq I \leq 4,\\
     \mathbb{S}_{I5} & \textrm{for}& 5 \leq I \leq 8,\end{array}\right.
\end{equation}
where $\phi^{(I)}(n)$ contain the subleading terms. Then, using the explicit coefficients $c_I$ given above and the fact that the only non vanishing entries of $\mathbb{S}_{I1}$ are $S_{i1}\neq 0$, we can see that the leading exponentials in $\Phi^{(I)}(n)$ cancel when summing over all sectors. Indeed
\begin{equation}
    \sum_I c_I s_I  =c_2 S_{21} +c_7 S_{31} = S_{21} - S_{31}= 0
\end{equation}
and then
\begin{equation}
\langle a_n \rangle  =  \sum_I c_I \Phi^{(I)}(N) =  \sum_I c_I \phi^{(I)}(N).
\end{equation}
This proves that misaligned supersymmetry is at work for the anti-D$p$-brane on top of an O$p$-plane in type II string theory and 
\begin{equation}
    C_{\mathrm{eff}} < C_{\mathrm{tot}}.
\end{equation}
We present now a heuristic argument for why moreover the conjecture $C_{\mathrm{eff}}=0$ is expected to be true. We can take advantage of the fact that the partition function of the anti-D$p$-brane differs just in one sign from that of the D$p$-brane, which is vanishing. Recalling that the D$p$-brane partition function is proportional to $V_8 - S_8 = 0$, we can organize its (infinite) series expansion as
\begin{equation}
\begin{aligned}
   0 & \equiv -\frac12 \frac{\left(V_8-S_8\right)}{\eta^8}\left[{\rm i} t +\frac12\right] \\
   & = - \frac{1}{2} \left(- 8 + 128 q - 1152 q^2 + \mathcal{O}(q)^{3}\right) + \frac{1}{2} \left(- 8 + 128 q - 1152 q^2 +\mathcal{O}(q)^{3}\right) \\
   & = - \frac{1}{2}\left[ \left( 8 + 128 q + 1152 q^2 +\mathcal{O}(q)^{3}\right)  - \left( 8 + 128 q + 1152 q^2 + \mathcal{O}(q)^{3}\right) \right] \\
   & = -\frac12 \frac{\left(V_8-S_8\right)}{\eta^8}\left[{\rm i} t \right]
\end{aligned}
\end{equation}
Therefore, we see that we have an exact cancellation between positive and negative contributions. For the D$p$-brane, this cancellation occurs not only in the functional forms, but precisely in the $a_n^{(i)}$ at each energy level. Now, the point is that, modulo overall factors, the coefficients of the anti-D$p$-brane matches precisely with those in either of the two parenthesis above. More precisely, calling $a_n^{(\mathrm{D}p,b)} = - a_n^{(\mathrm{D}p,f)}$ the coefficients of one of the two identical parenthesis in the third line above, we can recover the coefficients of the anti-D$p$-brane as
\begin{equation}
\begin{array}{lcl}
    a_{2n}^{(\overline{\mathrm{D}p}, b)} = 0, & \qquad \qquad & a_{2n+1}^{(\overline{\mathrm{D}p}, b)} = a^{(\mathrm{D}p,b)}_{2n+1}; \\[1.0ex]
    a_{2n}^{(\overline{\mathrm{D}p}, f)} = a^{(\mathrm{D}p,f)}_{2n}, & & a_{2n+1}^{(\overline{\mathrm{D}p}, f)} = 0.
\end{array}
\end{equation}
Due to this map, the functional forms $\Phi^{(I)}$ for the D$p$-brane and for the anti-D$p$-brane should be exactly the same, and therefore we can expect that they cancel for the latter case, due to the vanishing partition function of the former (see Figs. \ref{D/O plot}, \ref{anti-D/O plot}).

We will show that cancellations do indeed occur also at subleading orders in the following sections, in a variety of models including an anti-D$p$-brane on top of an O$p$-plane and the heterotic $\mathrm{SO}(16)\!\times\!\mathrm{SO}(16)$-theory as well. In particular, we will show fairly generically that $C_{\mathrm{eff}}=0$.

\section{Going beyond leading order}
\label{sec:sussman}

In the previous sections, we discussed the presence of misaligned supersymmetry just by looking at the leading exponentials in the asymptotic expansion of the net state degeneracies $a_n^{(i)}$. This was enough to prove that $C_{\mathrm{eff}}< C_{\mathrm{tot}}$. However, one can wonder whether cancellations occur also at subleading orders. Answering this question would prove the conjecture $C_{\mathrm{eff}}=0$.

As explained in subsection \ref{sec:gendisc}, the formalism of the functional forms $\Phi^{(i\bar\jmath)}(n)$ is not well suited when going beyond leading order in the Hardy-Ramanujan-Rademacher expansion. The core of the problem is that it is not clear how to define the terms $Q(\alpha;n)_{ij}$ appearing in the general formula \eqref{hardy-ramanujan formula} when promoting the variable $n$ to be continuous. Indeed, such functions become typically complex, while $\Phi^{(i\bar\jmath)}(n)$ should be a real quantity, since it is counting physical degrees of freedom. In particular, a general prescription for analyzing an arbitrary order in $\alpha$ is extremely complicated to implement due to the intricacies in the definition of the function $Q(\alpha;n)_{ij}$.

For a particular class of partition functions, in \cite{sussman2017rademacher} a more explicit Hardy-Ramanujan-Rademacher formula has been derived for the state degeneracies at all orders. Below, we will review such a result and then show how to recast the partition functions of the heterotic $\mathrm{SO(16)} \! \times \! \mathrm{SO(16)}$-theory and of anti-D$p$-branes on O$p$-planes in the form needed to apply the results of \cite{sussman2017rademacher}. In this framework, in the next section we will be able to provide a general procedure to study cancellations beyond leading order.

\subsection[Rademacher series for eta-quotients]{Rademacher series for \texorpdfstring{$\boldsymbol{\eta}$}{$eta$}-quotients} \label{sussman method}

In \cite{sussman2017rademacher}, an exact expression for the state degeneracies $a_n$ is calculated for partition functions that can be written as $\eta$-quotients, i.e. of the form
\begin{equation}
    Z(\tau) = \prod_{m=1}^\infty \bigl[ \eta (m \tau) \bigr]^{\delta_m},
\end{equation}
where $\{\delta_m\}_{m=1}^\infty$ is a sequence of (positive or negative) integers of which only finitely many are non-vanishing. Before giving such a result, a few definitions are in order. First, one defines the constants $n_0$, $c_1$ and the functions $c_2=c_2(\alpha)$, $c_3=c_3(\alpha)$ as
\begin{align}
    n_0 &= - \dfrac{1}{24} \sum_{m=1}^\infty m \, \delta_m,\\
    c_1 & = - \dfrac{1}{2} \sum_{m=1}^\infty \delta_m, \\
    c_2(\alpha) & = \prod_{m=1}^\infty \biggl[ \dfrac{\mathrm{gcd} (m,\alpha)}{m} \biggr]^{\frac{\delta_m}{2}}, \\
    c_3(\alpha) & = - \sum_{m=1}^\infty \delta_m \, \dfrac{[\mathrm{gcd} (m,\alpha)]^2}{m}.
\end{align}
Then, given the Dedekind sum
\begin{equation}
    s(\beta,\alpha) = \sum_{n=0}^{\alpha-1} \dfrac{n}{\alpha} \biggl( \dfrac{\beta n}{\alpha} - \biggl\lfloor \dfrac{\beta n}{\alpha} \biggr\rfloor - \dfrac{1}{2} \biggr)
\end{equation}
and the function
\begin{equation}
    \varphi(\beta,\alpha) = \mathrm{e}^{-\rmi \pi \, \sum_{m=1}^\infty \delta_m \, s \left( \frac{m \beta}{\mathrm{gcd} \, (m,\alpha)}, \frac{\alpha}{\mathrm{gcd} \, (m,\alpha)} \right)},
\end{equation}
let the function $P_\alpha=P_\alpha(n)$ be
\begin{equation} \label{A_k(n)}
    P_\alpha(n) = \sum_{\substack{0 \leq \beta < \alpha, \\ \mathrm{gcd} \, (\beta,\alpha) = 1}} \mathrm{e}^{- 2 \pi \rmi n \frac{\beta}{\alpha}} ~ \varphi(\beta,\alpha).
\end{equation}
Finally, let the function $G=G(\alpha)$ be
\begin{equation}
\label{G(k)}
    G(\alpha) = \underset{ m \in \mathbb{N}: \; \delta_m \neq 0 }{\mathrm{min}} \,\biggl\lbrace \dfrac{[\mathrm{gcd} \, (m,\alpha)]^2}{m} \biggr\rbrace - \dfrac{c_3(\alpha)}{24}.
\end{equation}
In this setup, the main result of \cite{sussman2017rademacher} is the following theorem.

\emph{Theorem.} If $c_1 > 0$ and $G(\alpha)$ is a non-negative function, then, for an arbitrary integer $n > n_0$, the coefficients $a_n$ in the series expansion
\begin{equation}
     Z(\tau) = q^{-n_0} \sum_{n=0}^\infty a_n q^n
\end{equation}
can be written as
\begin{equation} \label{d(n)}
    a_n = \dfrac{2\pi}{[24(n-n_0)]^{\frac{c_1+1}{2}}} \sum_{\substack{\alpha \in \mathbb{N}^+, \\ c_3(\alpha) > 0}} c_2(\alpha) \, [c_3(\alpha)]^{\frac{c_1+1}{2}} \, \dfrac{P_\alpha(n)}{\alpha} \, I_{c_1+1} \biggl[ \biggl( \dfrac{2 \pi^2}{3 \alpha^2} \, c_3(\alpha) (n-n_0) \biggr)^{\frac{1}{2}} \biggr],
\end{equation}
where $I_\nu$ represents the modified Bessel function of the first kind.

This formula allows us to have control over each of the various contributions (leading and all of the subleading) to a given state degeneracy $a_n$. Because of the asymptotic expansion $\smash{I_\nu(x) \overset{x \to \infty}{\approx} \mathrm{e}^x/(2 \pi x)^{\frac{1}{2}}}$, each value $c_3(\alpha)/\alpha^2$, for $\alpha \in \mathbb{N}^+$, represents a successively subleading exponential correction to the coefficient $a_n$. The asymptotic expression of $a_n$ is 
\begin{equation} \label{asymptotic d(n)}
a_n \overset{n \to \infty}{\approx} \dfrac{1}{8^{\frac{1}{2}}} \dfrac{1}{n^{\frac{2c_1+3}{4}}} \, \biggl[\dfrac{2 c_0}{3}\biggr]^{\frac{2c_1+1}{4}} c_2(\alpha_0) \, P_{\alpha_0}(n) \, \biggl[\dfrac{\alpha_0}{4}\biggr]^{c_1} \, \mathrm{e}^{\pi \bigl[\frac{2 c_0}{3} \, n\bigr]^{\frac{1}{2}}},
\end{equation}
where $c_0=c_3(\alpha_0)/\alpha_0^2$, $\alpha_0$ being the integer maximising $c_3(\alpha)/\alpha^2$.

Compared to the general Hardy-Ramanujan-Rademacher formula \eqref{hardy-ramanujan formula}, we see that in the case of \eqref{d(n)} there is no mixing between different sectors and that all contributions are in the form of the modified Bessel functions of the first kind.

\subsubsection[A lemma for the function P]{A lemma for the function \texorpdfstring{$\boldsymbol{P_\alpha(n)}$}{$P_\alpha(n)$}}
The series coefficients $a_n$ in equation \eqref{d(n)} involve $n$-dependent functions $P_\alpha(n)$ that are defined above in equation \eqref{A_k(n)} and that are sums over phases. Being invariant under the shift $n \to n + m \alpha$ for any $m \in \mathbb{Z}$, namely $P_\alpha(n) = P_\alpha(n + m \alpha)$, these terms $P_\alpha(n)$ can take only $\alpha$ different values, at fixed order $\alpha$. We denote these values as $P_{\alpha}(k)$, with $k=1,\dots,\alpha$. An important property for our forthcoming discussion is that the sum over $k$ of $P_{\alpha}(k)$ is vanishing. This is a consequence of the following lemma.

\emph{Lemma.} Given the integers $m$, $\alpha\in \mathbb{N}$, $n\in \mathbb{N}_0$ and $\gamma = \mathrm{gcd}  (\alpha,m)$, if $\nexists \, p \in \mathbb{N}: \; m = p\alpha$, i.e. if $m$ is not a multiple of $\alpha$ and if $\alpha>1$, then
\begin{equation} \label{closure}
        \sum_{k=0}^{\frac{\alpha}{\gamma}-1} P_\alpha(n+mk) = 0.
\end{equation}
    
The proof is straightforward:
\begin{equation}
\begin{split}
    \sum_{k=0}^{\frac{\alpha}{\gamma}-1} P_\alpha(n+mk) & = \sum_{k=0}^{\frac{\alpha}{\gamma}-1} \sum_{\substack{0 \leq \beta < \alpha, \\ \mathrm{gcd} \, (\beta,\alpha) = 1}} \mathrm{e}^{- 2 \pi \rmi (n+mk) \frac{\beta}{\alpha}} \varphi(\beta,\alpha) \\
    & = \sum_{\substack{0 \leq \beta < \alpha, \\ \mathrm{gcd} \, (\beta,\alpha) = 1}} \mathrm{e}^{-2 \pi \rmi n \frac{\beta}{\alpha}} \varphi(\beta,\alpha) \sum_{k=0}^{\frac{\alpha}{\gamma}-1} \mathrm{e}^{- 2 \pi \rmi m k \frac{\beta}{\alpha}} \\
    & = \sum_{\substack{0 \leq \beta < \alpha, \\ \mathrm{gcd} \, (\beta,\alpha) = 1}} \mathrm{e}^{-2 \pi \rmi n \frac{\beta}{\alpha}} \varphi(\beta,\alpha) \, \biggl[\dfrac{1-\mathrm{e}^{- 2 \pi \rmi m \frac{\beta}{\gamma}}}{1 - \mathrm{e}^{- 2 \pi \rmi m \frac{\beta}{\alpha}}} \biggr]=0,
\end{split}
\end{equation}
where we used the geometric sum $\sum_{n=0}^{s-1} r^n = (1-r^{s})/(1-r)$, with $\smash{r=\mathrm{e}^{- 2 \pi \rmi m \frac{\beta}{\alpha}}}$ and $s=\alpha/\gamma$, and the fact that $\alpha/\gamma, m \beta/\gamma\in \mathbb{N}$. An important subcase is for $n=0$ and $m=1$, giving
\begin{equation}
\label{sublemma}
    \sum_{k=0}^{\alpha-1} P_\alpha(k) = 0.
\end{equation}

This lemma will be used explicitly to prove the cancellations among the various sectors beyond leading order.

\subsection[Heterotic SO(16)xSO(16)-theory in terms of eta-quotients]{Heterotic \texorpdfstring{$\boldsymbol{\mathrm{SO}(16) \!\times\! \mathrm{SO}(16)}$}{$\mathrm{SO}(16) \!\times\! \mathrm{SO}(16)$}-theory in terms of \texorpdfstring{$\boldsymbol{\eta}$}{$eta$}-quotients} \label{SO(16)xSO(16) in terms of eta-quotients}

In this subsection, we show explicitly how to recast the partition function of the heterotic  $\mathrm{SO}(16) \!\times\! \mathrm{SO}(16)$-theory in a form that is suitable for applying formula \eqref{d(n)}. To this purpose, one  essentially needs to employ standard identities for modular functions, but there is also a further subtlety concerning the sign of the function $G(\alpha)$ defined in \eqref{G(k)}, as will be explained.

One can start from the partition function in \eqref{ZSO16}. Suppressing the factor coming from the spacetime momentum integration, i.e. concentrating on the quantity $Z(\tau,\bar \tau)$ defined by $Z(\tau, \bar{\tau})_{\mathrm{SO}(16) \!\times\! \mathrm{SO}(16)} = (\mathrm{Im} \, \tau)^{-4} Z(\tau, \bar{\tau})$, we can express it in terms of Jacobi $\vartheta$-functions as
\begin{equation} \label{SO(16)xSO(16)}
    \begin{split}
        Z(\tau, \bar{\tau}) = \dfrac{1}{2 \eta^{12}(\tau)\bar{\eta}^{24}(\bar{\tau})} \bigl[\vartheta_2^4(\tau) \bar{\vartheta}_3^8(\bar{\tau}) \bar{\vartheta}_4^8(\bar{\tau}) - \vartheta_3^4(\tau) \bar{\vartheta}_2^8(\bar{\tau}) \bar{\vartheta}_4^8(\bar{\tau}) + \vartheta_4^4(\tau) \bar{\vartheta}_2^8(\bar{\tau}) \bar{\vartheta}_3^8(\bar{\tau}) & \bigr].
    \end{split}
\end{equation}
It is convenient to separate the three terms in the sum and factorize the contributions from left and right-movers by writing
\begin{equation}
\label{Z3terms}
    Z(\tau, \bar{\tau}) = \sum_{i=1}^3 Z_i(\tau, \bar{\tau}), \qquad \qquad Z_i (\tau, \bar{\tau}) = \dfrac{1}{2} \, R_i(\tau) \bar{L}_i(\bar{\tau}).
\end{equation}
For completeness, we observe that in terms of the variable $q=\mathrm{e}^{2\pi i \tau}$ one has the expansions
\begin{subequations}
\begin{align}
    R_1(\tau) & = 16 + 256 q + 2304 q^2 + 15360 q^3 + 84224 q^4 + \mathcal{O}(q)^5, \\
    L_1(\tau) & = q^{-1} \bigl[ 1 - 8 q + 36 q^2 - 128 q^3 + 402 q^4 + \mathcal{O}(q)^5 \bigr], \\
    R_2(\tau) & = -q^{-\frac{1}{2}} \bigl[ 1 + 8 q^{\frac{1}{2}} + 36 q + 128 q^{\frac{3}{2}} + \mathcal{O}(q)^{2} \bigr], \\
    L_2(\tau) & = 256 - 4096 q^{\frac{1}{2}} + 36864 q - 245760 q^{\frac{3}{2}} + 1347584 q^2 + \mathcal{O}(q)^{\frac{5}{2}}, \\
     R_3(\tau) & = q^{-\frac{1}{2}} \bigl[ 1 - 8 q^{\frac{1}{2}} + 36 q - 128 q^{\frac{3}{2}} + \mathcal{O}(q)^{2} \bigr], \\
    L_3(\tau) & = 256 + 4096 q^{\frac{1}{2}} + 36864 q + 245760 q^{\frac{3}{2}} + 1347584 q^2 + \mathcal{O}(q)^{\frac{5}{2}}.
\end{align}
\end{subequations}
Now, one can express each Jacobi $\vartheta$-function as a product of Dedekind $\eta$-functions, as reviewed in appendix \ref{app:thetafunc}. Below, we discuss the three terms in \eqref{Z3terms} separately.
\begin{itemize}
    \item In the first product, one has
    \begin{subequations}
        \begin{align}
            R_1(\tau) & = \dfrac{\vartheta_2^4(\tau)}{\eta^{12}(\tau)} = \dfrac{16 \, \eta^{8}(2 \tau)}{\eta^{16}(\tau)}, \\
            L_1(\tau) & = \dfrac{\vartheta_3^8(\tau) \vartheta_4^8(\tau)}{\eta^{24}(\tau)} = \dfrac{\eta^{8}(\tau)}{\eta^{16}(2\tau)}.
        \end{align}
    \end{subequations}
    In both cases, $c_1=4$ and $G(\alpha) \geq 0$, therefore \eqref{d(n)} applies and gives us complete knowledge over all of the subleading contributions. In particular, for $R_1$ one finds $n_0=0$, $c_2(2\alpha+1)=1/16$ and $c_3(2\alpha+1)=12$, with $c_3(2\alpha)=0$, while for $L_1$ one finds $n_0=1$, $c_2(2\alpha)=1$ and $c_3(2\alpha)=24$, with $c_3(2\alpha+1)=0$.
    
    Further, one can easily evaluate the asymptotic forms. For $R_1$, one has $c_0 = 12$ for $\alpha_0=1$, with $c_2(1)=1/16$ and $P_1(n) = 1$ (with an overall factor $16$), while for $L_1$ one finds $c_0 = 6$ for $\alpha_0=2$, with $c_2(2)=1$ and $P_2(n) = (-1)^n$, therefore
    \begin{subequations}
    \begin{align}
        a_n^{R_1} & \overset{n \to \infty}{\approx} \dfrac{1}{4 \cdot 8^{\frac{1}{4}}} \, \dfrac{1}{n^{\frac{11}{4}}} \mathrm{e}^{(8 \pi^2 n)^{\frac{1}{2}}},\\
        a_n^{L_1} & \overset{n \to \infty}{\approx} \dfrac{1}{2} \, \dfrac{(-1)^n}{n^{\frac{11}{4}}} \mathrm{e}^{(4 \pi^2 n)^{\frac{1}{2}}}.
    \end{align}
    \end{subequations}
    The growth of the coefficients of the first term in the partition function reads then
    \begin{equation}
        a^{(1)}_{nn} =  a_{n+n_0}^{R_1} \,  \bar{a}_{n+n_0}^{L_1} \overset{n \to \infty}{\approx} \dfrac{1}{8
        \cdot 8^{\frac{1}{4}}} \, \dfrac{(-1)^{n+1}}{n^{\frac{11}{2}}} \mathrm{e}^{[(8 \pi^2)^{\frac{1}{2}} + (4 \pi^2)^{\frac{1}{2}}] \, n^{\frac{1}{2}}}.
    \end{equation}

    \item In the second product, one has 
    \begin{subequations}
        \begin{align}
            R_2(\tau) & = -\dfrac{\vartheta_3^4(\tau)}{\eta^{12}(\tau)} = -\dfrac{\eta^{8}(\tau)}{\eta^{8}(\tau/2) \eta^{8}(2\tau)}, \\
            L_2(\tau) & = \dfrac{\vartheta_2^8(\tau) \vartheta_4^8(\tau)}{\eta^{24}(\tau)} = \dfrac{256 \, \eta^{16}(\tau/2) \eta^{16}(2\tau)}{\eta^{40}(\tau)}.
        \end{align}
    \end{subequations}
    Clearly, we cannot apply directly \eqref{d(n)} to these expressions. Since it just amounts to an index relabelling, we can consider the argument $\tau'=2\tau$, which gives the functions
    \begin{subequations}
        \begin{align}
            R'_2(\tau) & = R_2(2\tau) = -\dfrac{\eta^{8}(2\tau)}{\eta^{8}(\tau) \eta^{8}(4\tau)}, \\
            L'_2(\tau) & = L_2(2\tau) = \dfrac{256 \, \eta^{16}(\tau) \eta^{16}(4\tau)}{\eta^{40}(2\tau)}.
        \end{align}
    \end{subequations}
    While $c_1=4$ and $G(\alpha) \geq 0$ for $-R'_2(\tau)$, $L'_2(\tau)$ does not have a non-negative function $G(\alpha)$, but it turns out that a further shift $\tilde{\tau} = \tau+1/2$, which amounts to flipping half of the signs in the series expansion (something we can keep track of), happens to have a positive semidefinite function $G(\alpha)$, along with $c_1=4$. So one has to consider the function 
    \begin{equation}
        {\tilde{L}}'_2(\tau) = L'_2(\tau+1/2) = \dfrac{256 \, \eta^{8}(2\tau)}{\eta^{16}(\tau)}.
    \end{equation}
    We can now apply \eqref{d(n)} to these expressions and obtain again a complete understanding of all of the subleading contributions. For $- R'_2$ one finds $n_0=1$, $c_2(2\alpha+1)=16$, $c_2(4\alpha+4)=1$,  $c_3(2\alpha+1)=6$ and $c_3(4\alpha+4)=24$, with $c_3(2 \, \mathrm{mod} \, 4)=0$, while for $\tilde{L}'_2$ one finds $n_0=0$, $c_2(2\alpha+1)=1/16$ and $c_3(2\alpha+1)=12$, with $c_3(2\alpha)=0$.
    
    One can easily evaluate the asymptotic forms. For $-R'_2$, one has $c_0=6$ for $\alpha_0=1$, with $c_2(1)=16$ and $P_1(n)=1$, while for $L'_2$ (the result for $L_2'(\tau)$ can be obtained by studying $L_2'(\tau+1/2)$ and inserting a factor $(-1)^n$) one finds $c_0=12$ for $\alpha_0=1$, $c_2(1)=1/16$ and $P_1(n) =(-1)^n$ (with an overall factor $256$), therefore
\begin{subequations}
\label{anR2}
    \begin{align}
        a^{R'_2}_n & \overset{n \to \infty}{\approx} -\dfrac{1}{2} \, \dfrac{1}{n^{\frac{11}{4}}} \mathrm{e}^{(4 \pi^2 n)^{\frac{1}{2}}},\\
        a^{L'_2}_n & \overset{n \to \infty}{\approx} \dfrac{2 \cdot 2^{\frac{1}{2}}}{2^{\frac{1}{4}}} \, \dfrac{(-1)^n}{n^{\frac{11}{4}}} \mathrm{e}^{(8 \pi^2 n)^{\frac{1}{2}}}.
    \end{align}
    \end{subequations}
    The coefficients of the original functions are actually $a^{R_2}_n=a^{R'_2}_{2n+n_0}$ and $a^{L_2}_n = a^{L'_2}_{2n+n_0}$, so that the growth of coefficients of the second term in the partition function reads
    \begin{equation}
        a^{(2)}_{nn} = a^{R_2}_n \, \bar{a}^{L_2}_n \overset{n \to \infty}{\approx} \dfrac{1}{32 \cdot 2^{\frac{1}{4}}} \, \dfrac{(-1)^{2n+1}}{n^{\frac{11}{2}}} \mathrm{e}^{[(16 \pi^2)^{\frac{1}{2}} + (8 \pi^2)^{\frac{1}{2}}] \, n^{\frac{1}{2}}}.
    \end{equation}

    \item In the third product, one has
    \begin{subequations}
        \begin{align}
            R_3(\tau) & = \dfrac{\vartheta_4^4(\tau)}{\eta^{12}(\tau)} = \dfrac{\eta^{8}(\tau/2)}{\eta^{16}(\tau)}, \\
            L_3(\tau) & = \dfrac{\vartheta_2^8(\tau) \vartheta_3^8(\tau)}{\eta^{24}(\tau)} = \dfrac{256 \, \eta^{8}(\tau)}{\eta^{16}(\tau/2)}.
        \end{align}
    \end{subequations}
    In order to apply \eqref{d(n)}, one can consider the functions
    \begin{subequations}
        \begin{align}
            R'_3(\tau) & = R_3(2\tau) = \dfrac{\eta^{8}(\tau)}{\eta^{16}(2\tau)}, \\
            L'_3(\tau) & = L_3(2\tau) = \dfrac{256 \, \eta^{8}(2\tau)}{\eta^{16}(\tau)}.
        \end{align}
    \end{subequations}
    Indeed, these have $c_1=4$ and $G(\alpha) \geq 0$ as required. Once more, formula \eqref{d(n)} gives us now the complete information on the subleading contributions of this sector. For $R'_3$, one has $n_0=1$, $c_2(2\alpha)=1$ and  $c_3(2\alpha)=24$, with $c_3(2\alpha+1)=0$, while for $L'_3$ one finds $n_0=0$, $c_2(2\alpha+1)=1/16$ and $c_3(2\alpha+1)=12$, with $c_3(2\alpha)=0$.
    
    One can easily evaluate the asymptotic forms. For $R'_3$, one has $c_0=6$ for $\alpha_0=2$, with $c_2(2)=1$ and $P_2(n)=(-1)^n$, while for $L'_3$ one finds $c_0=12$ for $\alpha_0=1$, with $c_2(1)=1/16$ and $P_1(n) =1$ (with an overall factor $256$), therefore
     \begin{subequations}
    \begin{align}
        a^{R'_3}_n & \overset{n \to \infty}{\approx} \dfrac{1}{2} \, \dfrac{(-1)^n}{n^{\frac{11}{4}}} \mathrm{e}^{(4 \pi^2 n)^{\frac{1}{2}}},\\
        a^{L'_3}_n & \overset{n \to \infty}{\approx} \dfrac{2 \cdot 2^{\frac{1}{2}}}{2^{\frac{1}{4}}} \, \dfrac{1}{n^{\frac{11}{4}}} \mathrm{e}^{(8 \pi^2 n)^{\frac{1}{2}}}.
    \end{align}
    \end{subequations}
    The coefficients of the original functions are $a^{R_3}_n=a^{ R'_3}_{2n+n_0}$ and $a^{L_3}_n = a^{L'_3}_{2n+n_0}$, so that the growth of coefficients of the third term in the partition function reads
    \begin{equation}
        a^{(3)}_n = a^{R_3}_n \, a^{L_3}_n \overset{n \to \infty}{\approx} \dfrac{1}{32 \cdot 2^{\frac{1}{4}}} \, \dfrac{(-1)^{2n+1}}{n^{\frac{11}{2}}} \mathrm{e}^{[(16 \pi^2)^{\frac{1}{2}} + (8 \pi^2)^{\frac{1}{2}}] \, n^{\frac{1}{2}}}.
    \end{equation}
    
\end{itemize}

\begin{figure} [H]
    \centering
    \begin{tikzpicture}[xscale=0.5,yscale=0.035,bos1/.style={draw,circle,minimum size=2mm,inner sep=0pt,outer sep=0pt,black,fill=green,solid},fer1/.style={draw,circle,minimum size=2mm,inner sep=0pt,outer sep=0pt,black,fill=magenta,solid},bos2/.style={draw,circle,minimum size=2mm,inner sep=0pt,outer sep=0pt,black,fill=cyan,solid},fer2/.style={draw,circle,minimum size=2mm,inner sep=0pt,outer sep=0pt,black,fill=orange,solid}]
    
   \draw[domain=2:20.5, smooth, thick, variable=\n, lightgray] plot ({\n}, {- ln(8*8^(1/4)) -(11/2)*ln(\n) + ((8*pi^2)^(1/2) + (4*pi^2)^(1/2))*\n^(1/2)}) node[right, black]{$\Phi_1(n)$};
   \draw[domain=2:20.5, smooth, thick, variable=\n, lightgray] plot ({\n}, {ln(8*8^(1/4)) + (11/2)*ln(\n) - ((8*pi^2)^(1/2) + (4*pi^2)^(1/2))*\n^(1/2)}) node[right, black]{$-\Phi_1(n)$};
   \draw[domain=1.25:20.5, smooth, thick, variable=\n, lightgray] plot ({\n}, {- ln(32*2^(1/4)) -(11/2)*ln(\n) + ((16*pi^2)^(1/2) + (8*pi^2)^(1/2))*\n^(1/2)}) node[right, black]{$\Phi_{2,3}(n)$};
   \draw[domain=1.25:20.5, smooth, thick, variable=\n, lightgray] plot ({\n}, {ln(32*2^(1/4)) + (11/2)*ln(\n) - ((16*pi^2)^(1/2) + (8*pi^2)^(1/2))*\n^(1/2)}) node[right, black]{$-\Phi_{2,3}(n)$};
    
    \draw (-1,0) -- (0,0) node[below left]{$0$};
    \draw[-|] (0,0) -- (10,0) node[below]{$10$};
    \draw[-|] (10,0) -- (20,0) node[below]{$20$};
    \draw[->] (20,0) -- (22,0) node[below]{$n$};
    \draw[-|] (0,-100) -- (0,-80) node[left]{$-80$};
    \draw[-|] (0,-80) -- (0,-40) node[left]{$-40$};
    \draw[-|] (0,-40) -- (0,40) node[left]{$40$};
    \draw[-|] (0,40) -- (0,80) node[left]{$80$};
    \draw[->] (0,80) -- (0,100) node[left]{$\pm \mathrm{log} \, (\pm g_n)$};

    \draw [dotted] (0,-6.9314718) node[fer2]{} -- (1/2,11.207866620454752) node[bos2]{} -- (1,-14.673875522834805) node[fer2]{} -- (3/2,17.715413592371306) node[bos2]{} -- (2,-20.46993148671032) node[fer2]{} -- (5/2,23.008379787683545) node[bos2]{} -- (3,-25.378465732978) node[fer2]{} -- (7/2,27.613127356433125) node[bos2]{} -- (4,-29.734854815814174) node[fer2]{} -- (9/2,31.760011535946926) node[bos2]{} -- (5,-33.70148554137352) node[fer2]{} -- (11/2,35.569622103695906) node[bos2]{} -- (6,-37.37266727278105) node[fer2]{} -- (13/2,39.117328368435835) node[bos2]{} -- (7,-40.80924462149305) node[fer2]{} -- (15/2,42.453212962797686) node[bos2]{} -- (8,-44.053314896855454) node[fer2]{} -- (17/2,45.61305898479412) node[bos2]{} -- (9,-47.135507243293574) node[fer2]{} -- (19/2,48.6233520327096) node[bos2]{} -- (10,-50.07896750286573) node[fer2]{} -- (21/2,51.504459964189195) node[bos2]{} -- (11,-52.90171313843709) node[fer2]{} -- (23/2,54.27242042570425) node[bos2]{} -- (12,-55.61810878291711) node[fer2]{} -- (25/2,56.940160755255995) node[bos2]{} -- (13,-58.239834283180365) node[fer2]{} -- (27/2,59.518278157398626) node[bos2]{} -- (14,-60.77654430265658) node[fer2]{} -- (29/2,62.01559886487886) node[bos2]{} -- (15,-63.23633214087178) node[fer2]{} -- (31/2,64.43956678247066) node[bos2]{} -- (16,-65.62606464188137) node[fer2]{} -- (33/2,66.79653290615381) node[bos2]{} -- (17,-67.95162959699068) node[fer2]{} -- (35/2,69.09196828606422) node[bos2]{} -- (18,-70.21812215605578) node[fer2]{} -- (37/2,71.33062764589826) node[bos2]{} -- (19,-72.42998773266476) node[fer2]{} -- (39/2,73.51667481371258) node[bos2]{} -- (20,-74.5911332445854) node[fer2]{};
    
    \draw [dotted] (0,-4.15888308) node[fer1]{} -- (1,8.434463543817241) node[bos1]{} -- (2,-11.901427510399932) node[fer1]{} -- (3,14.942869344934364) node[bos1]{} -- (4,-17.697302948335174) node[fer1]{} -- (5,20.235799788077216) node[bos1]{} -- (6,-22.60588095548532) node[fer1]{} -- (7,24.840534869190503) node[bos1]{} -- (8,-26.962267063205697) node[fer1]{} -- (9,28.98742331154802) node[bos1]{} -- (10,-30.928896257603647) node[fer1]{} -- (11,32.797033537591695) node[bos1]{} -- (12,-34.60007864369988) node[fer1]{} -- (13,36.34473954237465) node[bos1]{} -- (14,-38.03665592860725) node[fer1]{} -- (15,39.68062425889047) node[bos1]{} -- (16,-41.28072615255874) node[fer1]{} -- (17,42.84047026953521) node[bos1]{} -- (18,-44.362918525228366) node[fer1]{} -- (19,45.85076330509255) node[bos1]{} -- (20,-47.30637878233386) node[fer1]{};
    
    \end{tikzpicture}
    \caption{The lightest string states in the heterotic $\mathrm{SO}(16) \!\times\! \mathrm{SO}(16)$-theory. The three interpolating functions $\Phi_i(n)$, for $i=1,2,3$, correspond to the three terms $Z_1$, $Z_2$ and $Z_3$ that combine into the total partition function, and in particular they are simply the degeneracies $a_n^{(i)}$ plotted for a continuous variable $n$. Notice that, although $\Phi_2(n) = \Phi_3(n)$, i.e. $Z_2$ and $Z_3$ contribute equally to physical states, the associated off-shell coefficients are different.}
    \label{SO(16)xSO(16) plot for eta-quotients}
\end{figure}
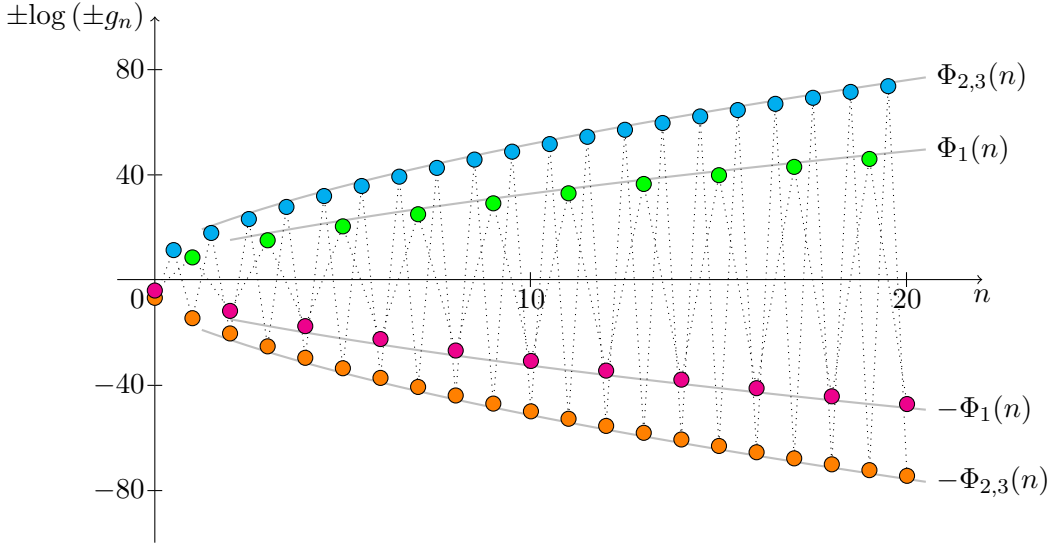

In Fig. \ref{SO(16)xSO(16) plot for eta-quotients} below we have reported a plot of the three different sectors that one individuates when writing the partition function in terms of Dedekind $\eta$-quotients.

It is interesting to highlight that, except for the scaling $\tau \to \tau'= 2 \tau$ and/or the shift $\tau \to \tilde{\tau}= \tau+1/2$, the heterotic $\mathrm{SO(16)} \!\times\! \mathrm{SO}(16)$-theory can be written entirely in terms of two functions. Indeed, one has the identities
\begin{subequations} \label{SO(16)xSO(16) identities}
    \begin{align}
        R'_3(\tau) & = L_1(\tau) = R'_2(\tau+1/2), \\
        L'_3(\tau) & = 16 \, R_1(\tau) = L'_2(\tau+1/2).
    \end{align}
\end{subequations}

Notice that a product of Dedekind $\eta$-functions does not necessarily satisfy the requirements of applicability of the formula \eqref{d(n)}. In the specific case at hand, nevertheless, some manipulations on $\tau$ allowed us to bypass the problem of a non-positive semi-definite function $G(\alpha)$ in \eqref{G(k)}.

\subsection[Anti-Dp-branes on Op-planes in terms of eta-quotients]{Anti-D\texorpdfstring{$\boldsymbol{p}$}{$p$}-branes on O\texorpdfstring{$\boldsymbol{p}$}{$p$}-planes in terms of \texorpdfstring{$\boldsymbol{\eta}$}{$eta$}-quotients} \label{anti-D/O in terms of eta-quotients}

As for the heterotic string in the previous subsection, we would like to rewrite the partition function of an anti-D$p$-brane on top of an O$p$-plane as an $\eta$-quotient, in order then to apply \eqref{d(n)} to study the subleading contributions to the state degeneracies.

The partition function \eqref{M_antiDp} can be simplified by exploiting the properties of the ${\rm so}(2n)$ characters. First, we recall that\footnote{Due to the fact that $\vartheta_1=0$, there is an ambiguity in the following formulae between $S_{2n}$ and $C_{2n}$ when evaluating the characters at $z \neq 0$. This will not play any role in our discussion.}
\begin{equation}
\begin{aligned}
    S_{2n}\left[{\rm i} t+\frac12\right] &= S_{2n}(O_{2n}-V_{2n})[2{\rm i} t],\\
    \eta^{2n}\left[{\rm i}t+\frac12\right] &=\eta^{2n}(O_{2n}+V_{2n})[2{\rm i}t].
\end{aligned}
\end{equation}
These follow for example from formula (9.80) of \cite{BLT} and allow to remove the dependence on the constant real part in the argument. Then, we can recast the Jacobi triple product identity as
\begin{equation}
    (S_{2n}+C_{2n})(O_{2n}+V_{2n})(O_{2n}-V_{2n})=2^{n}.
\end{equation}
In the particular case $n=4$, we have a further simplification since $V_8=S_8$, giving
\begin{equation}
    V_8\left[{\rm i} t+\frac12\right] = V_{8}(O_{8}-V_{8})[2{\rm i} t], \qquad V_8(O_8+V_8)(O_8-V_8)=8.
\end{equation}
Using these relations, we have
\begin{equation}
\label{V8eta8}
    \frac{V_8}{\eta^8}\left[{\rm i}t+ \frac12 \right] = \frac{2^{-3}}{\eta^8}(V_8(O_8-V_8))^2[2{\rm i}t] = \frac{2^3}{\eta^8}(O_8+V_8)^{-2}[2{\rm i}t]=8 \vartheta_3[2{\rm i}t]^{-8}
\end{equation}
and the M\"{o}bius strip amplitude becomes 
\begin{equation}\begin{aligned} \label{Moebius strip amplitude}
    \mathcal{M}_{\overline{\mathrm{D}p}} &=\pm \mathcal{V}_{p+1}\int_0^\infty \frac{dt}{t}(2 t)^{-\frac{p+1}{2}} \frac{V_8}{\eta^8}\left[{\rm i}t+ \frac12 \right]\\
    &= \pm \mathcal{V}_{p+1}\int_0^\infty \frac{dt}{t}(2 t)^{-\frac{p+1}{2}} 8\vartheta_3[2{\rm i}t]^{-8}.
    \end{aligned}
\end{equation}
Therefore, we are interested in the quantity (defined already in \eqref{M(t)})
\begin{equation} \label{M}
    M(\tau) =  - \dfrac{8}{\vartheta_3^{8}(2\tau)} =  - \dfrac{8 \, \eta^{16}(\tau) \, \eta^{16}(4 \tau)}{\eta^{40}(2\tau)}.
\end{equation}
We cannot apply directly formula \eqref{d(n)} to this expression, since the condition $G(\alpha)\geq0$ in \eqref{G(k)} is not satisfied. As for the heterotic string, we can overcome the problem by shifting $\tau \to \tau +1/2$ (this amounts to have an expansion in powers of $q$ without alternating signs), leading to
\begin{equation} \label{M'}
   \tilde M (\tau) = M (\tau+1/2) =  - \dfrac{8 \, \eta^8(2 \tau)}{\eta^{16} (\tau)}.
\end{equation}
For this function, the condition $G(\alpha)\geq0$ is satisfied and therefore \eqref{d(n)} applies and allows us to understand all subleading contributions. In particular, one finds $n_0=0$, $c_1=4$, $c_2(2\alpha+1)=1/16$, $c_3(2\alpha)=0$ and $c_3(2\alpha+1)=12$. One can easily evaluate the asymptotic form. Indeed, one has $c_0 = 12$ for $\alpha_0=1$, $c_2(1)=1/16$ and $ P_1(n) = (-1)^{n+1}$, with an overall factor $8$ (the factor $(-1)^{n+1}$ has been inserted along with the same logic as explained above formula \eqref{anR2}), and the asymptotic form of $a_n$ is
\begin{equation}
    a_n \overset{n \to \infty}{\approx} \dfrac{1}{8 \cdot 8^{\frac{1}{4}}} \, \dfrac{(-1)^{n+1}}{n^{\frac{11}{4}}} \mathrm{e}^{(8 \pi^2 n)^{\frac{1}{2}}}.
\end{equation}

Interestingly, the function \eqref{M'} is related to the heterotic $\mathrm{SO}(16) \!\times\! \mathrm{SO}(16)$-theory by the following identities
\begin{equation}
     - M(\tau+1/2) = \dfrac{1}{2} \, R_1(\tau) = \dfrac{1}{32} \, \tilde L'_2(\tau+1/2) = \dfrac{1}{32} \, L'_3(\tau).
\end{equation}

\section[Cancellations at all orders and Ceff=0]{Cancellations at all orders and \texorpdfstring{$\boldsymbol{C_{\textrm{eff}}=0}$}{$C_{\textrm{eff}}=0$}} \label{misaligned SUSY beyond leading order}

In this section, we finally show that, in the class of models for which the tools presented in section \ref{sec:sussman} can be employed, the cancellations implied by misaligned supersymmetry occur at all orders in the Hardy-Ramanujan-Rademacher expansions and the conjecture $C_{\mathrm{eff}}=0$ holds. The result will rely crucially on the lemma \eqref{closure}. First, we give a general prescription to construct the functional forms $\Phi^{(i)}(n)$ in such a way that they remain explicitly real when $n$ is continuos even beyond leading order, thus overcoming the problem mentioned in subsection \ref{sec:gendisc}. Then, we specialize our discussion to the two systems we analyzed explicitly in the present work: the heterotic $\mathrm{SO}(16) \!\times\! \mathrm{SO}(16)$-theory and an anti-D$p$-brane on top of an O$p$-plane.

\subsection{General procedure}
\label{genproc}
The starting point is the expression for the Laurent coefficients \eqref{d(n)}, which we rewrite here for convenience as
\begin{equation} \label{d(n)2}
\begin{aligned}
    a_n &= \sum_{\substack{\alpha \in \mathbb{N}^+, \\ c_3(\alpha) > 0}} a_n(\alpha)= \sum_{\substack{\alpha \in \mathbb{N}^+, \\ c_3(\alpha) > 0}} P_{\alpha}(n) f_n(\alpha),
    \end{aligned}
\end{equation}
where
\begin{equation}
  f_n(\alpha)=\dfrac{2\pi \, c_2(\alpha) \, [c_3(\alpha)]^{\frac{c_1+1}{2}}}{\alpha [24(n-n_0)]^{\frac{c_1+1}{2}}} \, I_{c_1+1} \biggl[ \biggl( \dfrac{2 \pi^2}{3 \alpha^2} \, c_3(\alpha) (n-n_0) \biggr)^{\frac{1}{2}} \biggr].
\end{equation}
We are again suppressing the indices $i,j$ labelling the various sectors, therefore this quantity should really read $a_n^{(i)}$.\footnote{Note that the number of sectors did change in our discussion. In particular, for the heterotic SO(16)$\times$SO(16) example we wrote the partition function in equation \eqref{ZSO16} using four sectors, while we wrote it in equation \eqref{Z3terms} in terms of three $\eta$-quotients. In this section, we are interested in the number of $\eta$-quotients.} We will restore these indices later on, when it will become necessary, as we will see. The prescription presented in subsection \ref{sec:gendisc} amounts to letting $n$ be a continuous variable, thus promoting $a_n$ to the functional forms $a_n \to \Phi(n)$. However, while the term with $\alpha=1$ in $\Phi(n)$ is real, the terms with $\alpha >1$ can be complex, due to the fact that $P_\alpha(n)$ is a complex number in general.\footnote{It might be helpful to think of the terms $\alpha=1$ and $\alpha>1$ as leading and subleading orders respectively. However, strictly speaking this identification could be misleading in the present context. Indeed, what governs the exponential growth in \eqref{d(n)} is the quantity $c_3(\alpha)/\alpha^2$ and it is not guaranteed that this is maximized at $\alpha=1$. It can also happen that $c_3(1)\leq 0$, and therefore the corresponding term with $\alpha=1$ would not appear in the sum. In fact, such cases did indeed appear already in subsection \ref{SO(16)xSO(16) in terms of eta-quotients} above. Nevertheless, when present, the contribution with $\alpha=1$ has $P_1(n)=1$ and there is only one subsector $k=1$, thus making this case somehow special. In \ref{sec:subtle}, we will analyze these subtleties in more detail, but for the time being we are keeping the discussion as plain as possible.} To overcome this problem and construct a real function $\Phi(n)$ we notice two facts.
\begin{enumerate}[(i)]
    \item First, in general the leading order contribution in $a_n$ can underestimate or overestimate the real value. This means that subleading corrections can come with either positive or negative signs.
    \item Second, as noticed in section \ref{sec:sussman}, for each fixed $\alpha$, there are only $\alpha$ independent real values of the function $P_\alpha(n)$ as $n \in \mathbb{N}$ varies. To stress when we will employ them, we introduce a hat notation
    \begin{equation}
      P_\alpha(n) \equiv \hat{P}_\alpha(k) \in \mathbb{R}, \qquad \forall n \in \mathbb{N}_\alpha^{(k)}, \qquad k=1,\dots,\alpha,
    \end{equation}
    where $\mathbb{N}_\alpha^{(k)} = \lbrace n \in \mathbb{N}: \; n = k \, \mathrm{mod} \, \alpha \rbrace$ are $\alpha$ subsets of $\mathbb{N}$. The lemma \eqref{closure} indicates that the sum of $\hat{P}_\alpha(k)$ over $k=1,\dots,\alpha$ is zero.
\end{enumerate}
Since our aim is to define functional forms that interpolate between the physical degeneracies at discrete $n$, we can define $\alpha$ different subsectors, depending on the value taken by the function $P_\alpha(n)$. In particular, for each of these $\alpha$ different values, we define functional forms $\Phi_k (n;\alpha)$ such that
\begin{equation}
\label{phikn}
   \Phi_k(n;\alpha) =\hat{P}_\alpha(k) \, f_n(\alpha) , \qquad k=1,\dots,\alpha.
\end{equation}
The crucial step here is that we replace the quantities $P_\alpha(n)$, which are complex when $n\in \mathbb{R},$ with the discrete and manifestly real ones $\hat{P}_\alpha(k)$, which are independent of $n$. Therefore, the functions \eqref{phikn} are now real when $n$ is assumed to be a continuous variable. The price we have to pay is that, for each $\alpha>1$ we are in fact introducing $\alpha$ different subsectors within the same sector $i$ (whose index had been understood up to now) 
\begin{equation}
    a_n^{(i)}(\alpha) \rightarrow \Phi^{(i)}_k(n;\alpha) , \qquad k=1,\dots,\alpha.
\end{equation}
The number of these subsectors increases with $\alpha$ in the Hardy-Ramanujan-Rademacher expansion, up to an infinite number of them. As explained above, these subsectors will be populated by positive and negative contributions in general, therefore we expect that cancellations can occur among them. That this is indeed the case is a consequence of the lemma on the functions $\hat{P}_\alpha(k)$. To see this explicitly, we have just to average over the $k=1,\dots,\alpha$, subsectors at fixed $\alpha$.  Then, since $f_n(\alpha)$ does not depend on $k$, at any fixed order $\alpha>1$ we immediately conclude that such an average is vanishing, i.e.
\begin{equation}
    \sum_{k=1}^{\alpha} \Phi_k(n;\alpha) = \left[ \, \sum_{k=1}^\alpha \hat{P}_\alpha(k) \, \right] f_n(\alpha) = 0, \qquad \alpha>1,
\end{equation}
where we used $\sum_{k=1}^\alpha \hat{P}_\alpha(k) = 0$. As shown in the proof of lemma \eqref{closure}, the result holds for every integer $\alpha>1$, including the limit $\alpha \to \infty$. Performing these cancellations for every order in the Hardy-Ramanujan-Rademacher expansions, we are left at most with $\alpha=1$, if it is present in the original expansion \eqref{d(n)}. This is special in some sense, since there are no subsectors associated to it and therefore the mechanisms outlined above cannot work. However, here comes to rescue the presence of other sectors, labelled by $i,j$. Indeed, cancellations among terms with $\alpha=1$ have to occur among different sectors, analogous to the original formulation of misaligned supersymmetry reviewed in subsection \ref{sec:gendisc}. Therefore, due to the cancellations between sectors $i,j$ for $\alpha=1$ and the cancellations between subsectors $k$ for $\alpha>1$, the result $C_{\mathrm{eff}}=0$ follows.

As an example, consider for instance the function $\tilde{M}(\tau)$, defined in \eqref{M'}, that is associated to the description of an anti-D$p$-brane on top of an O$p$-plane. As discussed in subsection \ref{anti-D/O in terms of eta-quotients}, this function is such that all the even values $\alpha = 2l$ in the Hardy-Ramanujan-Rademacher expansions give zero, the only contributions to the Laurent coefficient being from odd values $\alpha=2l+1$. Focusing on the first correction beyond leading order, i.e. $\alpha=3$, we have the complex-valued function
\begin{equation}
    \tilde{P}_3(n) = \mathrm{e}^{-\frac{2 \pi \rmi}{3}(n-2)} + \mathrm{e}^{-\frac{4 \pi \rmi}{3}(n+1)}.
\end{equation}
However, restricting to integer values of $n$, one finds the three possible real values
\begin{equation}
    \tilde{P}_3(1)=-1, \qquad \tilde{P}_3(2)=+2, \qquad \tilde{P}_3(3)=-1,
\end{equation}
and these add up to zero, as expected. All values of $\alpha$ behave in a similar way.

\subsubsection{Refining the argument}
\label{sec:subtle}
There are some subtle points in the previous reasoning that we omitted for convenience of presentation and that we address below. We will also discuss explicit examples later on.

In sections \ref{misaligned supersymmetry in closed strings} and \ref{misaligned supersymmetry in open strings}, a defining feature of misaligned supersymmetry was identified in the presence of a bosonic-fermionic oscillation at leading order in the Hardy-Ramanujan-Rademacher expansion. Therefore, we will assume that the partitions functions $Z(\tau)$ we work with have this property. However, this per se is not enough to guarantee the presence of misaligned supersymmetry in full generality, as we are going to explain.

For simplicity, we start by considering the case in which the partition function is given by a single term which is also an $\eta$-quotient. This corresponds to the open-string system we are interested in. If more $\eta$-quotients are present, one can just repeat the analysis for each of them separately. The heterotic model will indeed be of this latter type.

In general, for a single $\eta$-quotient, we can distinguish the following situations.

\begin{enumerate}
    \item The conditions of applicability of \eqref{d(n)}, i.e. $c_1>0$ and $G(\alpha)>0$, are met either by the function $Z(\tau)$ or by the function $\tilde{Z}(\tau) = Z(\tau+1/2)$. The two subcases must be distinguished.
    \begin{enumerate}
        \item The conditions of applicability of \eqref{d(n)} are met by $\tilde{Z}(\tau) = Z(\tau+1/2)$, which corresponds to a Laurent series with positive-semidefinite coefficients (as the original $Z(\tau)$ has oscillating coefficients by assumption). In this case, we can work out the alternating coefficients $a_n$ of the original series from the positive coefficients $\tilde{a}_n$ of the new series by just noticing that $a_n = (-1)^n \tilde{a}_n$. We will have then to keep track of which states had positive/negative coefficients before the shift of $\tau$ was performed. A concrete example is the partition function $M(\tau)$ of an anti-D$p$-brane on top of an O$p$-plane, as in subsection \ref{anti-D/O in terms of eta-quotients}.
        \item  The conditions of applicability of \eqref{d(n)} are met by $Z(\tau)$, which has oscillating coefficients $a_n$. A concrete example are the functions $L_1(\tau)$ and $R_3'(\tau)$ in subsection \ref{SO(16)xSO(16) in terms of eta-quotients}.
    \end{enumerate}
    \item The conditions of applicability of \eqref{d(n)} are met neither by the function $Z(\tau)$ nor by the function $\tilde{Z}(\tau) = Z(\tau+1/2)$.
\end{enumerate}
We do not consider the case 2.~in this paper and leave its investigation for future work. Case 1.(b) happens to be trivially described as in the general treatment above, while case 1.(a) is more subtle. We now focus on it, since it corresponds to the open-string system of our interest. We will also discuss extensions of the reasoning to the closed-string case, focusing again on the heterotic $\mathrm{SO}(16) \!\times\! \mathrm{SO}(16)$-model.

\subsection[Open Strings: Anti-Dp-brane on top of an Op-plane]{Open Strings: Anti-D\texorpdfstring{$\boldsymbol{p}$}{$p$}-brane on top of an O\texorpdfstring{$\boldsymbol{p}$}{$p$}-plane} \label{Ceff=0 for anti-Dp/Op}
Here we discuss in detail the case 1.(a) introduced above and then specialise it to the case of an anti-D$p$-brane sitting on top of an O$p$-plane.

In the function $\tilde{Z}(\tau)=Z(\tau+1/2)$, the shift in $\tau+1/2$ flips the signs of the state degeneracies and leads to coefficients $\tilde a_n$ which are all positive. Therefore, we cannot distinguish anymore which states are fermions in $\tilde Z(\tau)$. In order to discuss cancellations among bosons and fermions for the original model $Z(\tau)$, we must treat separately the values of $n$ that correspond to original bosonic degeneracies, namely $a_n = \tilde{a}_n$, and those that correspond to original fermionic degeneracies, namely $a_n = - \tilde{a}_n$. For definiteness, let us focus on the bosonic ones. Using a tilde notation to stress that we are dealing with the theory $\tilde Z(\tau)$, we have
\begin{equation} \label{bosonic degeneracies}
    \tilde{a}_n = \sum_{\substack{\alpha \in \mathbb{N}^+, \\ c_3(\alpha) > 0}} \tilde{a}_n(\alpha) = \sum_{\substack{\alpha \in \mathbb{N}^+, \\ c_3(\alpha) > 0}} \tilde{\hat P}_\alpha(n) \tilde{f}_n(\alpha), \qquad n \in \mathbb{N}_b,
\end{equation}
where $\mathbb{N}_b \subset \mathbb{N}$ represents the subset of values of $n$ with a bosonic degeneracy $a_n = \tilde{a}_n$. Although this formally looks the same as the generic case above, the fact that $n$ only takes values in a subset of $\mathbb{N}$ is crucial. Indeed, since the periodicity of $\tilde{P}_\alpha(n)$ is $\tilde{P}_\alpha(n) = \tilde{P}_\alpha(n+\alpha)$, we are no longer guaranteed that all the values of $n$ at our disposal in $\mathbb{N}_b$ are all the $\alpha$ distinct values that $\tilde{P}_\alpha(n)$ would assume if its domain was  $\mathbb{N}$. In particular, if $\mathbb{N}_f \subset \mathbb{N}$ represents the subset of fermionic states with $a_n = - \tilde{a}_n$, we may write
\begin{equation}
    \begin{array}{lcl}
        a_n(\alpha) = \tilde{a}_n (\alpha) = \tilde{\hat P}_\alpha(k) \tilde{f}_n (\alpha), & \qquad \qquad & \forall n \in \mathbb{N}_\alpha^{(k)} \cap \mathbb{N}_b; \\[1.0ex]
        a_n(\alpha) = - \tilde{a}_n (\alpha) = \tilde{\hat P}_\alpha(k) \tilde{f}_n (\alpha), & & \forall n \in \mathbb{N}_\alpha^{(k)} \cap \mathbb{N}_f.
    \end{array}
\end{equation}
It is therefore manifest that the value $\tilde{P}_\alpha (n) = \tilde{\hat{P}}_\alpha(k)$ is only found e.g. in the bosonic sector if $\mathbb{N}_\alpha^{(k)} \cap \mathbb{N}_b \neq \emptyset$. Of course, the missing values of $\tilde{\hat{P}}_\alpha(k)$ would be found in the fermionic sector, and vice versa, but this means that they would contribute with an extra $(-1)$-factor, invalidating the cancellation based on \eqref{closure}. Take for definiteness the case where $\mathbb{N}_b = 2 \mathbb{N}_0 + 1$. In this case, the degeneracies that appear in the corrections to the bosonic degeneracies \eqref{bosonic degeneracies} are
\begin{equation}
    \dots, \, \tilde{a}_{2n-1}(\alpha), \, \overbracket{\tilde{a}_{2n+1}(\alpha)}^{\tilde{P}_\alpha(2n+1)}, \, \overbracket{\tilde{a}_{2n+3}(\alpha)}^{\tilde{P}_\alpha(2n+3)}, \, \dots, \, \overbracket{\tilde{a}_{2n+2\alpha-1}(\alpha)}^{\tilde{P}_\alpha(2n+2\alpha-1)}, \, \underbracket{\tilde{a}_{2n+2\alpha+1}(\alpha)}_{\tilde{P}_\alpha(2n+1)}, \, \dots
\end{equation}
and one can observe how the periodicity $\mathrm{mod} \, 2\alpha$ in the functions $\tilde{P}_\alpha(n)$ allows to recognise all the sectors $\tilde{P}_\alpha(2n+1+2l)$, with $l=0,\dots,\alpha-1$ (we do not use the periodicity $\mathrm{mod} \, \alpha$ since if $\tilde{a}_{2n+1}(\alpha)$ is in the spectrum, then $\tilde{a}_{2n+1+2\alpha}(\alpha)$ is also always there, unlike $\tilde{a}_{2n+1+\alpha}(\alpha)$). Now one needs to understand whether the values $\tilde{P}_\alpha(2n+1), \tilde{P}_\alpha(2n+3), \dots, \tilde{P}_\alpha(2n+2\alpha-1)$ suffice to individuate all the $\alpha$ terms $\tilde{\hat P}_\alpha(k)$ that add up to zero. The answer is affirmative if $\alpha$ is odd, as one can see by direct inspection.

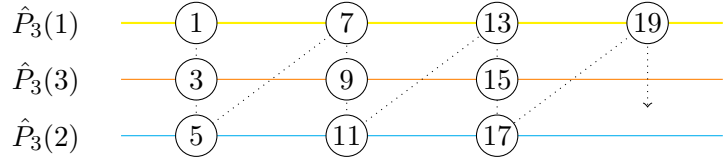
\begin{figure}
\centering
    \begin{tikzpicture}[xscale=2,yscale=0.75,a/.style={draw,circle,minimum size=5.5mm,inner sep=0pt,outer sep=0pt,black, fill=white,solid}]
        
    \draw[yellow, thick] (-0.5,1) -- (3.5,1);
    \draw[orange] (-0.5,0) -- (3.5,0);
    \draw[cyan] (-0.5,-1) -- (3.5,-1);
        
    \node[black,fill=white] at (-1,1){$\hat{P}_3(1)$};
    \node[black,fill=white] at (-1,0){$\hat{P}_3(3)$};
    \node[black,fill=white] at (-1,-1){$\hat{P}_3(2)$};

    \draw[dotted,->] (0,1) node[a]{$1$} -- (0,0) node[a]{$3$} -- (0,-1) node[a]{$5$} -- (1,1) node[a]{$7$} -- (1,0) node[a]{$9$} -- (1,-1) node[a]{$11$} -- (2,1) node[a]{$13$} -- (2,0) node[a]{$15$} -- (2,-1) node[a]{$17$} -- (3,1) node[a]{$19$} -- (3,-0.5);
        
    \end{tikzpicture}

    \caption{Periodicity of the function $P_\alpha(n)$ for $\alpha=3$, with odd argument $n \in 2 \mathbb{N}_0 + 1$. Each circle contains increasing odd integers $n=2l+1$, while the horizontal lines represent the associated term $P_3(n)$. The periodicity $P_3 (n) = P_3 (n \, \mathrm{mod} \, 3)$ permits to group all odd numbers $n$ in $\alpha=3$ different groups. All different values of $\hat P_\alpha(k)$ can be populated by $P_\alpha(n)$ for odd values of $\alpha$. Even arguments $n \in 2 \mathbb{N}_0$ behave in the same way for odd values of $\alpha$.}
    \label{odd alpha plot}
\end{figure}

\begin{figure}
\centering
    \begin{tikzpicture}[xscale=2,yscale=0.75,a/.style={draw,circle,minimum size=5.5mm,inner sep=0pt,outer sep=0pt,black, fill=white,solid}]
        
    \draw[green] (-0.5,2) -- (5,2);
    \draw[purple] (-0.5,1) -- (5,1);
    \draw[cyan] (-0.5,0) -- (5,0);
    \draw[magenta] (-0.5,-1) -- (5,-1);

    \node[black,fill=white] at (-1,2){$\hat{P}_4(2)$};
    \node[black,fill=white] at (-1,1){$\hat{P}_4(1)$};
    \node[black,fill=white] at (-1,0){$\hat{P}_4(3)$};
    \node[black,fill=white] at (-1,-1){$\hat{P}_4(4)$};

    \draw[dotted,->] (0,1) node[a]{$1$} -- (0,0) node[a]{$3$} -- (1,1) node[a]{$5$} -- (1,0) node[a]{$7$} -- (2,1) node[a]{$9$} -- (2,0) node[a]{$11$} -- (3,1) node[a]{$13$} -- (3,0) node[a]{$15$} -- (4,1) node[a]{$17$} -- (4,0) node[a]{$19$} -- (4.5,0.5);

    \end{tikzpicture}

    \caption{Periodicity of the function $P_\alpha(n)$ for $\alpha=4$, with odd argument $n \in 2 \mathbb{N}_0 + 1$. Each circle contains increasing odd integers $n=2l+1$, while the horizontal lines represent the corresponding value $P_4(n)$. The periodicity $P_4(n) = P_4(n \, \mathrm{mod} \, 4)$ necessarily leaves out half of the possible values $\hat P_\alpha(k)$ in the first column. Even arguments $n \in 2 \mathbb{N}_0$ behave in the same way for odd values of $\alpha$.}
    \label{even alpha plot}
\end{figure}
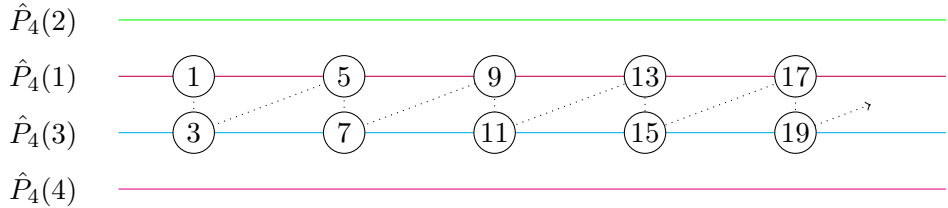

A more explicit treatment is below. See also Figs. \ref{odd alpha plot} and \ref{even alpha plot} for an explicit example.
\begin{itemize}
    \item In the bosonic sector, one has the sequence
    \begin{equation}
        \underbracket{\tilde{a}_1(\alpha)}_{\tilde{P}_\alpha(1)}, \, \tilde{a}_3(\alpha), \, \tilde{a}_5(\alpha), \, \dots, \, \tilde{a}_{2\alpha-5}(\alpha), \, \underbracket{\tilde{a}_{2\alpha-3}(\alpha)}_{\tilde{P}_\alpha(\alpha-3)}, \, \underbracket{\tilde{a}_{2\alpha-1}(\alpha)}_{\tilde{P}_\alpha(\alpha-1)}, \, \overbracket{\tilde{a}_{2\alpha+1}(\alpha)}^{\tilde{P}_\alpha(1)}, \, \dots
    \end{equation}
    and therefore:
    \begin{enumerate}[(a)]
        \item for odd $\alpha$, every $\alpha$ consecutive terms contain the $\alpha$ different terms $\tilde{\hat{P}}_\alpha(k)$ (this happens just because the difference of two odd numbers is even);
        \item for even $\alpha$, half of the terms $\tilde{\hat{P}}_\alpha(k)$ are never hit by the degeneracies (in fact, $\alpha-(2l+1)$, for any $l$, is never even if $\alpha$ is even).
    \end{enumerate}
    \item In the fermionic sector, one has the sequence
    \begin{equation}
        \underbracket{\tilde{a}_0(\alpha)}_{\tilde{P}_\alpha(0)}, \, \tilde{a}_2(\alpha), \, \tilde{a}_4(\alpha), \, \dots, \, \tilde{a}_{2\alpha-6}(\alpha), \, \underbracket{\tilde{a}_{2\alpha-4}(\alpha)}_{\tilde{P}_\alpha(\alpha-4)}, \, \underbracket{\tilde{a}_{2\alpha-2}(\alpha)}_{\tilde{P}_\alpha(\alpha-2)}, \, \overbracket{\tilde{a}_{2\alpha}(\alpha)}^{\tilde{P}_\alpha(0)}, \, \dots
    \end{equation}
    and therefore:
    \begin{enumerate}[(a)]
        \item for odd $\alpha$, every $\alpha$ consecutive terms contain the $\alpha$ different terms $\tilde{\hat{P}}_\alpha(k)$ (this happens just because the difference of two odd numbers is even);
        \item for even $\alpha$, half of the terms $\tilde{\hat{P}}_\alpha(k)$ are never hit by the degeneracies (in fact, $\alpha-2l$, for any $l$, is never odd if $\alpha$ is even).
    \end{enumerate}
\end{itemize}

\vspace{4pt}

The outcome of this analysis is that models exhibiting misaligned supersymmetry experience a net cancellation, at all odd subleading orders $\alpha=2l+1$, taking place among the pure bosonic and pure fermionic sectors individually. On the other hand, at all even subleading orders $\alpha=2l$, the pure bosonic and fermionic sector do still combine together into nonzero values. In short, we have seen that the even orders $\alpha=2l$ may be problematic within the case 1.(a) under consideration and we cannot draw any general conclusion for them at this stage, but we do not need them in the open-string case example below. Notice that they will be tractable easily for the closed-string case of interest below.

\begin{figure}[h]
    \centering
    
    \begin{tikzpicture}[xscale=0.5,yscale=0.08,bos/.style={draw,circle,minimum size=2mm,inner sep=0pt,outer sep=0pt,black,fill=green,solid},fer/.style={draw,circle,minimum size=2mm,inner sep=0pt,outer sep=0pt,black,fill=magenta,solid},gbos/.style={draw,circle,minimum size=2mm,inner sep=0pt,outer sep=0pt,thick,green,solid},gfer/.style={draw,circle,minimum size=2mm,inner sep=0pt,outer sep=0pt,thick,magenta,solid},zer/.style={draw,circle,minimum size=2mm,inner sep=0pt,outer sep=0pt,black,fill=orange,solid},corr1/.style={draw,circle,minimum size=1.5mm,inner sep=0pt,outer sep=0pt,black,fill=yellow,solid},corr2/.style={draw,circle,minimum size=1.5mm,inner sep=0pt,outer sep=0pt,black,fill=orange,solid},corr3/.style={draw,circle,minimum size=1.5mm,inner sep=0pt,outer sep=0pt,black,fill=cyan,solid}]
    
    \draw[domain=2:22.5, smooth, thick, variable=\n, gray] plot ({\n}, {- ln(8*8^(1/4)) - (11/4)*ln(\n) + (8*pi^2)^(1/2)*\n^(1/2)}) node[right, black]{$\Phi(n)$};
    \draw[domain=2:22.5, smooth, thick, variable=\n, gray] plot ({\n}, {ln(8*8^(1/4)) + (11/4)*ln(\n) - (8*pi^2)^(1/2)*\n^(1/2)}) node[right, black]{$-\Phi(n)$};
    
    \draw[white] (0,0) -- (34,0);
    \draw[white] (0,0) -- (0,44);
    
    \draw (-1,0) -- (0,0) node[below left]{$0$};
    \draw[-|] (0,0) -- (10,0) node[below]{$10$};
    \draw[-|] (10,0) -- (20,0) node[below]{$20$};
    \draw[->] (20,0) -- (22,0) node[below]{$n$};
    \draw[-|] (0,-35) -- (0,-20) node[left]{$-20$};
    \draw[-|] (0,-20) -- (0,20) node[left]{$20$};
    \draw[->] (0,20) -- (0,35) node[left]{$\pm \mathrm{log} \, (\pm g_n)$};
    
    \draw [dotted] (0,-4.15888308) node[fer]{} -- (1,4.852030263919617) node[bos]{} -- (2,-7.049254841255837) node[fer]{} -- (3,8.946374826141717) node[bos]{} -- (4,-10.648088014684989) node[fer]{} -- (5,12.208310140470365) node[bos]{} -- (6,-13.659502153634902) node[fer]{} -- (7,15.023099217200523) node[bos]{} -- (8,-16.31417809100231) node[fer]{} -- (9,17.543847620246574) node[bos]{} -- (10,-18.720586679634717) node[fer]{} -- (11,19.851049252007453) node[bos]{} -- (12,-20.940576396593297) node[fer]{} -- (13,21.99353602270788) node[bos]{} -- (14,-23.013556682034682) node[fer]{} -- (15,24.003693365249816) node[bos]{} -- (16,-24.966548083601058) node[fer]{} -- (17,25.90435950977719) node[bos]{} -- (18,-26.81907090081016) node[fer]{} -- (19,27.71238242905967) node[bos]{} -- (20,-28.585792100987767) node[fer]{};
    
    \draw[dotted, red] (1,4.852030263919617-1.5) node[corr1]{} -- (3,8.946374826141717-1.5*1.73) node[corr2]{} -- (5,12.208310140470365+3*2.24) node[corr3]{} -- (7,15.023099217200523-1.5*2.64) node[corr1]{} -- (9,17.543847620246574-1.5*3) node[corr2]{} -- (11,19.851049252007453+3*3.32) node[corr3]{} -- (13,21.99353602270788-1.5*3.61) node[corr1]{} -- (15,24.003693365249816-1.5*3.87) node[corr2]{} -- (17,25.90435950977719+3*4.12) node[corr3]{} -- (19,27.71238242905967-1.5*4.36) node[corr1]{};
    
    \draw[dotted, red] (2,-7.049254841255837-3*1.41) node[corr3]{} -- (4,-10.648088014684989+1.5*2) node[corr1]{} -- (6,-13.659502153634902+1.5*2.45) node[corr2]{} -- (8,-16.31417809100231-3*2.83) node[corr3]{} -- (10,-18.720586679634717+1.5*3.16) node[corr1]{} -- (12,-20.940576396593297+1.5*3.46) node[corr2]{} -- (14,-23.013556682034682-3*3.74) node[corr3]{} -- (16,-24.966548083601058+1.5*4) node[corr1]{} -- (18,-26.81907090081016+1.5*4.24){} node[corr2]{} -- (20,-28.585792100987767-3*4.47) node[corr3]{};
    
    \node[corr1] at (22,56){};
    \node[right] at (22,56){\, $\tilde{P}_3(1)=-1$};
    \node[corr3] at (22,49){};
    \node[right] at (22,49){\, $\tilde{P}_3(2)=+2$};
    \node[corr2] at (22,42){};
    \node[right] at (22,42){\, $\tilde{P}_3(3)=-1$};

    \end{tikzpicture}
    
    \caption{A schematic plot representing the spectrum of an anti-D$p$-brane on top of an O$p$-plane, including the terms at leading order, for $\alpha=1$, and the (magnified) corrections at next-to-leading order, for $\alpha=3$. One has to consider bosons (odd $n$) and fermions (even $n$) separately, since the corrections to the coefficients of the partition function $M(\tau)$ are computed with the dual function $\tilde{M}(\tau)$. Then, levels $n = 1 \, \mathrm{mod} \, 3$ have corrections multiplied by the value $\tilde{P}_3(1)=-1$, levels $n = 2 \, \mathrm{mod} \, 3$ have corrections multiplied by the value $\tilde{P}_3(2)=+2$ and levels $n = 3 \, \mathrm{mod} \, 3$ have corrections multiplied by the value $\tilde{P}_3(3)=-1$. For each different value the function $\tilde{P}_\alpha(n)$ can take, one can individuate a different interpolating function. Evidently, the average of such interpolating functions vanishes, independently from each other, both in the bosonic and in the fermionic sector.}
    
    \label{corrections to anti-D/O plot}
    
\end{figure}
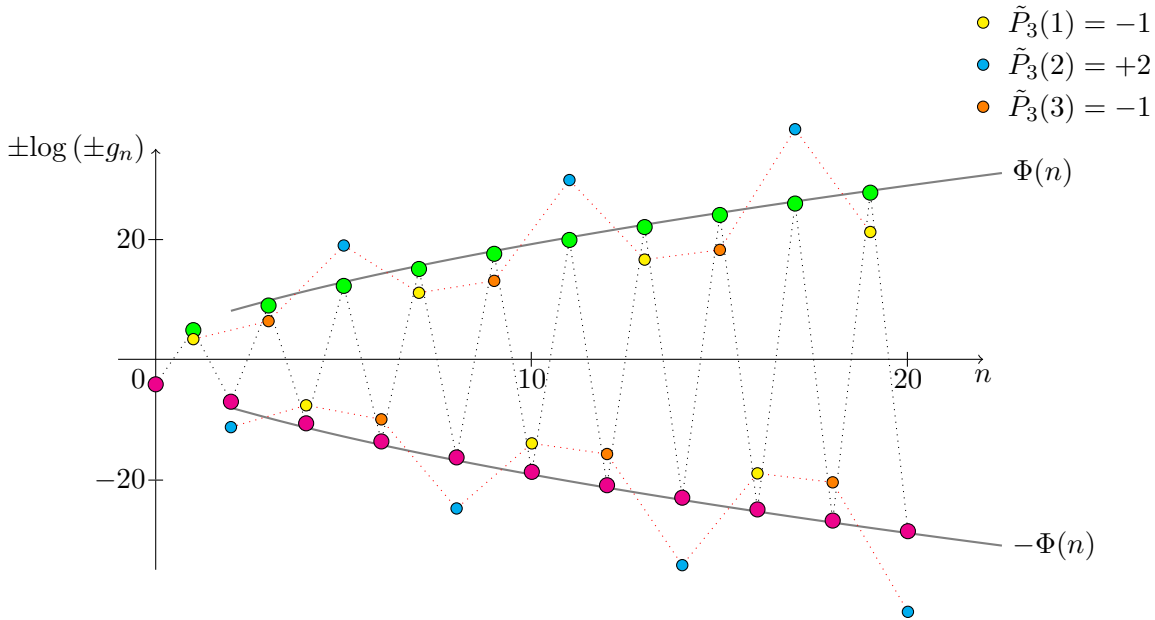

Fortunately, for the partition function of an anti-D$p$-brane on top of an O$p$-plane, given in \eqref{M}, the shifted function \eqref{M'} that we need to consider is such that all the even orders in the Hardy-Ramanujan-Rademacher expansions are vanishing, the only contributions being from $\alpha=2l+1$. This is discussed in subsection \ref{anti-D/O in terms of eta-quotients}. Therefore, we do not have to deal with additional complications and we can apply directly our machinery, which shows that the interpolating functions cancel at all subleading orders, with bosonic and fermionic corrections averaging out to zero independently from each other. This proves that
\begin{equation}
    C_\textrm{eff}=0
\end{equation}
for the anti-D$p$-brane on top of an O$p$-plane. Below, Fig. \ref{corrections to anti-D/O plot} reports a schematic representation of this.

\vspace{-10pt}

\subsection[Closed Strings: Heterotic SO(16)xSO(16)-theory]{Closed Strings: Heterotic \texorpdfstring{$\boldsymbol{\mathrm{SO}(16) \!\times\! \mathrm{SO}(16)}$}{$\mathrm{SO}(16) \!\times\! \mathrm{SO}(16)$}-theory} \label{Ceff=0 for het-SO(16)xSO(16)}

We turn now to closed strings, for which one has to consider products of two functions, one for the right- and one for the left-moving sector.  As in section \ref{sec:sussman}, let the generic one-loop closed-string partition function be $Z(\tau, \bar{\tau}) = R(\tau) \bar{L}(\bar{\tau})$. For the physical state degeneracies, one can generally write
\begin{equation}
    a_{nn} = r_{n-H_R} \bar{l}_{n-H_L} = \biggl[\sum_\alpha r_{n-H_R}(\alpha) \biggr] \biggl[\sum_\beta \bar{l}_{n-H_L}(\beta) \biggr],
\end{equation}
where $r_{n-H_R}$ and $l_{n-H_L}$ are the Laurent coefficients of the series $R(\tau)$ and $L(\tau)$, respectively, and the summations over $\alpha$ and $\beta$ represent their Hardy-Ramanujan-Rademacher expansions. Again, we refrain from diving into a general and all-encompassing analysis, but rather focus on an example. This should suffice to give all of the elements that one may need to take care of in the investigation of a given theory.

Let us consider the heterotic $\mathrm{SO}(16) \!\times\! \mathrm{SO}(16)$-theory. As discussed in subsection \ref{SO(16)xSO(16) in terms of eta-quotients}, its partition function can be written as a sum of three terms that are separate products of right- and left-moving Dedekind $\eta$-quotients. For definiteness, here we start with the product $Z_1(\tau,\bar{\tau}) = R_1(\tau) \bar{L}_1(\bar \tau)$.
Both these functions can be treated in the way discussed in subsection \ref{sussman method} and the full state degeneracies read
\begin{equation}
    a_{nn}^{(1)} = r_{n+n_0^{R_1}}^{R_1} \bar{l}_{n+n_0^{L_1}}^{L_1} = \Biggl[\sum_{\alpha \in \Gamma_{R_1}} r_{n+n_0^{R_1}}^{R_1}(\alpha) \Biggr] \Biggl[\sum_{\beta \in \Gamma_{L_1}} \bar{l}^{L_1}_{n+n_0^{L_1}}(\beta) \Biggr],
\end{equation}
where for brevity we defined the two sets containing the contributions to the Laurent coefficients, i.e. $\Gamma_{R_1} = \lbrace \alpha \in \mathbb{N}^+: c_3^{R_1}(\alpha)>0 \rbrace$ and $\Gamma_{L_1} = \lbrace \beta \in \mathbb{N}^+: c_3^{L_1}(\beta)>0 \rbrace$, according to the general formula \eqref{d(n)}. In particular, in an obvious notation, one can write
\begin{equation}
    \begin{split}
        a_{nn}^{(1)} & = \sum_{\alpha \in \Gamma_{R_1}} \sum_{\beta \in \Gamma_{L_1}} a^{(1)}_{nn}(\alpha,\beta) \\
        & = \sum_{\alpha \in \Gamma_{R_1}} \sum_{\beta \in \Gamma_{L_1}} P^{R_1}_\alpha (n+n_0^{R_1}) \bar{P}^{L_1}_\beta (n+n_0^{L_1}) f^{R_1}_{n+n_0^{R_1}}(\alpha) \bar{f}^{L_1}_{n+n_0^{L_1}} (\beta).
    \end{split}
\end{equation}
For any given value $\smash{\alpha \in \Gamma_{R_1}}$ and $\smash{\beta \in \Gamma_{L_1}}$, the series of contributions $\smash{a^{(1)}_{nn}(\alpha,\beta)}$ can be associated to some envelope functions $\smash{\Phi^{(1)}_{\ell_{\alpha \beta}}(n;\alpha,\beta)}$, for some $\ell_{\alpha \beta}=1,\dots, \rm{lcm}(\alpha,\beta)$, with the least common multiple being $\rm{lcm}(\alpha,\beta) = \alpha \beta/\gcd(\alpha,\beta)$. Indeed, the series of contributions $\smash{a^{(1)}_{nn}(\alpha,\beta)}$ allows us to define $\rm{lcm}(\alpha,\beta)$ continuous functions of $n \in \mathbb{R}$
\begin{equation}
    \Phi^{(1)}_{\ell_{\alpha \beta}}(n;\alpha,\beta) = P^{R_1}_\alpha (\ell_{\alpha \beta}+n_0^{R_1}) \bar{P}^{L_1}_\beta (\ell_{\alpha \beta}+n_0^{L_1}) f^{R_1}_{n+n_0^{R_1}}(\alpha) \bar{f}^{L_1}_{n+n_0^{L_1}} (\beta),
\end{equation}
which, taking into account the two different periodicities $P_\alpha^{R_1} (m) = P_\alpha^{R_1} (m \, \mathrm{mod} \, \alpha)$ and $\bar{P}_\beta^{L_1} (m) = \bar{P}_\beta^{L_1} (m \, \mathrm{mod} \, \beta)$, correspond to values $n$ in $\smash{a^{(1)}_{nn}(\alpha,\beta)}$ for which one has $\ell_{\alpha \beta} = n \, \mathrm{mod} \, \mathrm{lcm}(\alpha,\beta)$.

In order to show that misaligned supersymmetry takes place at any order in the Hardy-Ramanujan-Rademacher series, we want to show that these envelope functions average out to zero, i.e. that $\smash{\sum_{\ell_{\alpha \beta} = 1}^{\rm{lcm} (\alpha,\beta)} \Phi^{(1)}_{\ell_{\alpha \beta}}(n;\alpha,\beta) = 0}$.
In order to do that we rewrite the sum over $\ell_{\alpha \beta}$ in terms of a double sum, defining $\gamma_{\alpha \beta} = \beta/\gcd(\alpha,\beta)$, as
\begin{equation}
    \begin{split}
        & \sum_{\ell_{\alpha \beta}=1}^{\rm{lcm}(\alpha,\beta)} \Phi^{(1)}_{\ell_{\alpha \beta}}(n;\alpha,\beta) = \\
        = \, & \sum_{k_\alpha=1}^\alpha \sum_{m=0}^{\gamma_{\alpha \beta}-1} \Phi^{(1)}_{k_\alpha+m\alpha}(n;\alpha,\beta) \\
        = \, & \sum_{k_\alpha=1}^\alpha \sum_{m=0}^{\gamma_{\alpha \beta}-1} P^{R_1}_\alpha (k_\alpha+m\alpha+n_0^{R_1}) \bar{P}^{L_1}_\beta (k_\alpha+m\alpha+n_0^{L_1}) f^{R_1}_{n+n_0^{R_1}}(\alpha) \bar{f}^{L_1}_{n+n_0^{L_1}} (\beta) \\
        = \, & \sum_{k_\alpha=1}^\alpha P^{R_1}_\alpha (k_\alpha+m\alpha+n_0^{R_1}) f^{R_1}_{n+n_0^{R_1}}(\alpha) \bar{f}^{L_1}_{n+n_0^{L_1}} (\beta) \Biggl[ \sum_{m=0}^{\gamma_{\alpha \beta}-1}  \bar{P}^{L_1}_\beta (k_\alpha+m\alpha+n_0^{L_1}) \Biggr] \\[1.5ex]
        = \, & 0,
    \end{split}
\end{equation}
where in the last line we used lemma \eqref{closure}. In order to use that lemma, we need the condition $\gamma_{\alpha \beta} = \beta/\mathrm{gcd}(\alpha,\beta) > 1$, which is indeed the case since the function $\bar{L}_1$ has only even $\beta$s and the function $R_1$ has only odd $\alpha$s. Notice that one could not draw the same conclusion by splitting the sum over $l_{\alpha \beta}$ as a sum over $k_\beta=1,\dots,\beta$ and $m=0,\dots,\alpha/\mathrm{gcd}(\alpha,\beta)-1$ since, in the case where $\beta=2r\alpha$, for $r \in \mathbb{N}$, we have $\alpha/\mathrm{gcd}(\alpha,\beta)=1$. To sum up, we have given a general proof that
\begin{equation}
    \sum_{\ell_{\alpha \beta} = 1}^{\rm{lcm} (\alpha,\beta)} \Phi^{(1)}_{\ell_{\alpha \beta}}(n;\alpha,\beta) = 0.
\end{equation}
Thanks to the dualities reported in \eqref{SO(16)xSO(16) identities}, it is apparent that the functions $Z_2 = R_2 \bar{L}_2$ and $Z_3 = R_3 \bar{L}_3$ contribute to physical states, which are what defines the envelope functions, in an identical way, just with a different index labelling and with different overall factors. Therefore, the total cancellation shown for $Z_1$ holds for $Z_2$ and $Z_3$, too. To conclude, even for the heterotic $\mathrm{SO}(16) \!\times\! \mathrm{SO}(16)$-theory we proved that 
\begin{equation}
    C_{\textrm{eff}}=0.
\end{equation}
The key to this result is just lemma \eqref{closure}, which is a general property of the Dedekind $\eta$-quotients for which \eqref{d(n)} holds.

Here we have just considered an explicit closed-string case, but the machinery that has been described is easy to adapt to a large variety of closed-string models. Cancellations beyond leading order depend on the specific details of the right- and left-moving factors that define their partition function.

Notice that the open-string case discussed in subsection \ref{Ceff=0 for anti-Dp/Op} can also be discussed with the tools presented above. Given the partition function $M=M(\tau)$ of an anti-D$p$-brane on top of an O$p$-plane, one can consider a closed-string theory where the right-moving sector is $R(\tau) =  - \tilde{M}(\tau) =  - M(\tau+1/2)$ and the left-moving sector has coefficients $l_n=(-1)^{n+1}$ for $\beta=2$ and vanishing for all other $\beta$s. In this case, we know the right-moving sector is defined only for odd values of $\alpha$, so the number of envelope functions is $\mathrm{lcm} (\alpha,2) = 2\alpha$. For $\ell_\alpha=1,\dots,2\alpha$, the envelope functions can be defined as
\begin{equation}
    \Phi_{\ell_\alpha} (n;\alpha) = (-1)^{\ell_\alpha+1} \tilde{P}_\alpha(\ell_\alpha+n_0) \tilde{f}_{n+n_0}(\alpha).
\end{equation}
Therefore we simply have
\begin{equation}
    \begin{split}
        \sum_{\ell_{\alpha}=1}^{2 \alpha} \Phi_{\ell_{\alpha}}(n;\alpha) & = \sum_{k=1}^2 \sum_{m=0}^{\alpha-1} \Phi_{k+2m}(n;\alpha) \\
        & = \sum_{k=1}^2  (-1)^{k+1} \tilde{f}_{n+n_0}(\alpha) \sum_{m=0}^{\alpha-1} \tilde{P}_\alpha(k+2m+n_0) \\[1.5ex]
        & = 0.
    \end{split}
\end{equation}

\section{One-loop cosmological constant and supertrace formulae} \label{open-string cosmological constant}

In previous sections we have shown how the non-supersymmetric heterotic $\mathrm{SO}(16) \!\times\! \mathrm{SO}(16)$-theory and the anti-D$p$-/O$p$-system in flat space provide a realisation of misaligned supersymmetry. In this section we explore the expected finiteness properties of the non-supersymmetric anti-D$p$-/O$p$-system. In particular, we compute the finite value of the one-loop cosmological constant and show that the first four mass supertraces vanish. The one-loop cosmological constant for the heterotic $\mathrm{SO}(16) \!\times\! \mathrm{SO}(16)$-theory was computed in \cite{AlvarezGaume:1986jb} and the general properties of the one-loop cosmological constant for closed strings are discussed in \cite{Dienes:1995pm, Dienes:2001se}.

Let us consider a $D$-dimensional quantum field theory with mass levels $\smash{M^2_n}$, degeneracies $\smash{g_n}$ and fermion parities $\smash{F_n}$. Given an arbitrary mass scale $\mu^2$, the one-loop cosmological constant reads \cite{Dienes:1995pm}
\begin{equation}
\Lambda = - \dfrac{1}{2} \, \biggl(\dfrac{\mu^2}{8 \pi^2}\biggr)^{\!\frac{D}{2}} \sum_{n} (-1)^{F_n} g_n \int_0^\infty \dfrac{\de t}{t^{1+\frac{D}{2}}} \, \mathrm{e}^{- 2 \pi M^2_n t / \mu^2}.
\end{equation}
It is convenient to rearrange this expression as
\begin{equation}
    \Lambda = - \Bigl(\dfrac{\mu}{2 \pi}\Bigr)^{\!D} \int_0^\infty \dfrac{\de t}{2t} (2t)^{\frac{1}{2}(p+1-D)} \hat{M}_p (t),
\end{equation}
where, for $\smash{q = \mathrm{e}^{-2\pi t}}$, the function $\hat{M}_p(t)$ has been defined as
\begin{equation} \label{hatMp}
    \hat{M}_p (t) = \dfrac{1}{(2t)^{\frac{1}{2}(p+1)}} \sum_n (-1)^{F_n} g_n \, q^{M_n^2 / \mu^2}.
\end{equation}
For the field theory of a $p$-brane one must consider a spacetime of dimension $\smash{D=p+1}$, so
\begin{equation}
    \Lambda_{p} = - \Bigl(\dfrac{\mu}{2 \pi}\Bigr)^{p+1} \int_0^\infty \dfrac{\de t}{2t} \, \hat{M}_{p} (t).
\end{equation}
This also gives a cosmological constant with the right mass dimension for the contribution of a $p$-brane, namely $\Lambda \sim \mu^{p+1}$.

For an anti-D$p$-brane in flat space, in the string frame, the mass spectrum in both the NS- and R-sectors follows the pattern $M^2_n = n/\alpha' $ for each mass level $n \in \mathbb{N}_0$ (see eqn. (8.33, \cite{BLT})), therefore it is convenient to set $\mu=1/\sqrt{\alpha'}$. One can thus recognise that the function defined in \eqref{hatMp} corresponds to the partition function in \eqref{M_antiDp}, i.e. $\hat{M}_p(t) = M_{\overline{\mathrm{D}p}}(t)$ (in the case of an O$p^-$-plane). One can therefore write
\begin{equation}
    \Lambda_{\overline{\mathrm{D}p}} = - \dfrac{g_s}{2 \pi} \,  \tau_{\mathrm{D}p} \int_0^\infty \dfrac{\de t}{2t} \, M_{\overline{\mathrm{D}p}} (t),
\end{equation}
where the tension of the anti-D$p$-brane is $\tau_{\mathrm{D}p} = 2 \pi / (g_s l_s^{p+1})$, with the string length $l_s = 2 \pi \sqrt{\alpha'}$. The string coupling is the vacuum expectation value $g_s = \mathrm{e}^{\langle \Phi \rangle}$, where $\Phi$ is the dilaton field. $\Lambda_{\overline{\mathrm{D}p}}$ can be calculated explicitly. We can define the integral $I_p$ via
\begin{equation}
    \begin{split}
         - \int_0^\infty \dfrac{\de t}{2t} \, M_{\overline{\mathrm{D}p}} (t) = \int_0^\infty \dfrac{\de t}{(2t)^{\frac{1}{2}(p+3)}} \, \dfrac{16}{\vartheta_3^8[2it]} = 8 \int_0^\infty \dfrac{\de t}{t^{\frac{1}{2}(p+3)}} \, \dfrac{1}{\vartheta_3^8[it]} = I_p
    \end{split}
\end{equation}
so that
\begin{equation}
    \Lambda_{\overline{\mathrm{D}p}} =  \dfrac{g_s}{2 \pi} \,  \tau_{\mathrm{D}p} \, I_p.
\end{equation}
The value $I_p$ is finite as long as $p=0,1,2,3,4,5,6$, as can be seen immediately thanks to the small-$t$ expansion
\begin{equation}\label{eq:theta3smallt}
    \vartheta_3^{-8}[{\rm i}t] \; \overset{\;\; t \to 0^+}{\approx} \; (2t)^4.
\end{equation}
One can evaluate the integral numerically and find
\begin{subequations} \label{eq:I0123456}
\begin{align}
    I_0 & = I_6 \simeq 16.65, \\
    I_1 & = I_5 \simeq 9.086, \\
    I_2 & = I_4 \simeq 6.984, \\
    I_3 & \simeq 6.461.
\end{align}
\end{subequations}
The equality for $S$-dual D$p$-/D$(6-p)$-branes follows from the modular transformation $\vartheta_3[{\rm i}{t^{-1}}] = t^\frac12 \vartheta_3[{\rm i}t]$. For $p>6$ the above integral diverges and our flat space calculation does not give a sensible answer. This can be attributed to the fact that the corresponding anti-D$p$-brane on top of an O$p$-plane strongly backreacts and no asymptotic flat space solution exists.

The full vacuum energy consists of the tree-level potential, which corresponds to the DBI-contribution, as well as this one-loop correction. Let the shifted dilaton be $\sdil = \Phi - \langle \Phi \rangle$. Then, in the string frame, the contribution to the action reads
\begin{equation}
    \begin{split}
        S_\Lambda^{\overline{\mathrm{D}p}} & = - \tau_{\mathrm{D}p} \int_{W_{1,p}} \!\! \d^{p+1} \xi \; \sqrt{- \mathrm{det} \, (\varphi_* G)} \; \mathrm{e}^{- \sdil} \biggl[ 1  + \dfrac{g_s}{2 \pi} \, I_p \, \mathrm{e}^{\sdil} \biggr],
    \end{split}
\end{equation}
where $\varphi: \; W_{1,p} \, \hookrightarrow \, X_{1,9}$ is the embedding function of the anti-D$p$-brane worldvolume $W_{1,p}$ into the $10$-dimensional spacetime $X_{1,9}$. As expected, the one-loop correction to the tree-level vacuum energy is suppressed by a factor $g_s$, which is the open-string coupling. The 10-dimensional Einstein frame is defined by the metric $g_{MN} = \mathrm{e}^{-\frac{\sdil}{2}} G_{MN}$ and gives
\begin{equation}
    \begin{split}
            S_\Lambda^{\overline{\mathrm{D}p}} & = - \tau_{\mathrm{D}p} \int_{W_{1,p}} \!\! \d^{p+1} \xi \; \sqrt{- \mathrm{det} \, (\varphi_* g)} \; \mathrm{e}^{\frac{(p-3)}{4} \phi}\biggl[ 1 + \dfrac{g_s}{2 \pi} \, I_p \, \mathrm{e}^{\sdil} \biggr].
    \end{split}
\end{equation}

We now turn our attention to the supertrace formulae for this system. From the previous analysis of the partition function and its expansion in powers of $q$, we learned that 
\begin{equation}
    -8 \vartheta_3^{-8}[{\rm i} t] \equiv \sum_{n=0}^\infty (-1)^{F_n} g_{n} \mathrm{e}^{-2\pi n t},
\end{equation}
where $g_0 = 8$, $g_1=128$, $g_2=1152$ and so on. The regularised string supertrace formula can be defined as 
\begin{equation}
   {\rm Str}\, {M}^{2 \beta} = \lim_{t \to 0} \sum_{n=0}^\infty (-1)^{F_n} g_{n} \, M_n^{2\beta} \, \mathrm{e}^{-2 \pi t M_n^2 / \mu^2},
\end{equation}
 where $\alpha' M^2_n = n$ and the scale $\mu$ can again be conveniently defined as $\mu = 1/\sqrt{\alpha^\prime}$. The supertraces can then be computed as
\begin{equation}
   {\rm  Str} \, {M}^{2 \beta} = \lim_{t\to 0}\left[\left(-\frac{\mu^2}{2\pi}\frac{d}{dt}\right)^{\beta}(-8\vartheta_3^{-8}[{\rm i}t])\right].
\end{equation}
From the small-$t$ expansion in \eqref{eq:theta3smallt}, we see immediately that the first non-zero supertrace arises for $\beta=4$, with all the lower ones vanishing, i.e.
\begin{equation}
    {\rm Str} \, M^0 =  {\rm Str} \, M^2= {\rm Str} \, M^4=  {\rm Str} \, M^6 = 0, \qquad  {\rm Str} \, M^8 \neq 0. 
\end{equation}
In \cite{Dienes:1995pm}, a similar result was proven for closed strings and interpreted as a consequence of misaligned supersymmetry.

\section{Discussion} \label{conclusions}

In this work, we showed that misaligned supersymmetry is a feature that can characterise non-supersymmetric theories for both closed and open strings. In particular, we extended the previous results on closed-strings to the open-string case for models in which an anti-D$p$-brane is placed on top of an O$p$-plane. Misaligned supersymmetry leads to cancellations between bosons and fermions at all different energy levels. Such cancellations are usually visualised by proving that the sector-averaged state degeneracies grow at an exponential rate governed by a coefficient $C_{\textrm{eff}}$ that is smaller than the inverse Hagedorn temperature, i.e. $C_{\textrm{eff}} < C_{\textrm{tot}}$. Here, we showed that in a large class of theories it is possible to prove that such a coefficient is actually zero, i.e. $C_{\textrm{eff}}=0$. This proves a total cancellation that previously was only conjectured.

Given the exact cancellation of continuous functions that we have proven by showing that $C_{\textrm{eff}}=0$ and the fact that the formula for the expansion coefficients $a_n$ in equation \eqref{d(n)} is exact, one might wonder how finite non-zero results arise for example for the cosmological constant or other quantities of interest. The finite answers do arise when performing discrete sums over the states instead of using the continuous functions we introduced. It would be very interesting to use our improved understanding of all the subleading corrections to show explicitly how discrete sums over the number of states lead to a finite non-zero answer \cite{misalignedSUSY2}.

We proved that misaligned supersymmetry is present in systems which in principle do not share any common feature. Indeed, the heterotic $\mathrm{SO}(16) \!\times\! \mathrm{SO}(16)$-theory is a non-supersymmmetric closed string model, while anti-D$p$-branes and O$p$-planes are open string states that spontaneously break the supersymmetry preserved by the type II closed-string sector they are coupled to. Moreover, while the presence of misaligned supersymmetry in closed strings can be interpreted as a consequence of the underlying modular invariance \cite{Dienes:1994np}, this reasoning cannot be directly applied to open strings. However, in our open string theory example the partition function turns out to be invariant under a congruence subgroup of $\mathrm{SL}(2,\mathbb{Z})$, which is crucial for using our generalized version of the Hardy-Ramanujan-Rademacher sum. Certainly, misaligned supersymmetry seems to be a general phenomenon that might be capable of explaining why string theory can give finite answers (at any loop) even without supersymmetry.
 
We devoted great attention to models with anti-D$p$-branes on top of O$p$-planes, which are examples of brane supersymmetry breaking. Our results thus point towards a relation between this scenario and misaligned supersymmetry: some models of brane supersymmetry breaking can give finite answers thanks to misaligned supersymmetry.

The case $p=3$ is of particular interest due to its relation to the KKLT and LVS constructions. We showed that the one-loop cosmological constant of these models is finite and such a finiteness can be explained thanks to the presence of misaligned supersymmetry. In addition, it is known that the worldvolume field theory living on an anti-D3-brane on top of an O3-plane is described by non-linear supersymmetry \cite{Volkov:1973ix, Kallosh:2014wsa}. In this sense, our work indicates that low-energy effective theories with non-linear supersymmetry are completed in the high-energy regime into string theories with misaligned supersymmetry. A key observation for this is that the mass scale of the non-linear realisation of supersymmetry is the anti-D3-brane tension, $m \sim \tau_{\mathrm{D}3}^{1/4}$, and similarly this is the scale that characterises the infinite tower of string states that define the realization of misaligned supersymmetry.
 
There are various directions in which one can extend our work. First, we always assumed to have a single anti-D$p$-brane in a flat ten-dimensional background modded out by an orientifold projection. It would be interesting to consider deviations from this, including multiple coincident or intersecting branes and involving possibly also compact dimensions. In these cases, a central question would be the stability of the resulting construction, see for example \cite{Uranga:1999ib}. The role played by the Kaluza-Klein towers of states should also be investigated. Second, we performed our analysis of the partition function only at one-loop level. Therefore, a natural development would be to understand whether or not our findings hold at higher loops. For the two-loop level, one can see for example \cite{Abel:2017rch}. Finally, recently a connection between misaligned supersymmetry and swampland conjectures has been pointed out  in \cite{Palti:2020tsy}. It would be interesting to pursue along this line of investigation.

\acknowledgments

We are thankful to B.~Aaronson, S.~Abel, K.~Dienes, E.~Gonzalo, S.~Murthy, V.~Reys, J.~Rogers, A.~Sagnotti and G.~Shiu for very useful discussions. The work of NC is supported by an FWF grant with the number P 30265. The work of TW is supported in part by the NSF grant PHY-2013988.

\appendix

\section[Dedekind eta-function and Jacobi theta-functions]{Dedekind \texorpdfstring{$\boldsymbol{\eta}$}{$\eta$}-function and Jacobi \texorpdfstring{$\boldsymbol{\vartheta}$}{$\vartheta$}-functions}
\label{app:thetafunc}

In terms of the variable $q=\mathrm{e}^{2\pi \rmi \tau}$, the Dedekind function is defined as
\begin{equation}
    \eta(\tau)=q^{\frac{1}{24}} \sum_{n=-\infty}^\infty (-1)^n q^{n\frac{(3n-1)}{2}} =q^{\frac{1}{24}}\prod_{n=1}^\infty(1-q^n),
\end{equation}
while the Jacobi $\vartheta$-functions can be defined as infinite sums
\begin{equation}
    \vartheta {\small{\left[\begin{array}{c}a\\b\end{array}\right]}} (z|\tau) = \sum_{n=-\infty}^\infty q^{\frac{1}{2}(n+a)^2}\mathrm{e}^{2\pi \rmi (n+a)(z+b)},
\end{equation} 
or equivalently as infinite products
\begin{equation}
\begin{aligned}
    \vartheta {\small{\left[\begin{array}{c}a\\b\end{array}\right]}} (z|\tau) = \mathrm{e}^{2\pi \rmi a(z+b)} \, q^{\frac{a^2}{2}} & \prod_{n=1}^\infty \left(1-q^n\right) \left(1+q^{n+a-\frac{1}{2}}\mathrm{e}^{2\pi \rmi (z+b)}\right)\left(1+q^{n-a-\frac{1}{2}}\mathrm{e}^{-2\pi \rmi (z+b)}\right).
\end{aligned}
\end{equation}
Particularly relevant for writing string theory amplitudes are the following four functions:
\begin{subequations}
\begin{align}
    &\vartheta_1 (q)\equiv-\vartheta {\small{\left[\begin{array}{c}\frac{1}{2}\\\frac{1}{2}\end{array}\right]}} (0|\tau)=\sum_{n=-\infty}^\infty q^{\frac12 (n+\frac12)^2}(-1)^{n-\frac12}=-\rmi q^{\frac18}\prod_{n=1}^\infty(1-q^n)^2(1-q^{n-1}),\\
    &\vartheta_2(q)\equiv-\vartheta {\small{\left[\begin{array}{c}\frac{1}{2}\\ 0\end{array}\right]}} (0|\tau)=\sum_{n=-\infty}^\infty q^{\frac12(n+\frac12)^2}
    =2 q^{\frac18}\prod_{n=1}^\infty\left(1-q^n\right)\left(1+q^n\right)^2,\\
    &\vartheta_3(q)\equiv-\vartheta {\small{\left[\begin{array}{c}0\\ 0\end{array}\right]}} (0|\tau)=\sum_{n=-\infty}^\infty q^{\frac{n^2}2}
    =\prod_{n=1}^\infty(1-q^n)(1+q^{n-\frac12})^2,\\
    &\vartheta_4 (q)\equiv-\vartheta {\small{\left[\begin{array}{c}0\\\frac{1}{2}\end{array}\right]}} (0|\tau)=\sum_{n=-\infty}^\infty (-1)^n q^{\frac{n^2}{2}}
    =\prod_{n=1}^\infty(1-q^n)(1-q^{n-\frac12})^2.
\end{align}
\end{subequations}
Some useful relations are 
\begin{align}
\label{t1=0}
\vartheta_1 &= 0,\\
\label{Jaceq}
    \vartheta_3^4-\vartheta_4^4-\vartheta_2^4&=0,\\
    \label{Jactrip}
    \vartheta_2 \vartheta_3\vartheta_4 &= 2\eta^3.
\end{align}
The second one is known as Jacobi equation, while the third one as Jacobi triple product identity.

Under the generating modular transformations $T$ and $S$, which act on the modular parameter as $T (\tau) = \tau+1$ and $S(\tau) = -1/\tau$, the Dedekind function transforms as
\begin{align}
    & T: \qquad \eta(\tau+1) = \mathrm{e}^{\frac{\rmi \pi}{12}} \eta(\tau),\\
    & S: \qquad \eta(-1/\tau) = \sqrt{-{\rm i} \tau} \, \eta(\tau).
\end{align}
The general form for the transformation of the Jacobi $\vartheta$-functions is
\begin{align}
    & T: \qquad \vartheta {\small{\left[\begin{array}{c}a\\b\end{array}\right]}} (z|\tau+1) = \mathrm{e}^{-\rmi\pi a(a-1)} \, \vartheta {\small{\left[\begin{array}{c}a\\a+b-\frac{1}{2}\end{array}\right]}} (z|\tau) \\
    & S: \qquad \vartheta {\small{\left[\begin{array}{c}a\\b\end{array}\right]}} \left(z|-1/\tau\right) =\sqrt{-\rmi\tau} \mathrm{e}^{2\pi \rmi ab+\rmi\pi\frac{z^2}{\tau}} \, \vartheta {\small{\left[\begin{array}{c}b\\-a\end{array}\right]}} (z|\tau),
\end{align}
but more explicitly we can write
\begin{subequations}
\begin{align}
    & \vartheta_2(\tau+1) = \mathrm{e}^{\frac{\rmi \pi}{4}} \vartheta_2(\tau), \\
    T: \qquad \qquad & \vartheta_3(\tau+1) = \vartheta_4(\tau), \\
    & \vartheta_4(\tau+1) = \vartheta_3(\tau),
\end{align}
\end{subequations}\\[-8.5ex]
\begin{subequations}
\begin{align}
    & \vartheta_2(-1/\tau) = \sqrt{-\rmi \tau} \, \vartheta_4(\tau), \\
    S: \qquad \qquad & \vartheta_3(-1/\tau) = \sqrt{-\rmi \tau} \, \vartheta_3(\tau), \\
    & \vartheta_4(-1/\tau) = \sqrt{-\rmi \tau} \, \vartheta_2(\tau).
\end{align}
\end{subequations}

Another useful identity is
\begin{equation} \label{shift tau+1/2}
    \eta(\tau+1/2) = \mathrm{e}^{\frac{\rmi \pi}{24}} \, \dfrac{\eta^3(2\tau)}{\eta(\tau) \eta(4\tau)}.
\end{equation}

It is possible to express the Jacobi $\vartheta$-functions in terms of the Dedekind $\eta$-function and vice versa via the identities
\begin{subequations} \label{thetas vs eta}
    \begin{align}
        \vartheta_2 (\tau) & = \dfrac{2 \eta^2(2\tau)}{\eta(\tau)}, \\
        \vartheta_3 (\tau) & = \dfrac{\eta^5(\tau)}{\eta^2(\tau/2) \eta^2(2\tau)}, \\
        \vartheta_4(\tau) & = \dfrac{\eta^2(\tau/2)}{\eta(\tau)}.
    \end{align}
\end{subequations}

It is also useful to introduce the characters of the $\mathrm{so}(2n)$ algebras, which are defined as
\begin{subequations}
\begin{align}
    & O_{2n}=\frac{\vartheta_3^n+\vartheta_4^n}{2\eta^n}, \\
    & V_{2n}=\frac{\vartheta_3^n-\vartheta_4^n}{2\eta^n}, \\
    & S_{2n}=\frac{\vartheta_2^n+{\rm i}^{-n}\vartheta_1^n}{2\eta^n}, \\
    & C_{2n}=\frac{\vartheta_2^n-{\rm i}^{-n}\vartheta_1^n}{2\eta^n}.
\end{align}
\end{subequations}
In particular, $O_{2n}$ and $V_{2n}$ correspond to the traces over the NS-sector, whereas $S_{2n}$ and $C_{2n}$ correspond to the traces over the R-sector.

\subsubsection*{Further properties of \texorpdfstring{$\boldsymbol{\vartheta_3(z,\tau)}$}{$\vartheta_3(z,\tau)$}}
Since the elliptic function $\vartheta_3(z,\tau)$ has an important role in our discussion, we collect here some useful properties.

The Jacobi elliptic function $\vartheta_3$ can be defined as a infinite sum
\begin{equation}
    \vartheta_3(z,\tau) = \sum_{n=-\infty}^\infty \mathrm{e}^{{\rm i}\pi n^2 \tau} \mathrm{e}^{2 n {\rm i}z}
\end{equation}
and it is a solution of the heat equation
\begin{equation}
     \frac14  {\rm i}\pi \frac{\partial^2  \vartheta_3}{\partial z^2} (z, \tau) + \frac{\partial \vartheta_3}{\partial \tau} (z, \tau) = 0.
\end{equation}
In this work, we mainly employ the Jacobi $\vartheta$-constant, defined as
\begin{equation}
    \vartheta_3 (\tau) \equiv \vartheta_3(z=0, \tau).
\end{equation}
Moreover, we often need to restrict our attention to the case in which the argument is purely imaginary, namely $\tau= \rmi t$, with $t>0$. In this case, $\vartheta_3(\rmi t)$ satisfies the functional equation
\begin{equation}
    \vartheta_3 (\rmi t^{-1}) = t^{\frac12} \vartheta_3 (\rmi t),
\end{equation}
which can be interpreted as a modular $S$-transformation. One can also show the asymptotic behaviours
\begin{align}
    &\vartheta_3 (\rmi t) \overset{\;\;t \to 0^+}{\approx} \frac{1}{\sqrt{2t}},\\
    &\vartheta_3 (\rmi t) \overset{t \to \infty}{\approx} 1.
\end{align}

\bibliographystyle{JHEP}
\bibliography{refs}

\providecommand{\href}[2]{#2}\begingroup\raggedright\begin{thebibliography}{10}

\bibitem{AlvarezGaume:1986jb}
L.~Alvarez-Gaume, P.~H. Ginsparg, G.~W. Moore and C.~Vafa, \emph{{An O(16) x
  O(16) Heterotic String}},
  \href{https://doi.org/10.1016/0370-2693(86)91524-8}{\emph{Phys. Lett.}
  {\bfseries B171} (1986) 155}.

\bibitem{Dixon:1986iz}
L.~J. Dixon and J.~A. Harvey, \emph{{String Theories in Ten-Dimensions Without
  Space-Time Supersymmetry}},
  \href{https://doi.org/10.1016/0550-3213(86)90619-X}{\emph{Nucl. Phys.}
  {\bfseries B274} (1986) 93}.

\bibitem{Sugimoto:1999tx}
S.~Sugimoto, \emph{{Anomaly cancellations in type I D-9 - anti-D-9 system and
  the USp(32) string theory}},
  \href{https://doi.org/10.1143/PTP.102.685}{\emph{Prog. Theor. Phys.}
  {\bfseries 102} (1999) 685}
  [\href{https://arxiv.org/abs/hep-th/9905159}{{\ttfamily hep-th/9905159}}].

\bibitem{Antoniadis:1999xk}
I.~Antoniadis, E.~Dudas and A.~Sagnotti, \emph{{Brane supersymmetry breaking}},
  \href{https://doi.org/10.1016/S0370-2693(99)01023-0}{\emph{Phys. Lett.}
  {\bfseries B464} (1999) 38}
  [\href{https://arxiv.org/abs/hep-th/9908023}{{\ttfamily hep-th/9908023}}].

\bibitem{Angelantonj:1999jh}
C.~Angelantonj, \emph{{Comments on open string orbifolds with a nonvanishing
  B(ab)}}, \href{https://doi.org/10.1016/S0550-3213(99)00662-8}{\emph{Nucl.
  Phys.} {\bfseries B566} (2000) 126}
  [\href{https://arxiv.org/abs/hep-th/9908064}{{\ttfamily hep-th/9908064}}].

\bibitem{Aldazabal:1999jr}
G.~Aldazabal and A.~M. Uranga, \emph{{Tachyon free nonsupersymmetric type IIB
  orientifolds via Brane - anti-brane systems}},
  \href{https://doi.org/10.1088/1126-6708/1999/10/024}{\emph{JHEP} {\bfseries
  10} (1999) 024} [\href{https://arxiv.org/abs/hep-th/9908072}{{\ttfamily
  hep-th/9908072}}].

\bibitem{Angelantonj:1999ms}
C.~Angelantonj, I.~Antoniadis, G.~D'Appollonio, E.~Dudas and A.~Sagnotti,
  \emph{{Type I vacua with brane supersymmetry breaking}},
  \href{https://doi.org/10.1016/S0550-3213(00)00052-3}{\emph{Nucl. Phys.}
  {\bfseries B572} (2000) 36}
  [\href{https://arxiv.org/abs/hep-th/9911081}{{\ttfamily hep-th/9911081}}].

\bibitem{Dudas:2000ff}
E.~Dudas and J.~Mourad, \emph{{Brane solutions in strings with broken
  supersymmetry and dilaton tadpoles}},
  \href{https://doi.org/10.1016/S0370-2693(00)00734-6}{\emph{Phys. Lett.}
  {\bfseries B486} (2000) 172}
  [\href{https://arxiv.org/abs/hep-th/0004165}{{\ttfamily hep-th/0004165}}].

\bibitem{Dudas:2000nv}
E.~Dudas and J.~Mourad, \emph{{Consistent gravitino couplings in
  nonsupersymmetric strings}},
  \href{https://doi.org/10.1016/S0370-2693(01)00777-8}{\emph{Phys. Lett.}
  {\bfseries B514} (2001) 173}
  [\href{https://arxiv.org/abs/hep-th/0012071}{{\ttfamily hep-th/0012071}}].

\bibitem{Pradisi:2001yv}
G.~Pradisi and F.~Riccioni, \emph{{Geometric couplings and brane supersymmetry
  breaking}}, \href{https://doi.org/10.1016/S0550-3213(01)00441-2}{\emph{Nucl.
  Phys.} {\bfseries B615} (2001) 33}
  [\href{https://arxiv.org/abs/hep-th/0107090}{{\ttfamily hep-th/0107090}}].

\bibitem{Blumenhagen:1998uf}
R.~Blumenhagen and L.~Gorlich, \emph{{Orientifolds of nonsupersymmetric
  asymmetric orbifolds}},
  \href{https://doi.org/10.1016/S0550-3213(99)00241-2}{\emph{Nucl. Phys. B}
  {\bfseries 551} (1999) 601}
  [\href{https://arxiv.org/abs/hep-th/9812158}{{\ttfamily hep-th/9812158}}].

\bibitem{Blumenhagen:1999ns}
R.~Blumenhagen, A.~Font and D.~Lust, \emph{{Tachyon free orientifolds of type
  0B strings in various dimensions}},
  \href{https://doi.org/10.1016/S0550-3213(99)00381-8}{\emph{Nucl. Phys. B}
  {\bfseries 558} (1999) 159}
  [\href{https://arxiv.org/abs/hep-th/9904069}{{\ttfamily hep-th/9904069}}].

\bibitem{Mourad:2017rrl}
J.~Mourad and A.~Sagnotti, \emph{{An Update on Brane Supersymmetry Breaking}},
  \href{https://arxiv.org/abs/1711.11494}{{\ttfamily 1711.11494}}.

\bibitem{Dienes:1994np}
K.~R. Dienes, \emph{{Modular invariance, finiteness, and misaligned
  supersymmetry: New constraints on the numbers of physical string states}},
  \href{https://doi.org/10.1016/0550-3213(94)90153-8}{\emph{Nucl. Phys.}
  {\bfseries B429} (1994) 533}
  [\href{https://arxiv.org/abs/hep-th/9402006}{{\ttfamily hep-th/9402006}}].

\bibitem{Dienes:1995pm}
K.~R. Dienes, M.~Moshe and R.~C. Myers, \emph{{String theory, misaligned
  supersymmetry, and the supertrace constraints}},
  \href{https://doi.org/10.1103/PhysRevLett.74.4767}{\emph{Phys. Rev. Lett.}
  {\bfseries 74} (1995) 4767}
  [\href{https://arxiv.org/abs/hep-th/9503055}{{\ttfamily hep-th/9503055}}].

\bibitem{Dienes:2001se}
K.~R. Dienes, \emph{{Solving the hierarchy problem without supersymmetry or
  extra dimensions: An Alternative approach}},
  \href{https://doi.org/10.1016/S0550-3213(01)00344-3}{\emph{Nucl. Phys.}
  {\bfseries B611} (2001) 146}
  [\href{https://arxiv.org/abs/hep-ph/0104274}{{\ttfamily hep-ph/0104274}}].

\bibitem{Angelantonj:2010ic}
C.~Angelantonj, M.~Cardella, S.~Elitzur and E.~Rabinovici, \emph{{Vacuum
  stability, string density of states and the Riemann zeta function}},
  \href{https://doi.org/10.1007/JHEP02(2011)024}{\emph{JHEP} {\bfseries 02}
  (2011) 024} [\href{https://arxiv.org/abs/1012.5091}{{\ttfamily 1012.5091}}].

\bibitem{Abel:2015oxa}
S.~Abel, K.~R. Dienes and E.~Mavroudi, \emph{{Towards a nonsupersymmetric
  string phenomenology}},
  \href{https://doi.org/10.1103/PhysRevD.91.126014}{\emph{Phys. Rev.}
  {\bfseries D91} (2015) 126014}
  [\href{https://arxiv.org/abs/1502.03087}{{\ttfamily 1502.03087}}].

\bibitem{Abel:2017rch}
S.~Abel and R.~J. Stewart, \emph{{Exponential suppression of the cosmological
  constant in nonsupersymmetric string vacua at two loops and beyond}},
  \href{https://doi.org/10.1103/PhysRevD.96.106013}{\emph{Phys. Rev.}
  {\bfseries D96} (2017) 106013}
  [\href{https://arxiv.org/abs/1701.06629}{{\ttfamily 1701.06629}}].

\bibitem{Abel:2018zyt}
S.~Abel, E.~Dudas, D.~Lewis and H.~Partouche, \emph{{Stability and vacuum
  energy in open string models with broken supersymmetry}},
  \href{https://doi.org/10.1007/JHEP10(2019)226}{\emph{JHEP} {\bfseries 10}
  (2019) 226} [\href{https://arxiv.org/abs/1812.09714}{{\ttfamily
  1812.09714}}].

\bibitem{Kutasov:1990sv}
D.~Kutasov and N.~Seiberg, \emph{{Number of degrees of freedom, density of
  states and tachyons in string theory and CFT}},
  \href{https://doi.org/10.1016/0550-3213(91)90426-X}{\emph{Nucl. Phys.}
  {\bfseries B358} (1991) 600}.

\bibitem{Niarchos:2000kw}
V.~Niarchos, \emph{{Density of states and tachyons in open and closed string
  theory}}, \href{https://doi.org/10.1088/1126-6708/2001/06/048}{\emph{JHEP}
  {\bfseries 06} (2001) 048}
  [\href{https://arxiv.org/abs/hep-th/0010154}{{\ttfamily hep-th/0010154}}].

\bibitem{Kachru:2003aw}
S.~Kachru, R.~Kallosh, A.~D. Linde and S.~P. Trivedi, \emph{{De Sitter vacua in
  string theory}},
  \href{https://doi.org/10.1103/PhysRevD.68.046005}{\emph{Phys. Rev.}
  {\bfseries D68} (2003) 046005}
  [\href{https://arxiv.org/abs/hep-th/0301240}{{\ttfamily hep-th/0301240}}].

\bibitem{Conlon:2005ki}
J.~P. Conlon, F.~Quevedo and K.~Suruliz, \emph{{Large-volume flux
  compactifications: Moduli spectrum and D3/D7 soft supersymmetry breaking}},
  \href{https://doi.org/10.1088/1126-6708/2005/08/007}{\emph{JHEP} {\bfseries
  08} (2005) 007} [\href{https://arxiv.org/abs/hep-th/0505076}{{\ttfamily
  hep-th/0505076}}].

\bibitem{Kallosh:2015nia}
R.~Kallosh, F.~Quevedo and A.~M. Uranga, \emph{{String Theory Realizations of
  the Nilpotent Goldstino}},
  \href{https://doi.org/10.1007/JHEP12(2015)039}{\emph{JHEP} {\bfseries 12}
  (2015) 039} [\href{https://arxiv.org/abs/1507.07556}{{\ttfamily
  1507.07556}}].

\bibitem{Garcia-Etxebarria:2015lif}
I.~Garc\'ia-Etxebarria, F.~Quevedo and R.~Valandro, \emph{{Global String
  Embeddings for the Nilpotent Goldstino}},
  \href{https://doi.org/10.1007/JHEP02(2016)148}{\emph{JHEP} {\bfseries 02}
  (2016) 148} [\href{https://arxiv.org/abs/1512.06926}{{\ttfamily
  1512.06926}}].

\bibitem{Cribiori:2020bgt}
N.~Cribiori, C.~Roupec, M.~Tournoy, A.~Van~Proeyen and T.~Wrase,
  \emph{{Non-supersymmetric branes}},
  \href{https://doi.org/10.1007/JHEP07(2020)189}{\emph{JHEP} {\bfseries 07}
  (2020) 189} [\href{https://arxiv.org/abs/2004.13110}{{\ttfamily
  2004.13110}}].

\bibitem{Cascales:2003wn}
J.~Cascales, M.~Garcia~del Moral, F.~Quevedo and A.~Uranga, \emph{{Realistic
  D-brane models on warped throats: Fluxes, hierarchies and moduli
  stabilization}},
  \href{https://doi.org/10.1088/1126-6708/2004/02/031}{\emph{JHEP} {\bfseries
  02} (2004) 031} [\href{https://arxiv.org/abs/hep-th/0312051}{{\ttfamily
  hep-th/0312051}}].

\bibitem{Parameswaran:2020ukp}
S.~Parameswaran and F.~Tonioni, \emph{{Non-supersymmetric String Models from
  Anti-D3-/D7-branes in Strongly Warped Throats}},
  \href{https://arxiv.org/abs/2007.11333}{{\ttfamily 2007.11333}}.

\bibitem{Abel:2017vos}
S.~Abel, K.~R. Dienes and E.~Mavroudi, \emph{{GUT precursors and entwined SUSY:
  The phenomenology of stable nonsupersymmetric strings}},
  \href{https://doi.org/10.1103/PhysRevD.97.126017}{\emph{Phys. Rev. D}
  {\bfseries 97} (2018) 126017}
  [\href{https://arxiv.org/abs/1712.06894}{{\ttfamily 1712.06894}}].

\bibitem{Nibbelink:2015vha}
S.~Groot~Nibbelink, O.~Loukas and F.~Ruehle, \emph{{(MS)SM-like models on
  smooth Calabi-Yau manifolds from all three heterotic string theories}},
  \href{https://doi.org/10.1002/prop.201500041}{\emph{Fortsch. Phys.}
  {\bfseries 63} (2015) 609}
  [\href{https://arxiv.org/abs/1507.07559}{{\ttfamily 1507.07559}}].

\bibitem{Faraggi:2020wej}
A.~E. Faraggi, V.~G. Matyas and B.~Percival, \emph{{Towards the Classification
  of Tachyon-Free Models From Tachyonic Ten-Dimensional Heterotic String
  Vacua}},  \href{https://arxiv.org/abs/2006.11340}{{\ttfamily 2006.11340}}.

\bibitem{Faraggi:2020fwg}
A.~E. Faraggi, V.~G. Matyas and B.~Percival, \emph{{Type 0 $\mathbb{Z}_2\times
  \mathbb{Z}_2$ Heterotic String Orbifolds and Misaligned Supersymmetry}},
  \href{https://arxiv.org/abs/2010.06637}{{\ttfamily 2010.06637}}.

\bibitem{Faraggi:2020hpy}
A.~E. Faraggi, V.~G. Matyas and B.~Percival, \emph{{Type $\mathbf{\bar{0}}$
  Heterotic String Orbifolds}},
  \href{https://arxiv.org/abs/2011.12630}{{\ttfamily 2011.12630}}.

\bibitem{Hardy:1919}
G.~H. Hardy and S.~Ramanujan, \emph{{Asymptotic Formulae in Combinatory
  Analysis}}, \href{https://doi.org/10.1112/plms/s2-17.1.75}{\emph{Proc.
  London. Math. Soc.} {\bfseries 17} (1918) 75}.

\bibitem{Kani:1989im}
I.~Kani and C.~Vafa, \emph{{Asymptotic Mass Degeneracies in Conformal Field
  Theories}}, \href{https://doi.org/10.1007/BF02096934}{\emph{Commun. Math.
  Phys.} {\bfseries 130} (1990) 529}.

\bibitem{Verlinde:1988sn}
E.~P. Verlinde, \emph{{Fusion Rules and Modular Transformations in 2D Conformal
  Field Theory}},
  \href{https://doi.org/10.1016/0550-3213(88)90603-7}{\emph{Nucl. Phys.}
  {\bfseries B300} (1988) 360}.

\bibitem{Gross:1984dd}
D.~J. Gross, J.~A. Harvey, E.~J. Martinec and R.~Rohm, \emph{{The Heterotic
  String}}, \href{https://doi.org/10.1103/PhysRevLett.54.502}{\emph{Phys. Rev.
  Lett.} {\bfseries 54} (1985) 502}.

\bibitem{Angelantonj:2002ct}
C.~Angelantonj and A.~Sagnotti, \emph{{Open strings}},
  \href{https://doi.org/10.1016/S0370-1573(02)00273-9,
  10.1016/S0370-1573(03)00006-1}{\emph{Phys. Rept.} {\bfseries 371} (2002) 1}
  [\href{https://arxiv.org/abs/hep-th/0204089}{{\ttfamily hep-th/0204089}}].

\bibitem{BLT}
R.~Blumenhagen, D.~L{\"u}st and S.~Theisen, \emph{{Basics Concepts of String
  Theory}}. Springer-Verlag Berlin Heidelberg, 2013.

\bibitem{Polchinski:1996na}
J.~Polchinski, \emph{{Tasi lectures on D-branes}},  in \emph{{Theoretical
  Advanced Study Institute in Elementary Particle Physics (TASI 96): Fields,
  Strings, and Duality}}, pp.~293--356, 11, 1996,
  \href{https://arxiv.org/abs/hep-th/9611050}{{\ttfamily hep-th/9611050}}.

\bibitem{Uranga:1999ib}
A.~M. Uranga, \emph{{Comments on nonsupersymmetric orientifolds at strong
  coupling}}, \href{https://doi.org/10.1088/1126-6708/2000/02/041}{\emph{JHEP}
  {\bfseries 02} (2000) 041}
  [\href{https://arxiv.org/abs/hep-th/9912145}{{\ttfamily hep-th/9912145}}].

\bibitem{Dudas:2001wd}
E.~Dudas, J.~Mourad and A.~Sagnotti, \emph{{Charged and uncharged D-branes in
  various string theories}},
  \href{https://doi.org/10.1016/S0550-3213(01)00552-1}{\emph{Nucl. Phys.}
  {\bfseries B620} (2002) 109}
  [\href{https://arxiv.org/abs/hep-th/0107081}{{\ttfamily hep-th/0107081}}].

\bibitem{sussman2017rademacher}
E.~Sussman, \emph{Rademacher series for $\eta$-quotients},
  \href{https://arxiv.org/abs/1710.03415}{{\ttfamily 1710.03415}}.

\bibitem{misalignedSUSY2}
N.~Cribiori, S.~Parameswaran, F.~Tonioni and T.~Wrase, \emph{{In progress
  (2021)}}.

\bibitem{Volkov:1973ix}
D.~V. Volkov and V.~P. Akulov, \emph{{Is the Neutrino a Goldstone Particle?}},
  \href{https://doi.org/10.1016/0370-2693(73)90490-5}{\emph{Phys. Lett.}
  {\bfseries B46} (1973) 109}.

\bibitem{Kallosh:2014wsa}
R.~Kallosh and T.~Wrase, \emph{{Emergence of Spontaneously Broken Supersymmetry
  on an Anti-D3-Brane in KKLT dS Vacua}},
  \href{https://doi.org/10.1007/JHEP12(2014)117}{\emph{JHEP} {\bfseries 12}
  (2014) 117} [\href{https://arxiv.org/abs/1411.1121}{{\ttfamily 1411.1121}}].

\bibitem{Palti:2020tsy}
E.~Palti, \emph{{Fermions and the Swampland}},
  \href{https://arxiv.org/abs/2005.08538}{{\ttfamily 2005.08538}}.

\end{thebibliography}\endgroup

\end{document}